\documentclass[paper, 12pt, letterpaper, epsf]{JHEP} 
\input{epsf.tex}
\usepackage{graphics} 
\usepackage{epsfig} 
\usepackage{amssymb}
\usepackage{graphics}
\usepackage{latexsym,amsmath,amssymb,amscd,cite}
\usepackage{pdfsync}
\usepackage{epsf}
\usepackage[all]{xy}
\usepackage{amsthm}

\newtheorem{Theorem}{Theorem}[section]
\newtheorem{Lemma}[Theorem]{Lemma}
\newtheorem{Proposition}[Theorem]{Proposition}
\theoremstyle{definition}
\newtheorem{Definition}[Theorem]{Definition}
\theoremstyle{remark}

\newcommand{\Hom}{{\rm Hom}}

\newcommand{\A}{{\Bbb{A}}}

\newcommand{\bp}{\begin{Proposition}}
\newcommand{\ep}{\end{Proposition}}
\newcommand{\bl}{\begin{Lemma}}
\newcommand{\el}{\end{Lemma}}
\newcommand{\bt}{\begin{Theorem}}
\newcommand{\et}{\end{Theorem}}
\newcommand{\bd}{\begin{Definition}}
\newcommand{\ed}{\end{Definition}}
\newcommand{\End}{\rm{End}}
\newcommand{\Aut}{\rm{Aut}}

\newcommand{\Mom}{\rm{Mom}}

\newcommand{\iEnd}{{\underline \End}}
\newcommand{\iHom}{{\underline \Hom}}

\DeclareFontFamily{U}{rsf}{}
\DeclareFontShape{U}{rsf}{m}{n}{<5> <6> rsfs5 <7> <8> <9> rsfs7 <10-> rsfs10}{}
\DeclareMathAlphabet\Scr{U}{rsf}{m}{n}

\def\N{{\Bbb N}}
\def\Z{{\Bbb Z}}
\def\C{{\Bbb C}}

\def\deg{{\rm deg}}
\def\Der{{\rm Der}}

\def\lp{\overrightarrow{\partial}}
\def\rp{\overleftarrow{\partial}}
\def\ld{\overrightarrow{\delta}}
\def\rd{\overleftarrow{\delta}}

\def\tf{{\tilde f}}
\def\tom{{\tilde \omega}}
\def\th{{\tilde \theta}}

\newcommand{\be}{\begin{equation}}
\newcommand{\ee}{\end{equation}}
\newcommand{\bea}{\end{eqnarray}}
\newcommand{\eea}{\end{eqnarray}}
\newcommand{\nn}{\nonumber}

\newcommand{\beql}[1]{\begin{eqnarray}\label{eq:#1}}

\newcommand{\eeql}{\end{eqnarray}}

\long\def\begdel#1\enddel{}
\newcommand{\id}{{\rm id}}

\title{On the non-commutative geometry of topological D-branes}
\author{~~~Calin Iuliu Lazaroiu\\Department of Mathematics\\Trinity College Dublin\\Dublin 2, Ireland\\calin@maths.tcd.ie\\}

\abstract{This is a noncommutative-geometric study of the
semiclassical dynamics of finite topological D-brane systems. Starting
from the formulation in terms of $A_\infty$ categories, I show that
such systems can be described by the noncommutative symplectic
supergeometry of $\Z_2$-graded quivers, and give a synthetic formulation of the 
boundary part of the generalized WDVV equations. In particular, a {\em faithful} generating
function for integrated correlators on the disk can be constructed as a
linear combination of quiver necklaces, i.e.  a function
on the noncommutative symplectic superspace defined by the quiver's
path algebra. This point of view allows one to construct extended
moduli spaces of topological D-brane systems as non-commutative
algebraic `superschemes'. They arise by imposing further relations on
a $\Z_2$-graded version of the quiver's preprojective algebra, and
passing to the subalgebra preserved by a natural group of symmetries. }

\preprint{}

\begin{document}

\tableofcontents

\pagebreak

\vskip .6in

\section{Introduction}

The extended moduli spaces \cite{Witten_mirror} of closed topological
strings are Frobenius supermanifolds\cite{Dubrovin}, certain types of
flat Riemannian supermanifolds whose metric is induced by the
generating functional of tree-level closed string amplitudes. This
description follows from the consistency constraints on closed string
amplitudes on the sphere, known as  the WDVV equations
\cite{WDVV}. The theory of Frobenius supermanifolds encodes many
interesting properties such as the generic lack of
obstructions\footnote{A manifestation of this is the
Bogomolov-Tian-Todorov theorem \cite{Tian, Todorov} on
unobstructedness of deformations of complex structure for Calabi-Yau
manifolds, and its generalization to extended deformations due to
Barannikov and Kontsevich \cite{BarKon}.  The extended moduli space of
complex structures for such manifolds coincides with the moduli space
of deformations of the associated topological B-type string. }  of
(not necessarily conformal) topological bulk deformations, and has
well-known applications to closed string mirror symmetry.

It is natural to ask if a similar description can be given for
topological deformations of open strings.  A boundary topological
string theory admits two types of deformations, which are induced by
bulk and boundary observables. As in the closed string case, the
open-closed tree-level amplitudes obey consistency conditions (the
so-called {\em generalized WDVV equations}), which were derived in
\cite{HLL} by worldsheet arguments. While bulk deformations have the
same character as in the boundary-free case, boundary deformations
behave quite differently.  As shown in \cite{HLL}, they are
constrained by a {\em homotopy} version of the associativity
conditions, leading to an intricate structure known as a cyclic and
unital weak $A_\infty$ category (see \cite{CIL5, Stasheff_Kajiura, Kajiura, Gaberdiel, Fukaya_mirror, KS, 
CIL4, CIL8, CIL9, CIL10, Costello1, Costello2} for related
work).  The complication arises due to the non-commutativity of
boundary insertions on the disk, and leads to difficulties
when analyzing the boundary moduli problem.  Among these is the observation that the 
the homotopy associativity constraints for the integrated boundary disk amplitudes
$W_{a_1\ldots a_n}$ cannot be `integrated' faithfully to an ordinary
generating function of the boundary deformation parameters.

In the present paper, we investigate a resolution to this problem, by
arguing quite generally that the semiclassical dynamics of
open strings in finite topological D-brane systems can be described in
the framework of supersymplectic noncommutative geometry. This
approach, which is already implicit in the $A_\infty$ constraints of
\cite{HLL}, leads us to consider the boundary deformation potential as
a function on a {\em noncommutative} space, and allows for a synthetic
formulation of the boundary WDVV equations. Moreover, it leads
naturally to a construction of boundary moduli spaces of topological
D-branes as noncommutative superspaces.

Let us explain this for the simple example of a single topological
D-brane.  In this situation, one could try the following naive proposal
for the generating function: \be \nn
W_{naive}=\sum_{n}{\frac{1}{n}W_{a_1\ldots a_n}\sigma^{a_1}\ldots
\sigma^{a_n}}~~, \ee where one views the boundary deformation
parameters as (super)commuting variables $\sigma^a$.  However,
supercommutativity of the parameters implies that $W_{naive}$ reduces
to: \be\nn \sum_{n}{\frac{1}{n}W_{(a_1\ldots a_n)}\sigma^{a_1}\ldots
\sigma^{a_n}} \ee where $W_{(a_1\ldots a_n)}$ is the (super)
symmetrization of the amplitudes.  Thus a generating function based on
(super) commuting deformation parameters does not faithfully encode
the topological tree-level data of the worldsheet theory, and cannot
generally be used to reconstruct the latter.

It was suggested in \cite{HLL} (see also \cite{Kajiura}) that this
problem might be overcome by viewing the boundary deformation
parameters as {\em non-commuting}.  While this might seem unusual at
first sight, it is in fact quite natural if one recalls that any
boundary theory admits Chan-Paton extensions, whose effect is to
promote the deformation parameters to (super)matrices $X^a$. As a
result, some of the information lost by $W_{naive}$ is preserved by
the matrix potential: \be \nn {\hat
W}=\sum_{n}{\frac{1}{n}W_{a_1\ldots a_n}{\rm str}(X^{a_1}\ldots
X^{a_n})}~~.  \ee However, supertraces of matrix monomials of finite
dimensions generally obey polynomial constraints. As a consequence,
the matrix potential of a fixed Chan-Paton extension can be reduced by
such relations, and again fails to faithfully encode the data of the
theory.  To completely resolve the issue, one has to remove all
constraints on $X^a$, which amounts to replacing them by free
(and in particular non-commuting) supervariables $s^a$. Hence one is lead to the {\em
non-commutative generating function} of \cite{HLL}:
\be
\nn
W=\sum_{n} \frac{1}{n}W_{a_1\ldots a_n}(s^{a_1}\ldots s^{a_n})_c~~,
\ee
where $(.)_c$ denotes the graded-cyclization operation, which gives an
abstract analogue of the supertrace. Notice that 
$W$ allows one to recover any Chan-Paton extension upon
replacing $s^a$ with supermatrices $X^a$, which amounts to considering
a finite-dimensional representation of the free associative
superalgebra $A=\C\langle\{s^a\} \rangle$.  In this way, one can study
at once {\em all} Chan Paton extensions of the theory, as well as more general representations 
obtained by taking morphisms from $A$ to an arbitrary associative superalgebra.

Remarkably, the procedure outlined above agrees with a key
principle of noncommutative algebraic geometry espoused in
\cite{Konts_formal} and developed further in
\cite{LB_lectures} (see \cite{Ginzburg_lectures} for an introduction). 
According to this ideology, `good' notions in
affine non-commutative algebraic geometry should induce the
corresponding classical notions on the moduli spaces of
finite-dimensional representations of the noncommutative coordinate
ring.  In the example above, the free superalgebra $A$ is the
coordinate ring of a noncommutative affine superspace $\A$, while
finite-dimensional representations of this algebra
(i.e. supermatrix-valued points of the `noncommutative scheme' $\A$) 
correspond to Chan-Paton extensions of the theory. By insisting
that the generating function should faithfully encode the information
of integrated amplitudes on the disk, we are lead to consider $W$ as
an element of the cyclic subspace $A_c$ of $A$
(namely the subspace of $A$ spanned by all graded-cyclic
monomials in the generators $s^a$). This matches the interpretation 
\cite{Konts_formal} of $A_c$ as the space of regular 
functions on $\A$.  Thus $W$ is a function on a 
non-commutative affine space, and we find that physics
reasoning agrees with the approach to non-commutative algebraic
geometry advocated in \cite{Konts_formal, LB_lectures}. Moreover, one
can show that the boundary topological metric induces an (even or odd)
noncommutative symplectic form on $\A$, which makes this affine superspace into 
a noncommutative symplectic supermanifold. As
in the supercommutative case, the symplectic structure
determines a bracket $\{.,.\}$ on $A_c$, and one finds that the homotopy associativity
constraints of \cite{HLL} are equivalent with the equation:
\be
\label{BV}
\{W,W\}=0~~.
\ee
Moreover, the unitality condition \cite{HLL, CIL5} on the underlying
$A_\infty$ algebra can also be formulated as
a constraint on $W$. This gives a
non-commutative geometric interpretation of the boundary part of the 
generalized WDVV equations.  As explained in \cite{HLL}, the
boundary topological metric of a general worldsheet theory can be even
or odd; as a consequence, $\{.,.\}$ is an even or odd Lie bracket. In
the latter case, the constraint (\ref{BV}) is a {\em non-commutative}
analogue of the classical master equation.

Non-commutativity of boundary observable insertions is also 
responsible for the fact that deformations of topological D-branes
are generally obstructed, which is reflected in
the typically singular nature of the boundary moduli space. An algebro-geometric approach to
boundary deformations was developed in relation to homological mirror
symmetry \cite{HMS} in \cite{Fukaya_mirror} and related to the deformation theory of open strings 
in \cite{CIL4} (see \cite{Polishchuk} for related work); these proposals
rely on constructing the moduli space as a commutative algebraic
or complex-analytic variety.

The observations made above suggest that the deformation theory of
topological D-branes can be considered as a problem in {\em
noncommutative} geometry.  In particular, the possibility of
Chan-Paton extensions implies that the boundary moduli space can be
viewed as a non-commutative algebraic variety.  This is obtained by
`extremizing' the noncommutative function $W$ and modding out via
appropriate symmetries.  More precisely, one can impose the algebraic
relations:
\be
\label{crit}
\ld_a W=0~~,
\ee 
where $\ld_a$ is a $\Z_2$-graded version of the so-called {\em cyclic
derivatives} of \cite{RSS, Voiculescu} (see also \cite{Konts_formal}).
If $J$ is the two-sided ideal generated by $\ld_a W$, then the
quotient algebra $\C[{\cal Z}]:=\C[\{s^a\}]/J$ can be viewed as the
coordinate ring of a `noncommutative affine scheme' ${\cal Z}$ sitting
inside the affine superspace $\A$. Moreover, one can show that the symmetries of 
the system
form a subgroup ${\cal G}$ of the group of noncommutative symplectomorphisms
of $\A$.  These symmetries also preserve $J$, and thus descend to
automorphisms of $\C[{\cal Z}]$.  They can be viewed as gauge
transformations acting along the `noncommutative vacuum space' ${\cal Z}$.
One can thus define a {\em non-commutative extended moduli space}
${\cal M}$ as the affine `noncommutative scheme' whose coordinate ring
$\C[{\cal M}]=\C[{\cal Z}]^{\cal G}$ is the ${\cal G}$-invariant part
of $\C[{\cal Z}]$.  The existence of a unit observable in the boundary
sector implies that one of the conditions (\ref{crit}) is a
non-commutative moment map constraint in the sense of \cite{BEV,
Bergh}. Therefore, the (invariant theory) quotient leading to ${\cal M}$ amounts to
modding out a zero-level 'symplectic reduction' of $A$ through the
ideal defined by the remaining relations.

It turns out that the construction outlined above can be carried out in much greater generality. 
In fact, as pointed out in \cite{CIL5, HLL}, the
homotopy-associativity constraints on disk boundary amplitudes 
generalize to {\em systems} of D-branes. In
this situation, boundary observables are either boundary-preserving or
boundary condition-changing, a decomposition which defines the
boundary sectors discussed in \cite{CIL1}.  The homotopy associativity constraints of \cite{HLL}
admit an obvious extension to this case, which can be formulated by
saying \cite{CIL5} that the D-brane system defines a (generally weak) cyclic and
unital $A_\infty$ category. The objects $u$ of this category are the
D-branes themselves, while the morphism spaces $\Hom(u,v)=E_{uv}$ are
the spaces of topological observables of strings stretching from $u$
to $v$.  In this case, the boundary deformation parameters $s^a$ are
replaced by $s^i_{uv}$, where $u,v$ run over the
topological D-branes, while
$i$ indexes a basis $\psi^i_{uv}$ of $E_{uv}$ (in fact, $\{s\}$ can
be viewed as a parity-changed dual basis to $\{\psi\}$). Treating
$s^i_{uv}$ as non-commuting supervariables leads one to replace the
free superalgebra $\C\langle \{s^a\}\rangle$ with the associative
superalgebra generated\footnote{Over the subalgebra spanned by the trivial paths.} 
by $s^i_{uv}$ with the relations:
\be
s^i_{uv}s^j_{v'w}=0~~{\rm unless~}v=v'~~.
\ee
This is the the path algebra $A_{\cal Q}$ of a quiver ${\cal Q}$
whose vertices are the D-branes $u$, and whose arrows from $u$ to $v$
are the index triples $(u,v,i)$ associated to $s^i_{uv}$. This quiver is $\Z_2$-graded since $s^i_{uv}$ can be
even or odd; as a consequence, $A_{\cal Q}$ is an
associative superalgebra.  The case of a single D-brane corresponds to
a quiver with a single vertex $u$, whose path algebra coincides with
the free superalgebra generated by the $\Z_2$-graded loops at that vertex.

It is known that path algebras of quivers are formally smooth in the
sense of \cite{CQ}, and in fact they provide good non-commutative
generalizations of the coordinate rings of smooth affine varieties
\cite{LB_lectures}.  One can formulate a non-commutative symplectic
geometry for such spaces \cite{Ginzburg, LB_quivers} by extending the
construction of \cite{Konts_formal}.  For a general topological D-brane system,
the space $\A_{\cal Q}$ with coordinate ring $A_{\cal Q}$ 
is a noncommutative {\em super}space, whose symplectic
form is induced by the parity change of the boundary topological
metrics.  As for affine superspaces, one finds an even or odd Lie
superbracket $\{.,.\}$ which acts on the space $C^0_R(A_{\cal Q})$ of
regular functions on $\A_{\cal Q}$.  The latter can be viewed as the
vector superspace spanned by necklaces of the quiver (i.e. cycles of
the quiver whose marking by the start=endpoint is forgotten). The
generating function $W$ of integrated disk boundary amplitudes is an
element of this space, i.e. a linear combination of such necklaces;
its parity is opposite that of the boundary topological metrics. The
categorical $A_\infty$ constraints for the entire collection of
boundary operators amount to equation (\ref{BV}). The existence of
unit observables in the boundary-preserving sectors can be expressed
as a differential constraint on $W$. Once again, a noncommutative
moduli space can be constructed as the invariant theory quotient of the
noncommutative critical variety ${\cal Z}$ through the symmetries of
the system.  The existence of boundary unit observables implies that
this quotient can be be viewed as a noncommutative 'symplectic
reduction', followed by imposing further constraints.

The paper is organized as follows. In Section 2, we describe the
algebraic structure of finite D-brane systems, starting from the
formulation in terms of $A_\infty$ categories found in \cite{HLL,
CIL5} (which is summarized in Appendix \ref{data}). Upon introducing a
finite-dimensional commutative semisimple algebra $R$, we show that
one can express the boundary sector decomposition of \cite{CIL1} as an
$R$-superbimodule structure on the total space
$E=\oplus_{u,v}{\Hom(u,v)}$ of boundary observables. As a consequence,
the cyclic and unital weak $A_\infty$ category determined by
integrated string correlators can be described as a cyclic and unital
weak $A_\infty$ algebra on the superbimodule $E$.  We also summarize
the strategy which will be followed in later sections in order to
encode this data into a noncommutative generating function. The main
step is passing from the superbimodule $E$ to the tensor algebra
$A=T_R (E[1]^{\rm v})$ of its parity-changed dual, which is a
superalgebra over $R$. The boundary topological metrics of the theory
induce an (even or odd) noncommutative symplectic form on $A$. This
leads us to study the noncommutative symplectic geometry of
$R$-superalgebras, a subject which is addressed in Section 3.  In
Section 4, we apply this to the tensor algebra $A$, discussing the
realization of the abstract objects introduced previously. Upon
picking appropriate bases in the space of boundary observables, we show that $A$
can be identified with the path algebra of a $\Z_2$-graded quiver, and
give coefficient expressions for the various quantities of
interest. We also discuss the quiver version of cyclic derivatives and
the so-called loop partial derivatives, two types of operators acting
on the space of noncommutative functions. In Section 5, we apply the
machinery developed in Sections 3 and 4 to finite D-brane
systems. Using the results of Section 2, we show that the cyclic and
unital weak $A_\infty$ structure on $E$ is encoded faithfully by a
noncommutative generating function $W$, which can be expressed as a
linear combination of quiver necklaces. After discussing the
constraints obeyed by $W$, we say a few words about deformations of
the underlying string theory (as opposed to deformations of the
boundary data). In Section 6, we discuss symmetries of the topological
D-brane system and give the algebraic construction of the
noncommutative moduli space. Section 7 gives the realization of our
construction in the case of a single D-brane. In Section 8,
we discuss the simplest examples with an even and odd boundary topological
metric, and a rather general class of examples with odd boundary metrics, some
particular cases of which appeared recently in \cite{Katz}. Section 9
contains our conclusions.

\paragraph{Conventions}
Unless specified otherwise, an algebra means an associative and unital algebra over the complex numbers. 
All morphisms of algebras are assumed to be unital. We will often encounter $\Z\times \Z_2$-graded algebras. To fix the
sign convention for such algebras, one must chose a $\Z_2$-valued pairing on the
Abelian group $\Gamma:=\Z\times \Z_2$, i.e.  a biadditive and
symmetric map $\cdot:\Gamma\times \Gamma\rightarrow \Z_2$ specifying
how the Koszul rule is applied to bigraded objects. Namely,  one agrees
that commuting two objects of bidegrees $(n,\alpha)$ and $(m,\beta)$
always produces the sign $(n,\alpha)\cdot (m,\beta)$.  In the present
paper, we work with the following choice of pairing:
\be
\label{pairing}
(n,\alpha)\cdot (m,\beta)=[mn]+\alpha\beta~~,
\ee
where here and in the rest of the paper the notation $[k]$ for $k\in
\Z$ stands for the ${\rm mod}~2$ reduction of $k$. If $\alpha$ is an
element of $\Z_2$, we let $(-1)^{\alpha}=+1$ if $\alpha=0$ and $-1$ if
$\alpha=1$. We let $[1]$ be the parity change functor on the category
of super-vector spaces and $\Sigma_U$ the suspension operator of
the supervector space $U$ (i.e. $\Sigma:U\rightarrow U[1]$ is the
identity operator of $U$, viewed as an odd map from $U$ to itself).
We have $\Sigma_U^2=\id_U$ and $[1]^2\approx {\rm Id}$, where ${\rm Id}$ in the
last relation is the identity functor. We will often write $\Sigma$
without indicating the vector space on which it acts, since the latter
is usually clear from the context. Given a ring $R$, we let ${\rm
Mod-R}$, ${\rm R-Mod}$ and ${\rm R-Mod-R}$ and  ${\rm
Smod-R}$, ${\rm R-Smod}$ and ${\rm R-Smod-R}$ 
be the categories of left-, right- and bi- modules, respectively supermodules over $R$.  
For a pair $U,V$ of
objects of any of these categories, we let $\Hom(U,V)$ be the space of
morphisms from $U$ to $V$ in that category. For supermodules, this consists of
degree zero $R$-linear maps and is an ordinary (i.e. ungraded) module.  For two supermodules,
we also let $\iHom(U,V)$ be the space of morphisms in the corresponding
category ${\rm Mod-R}$, ${\rm R-Mod}$ or ${\rm R-Mod-R}$ of {\em
ungraded} objects, obtained by forgetting the $\Z_2$-grading of
$U,V$. The latter consists of all linear maps, without degree
conditions; it is known as the {\em inner} morphism space and is
$\Z_2$-graded. In fact, we have $\iHom(U,V)=\iHom^0(U,V)\oplus
\iHom^1(U,V)$, where the degree zero component coincides with the
space of degree zero maps, $\iHom^0(U,V)=\Hom(U,V)$. We will use the same notational 
convention for the endomorphism and automorphism spaces, for example $\End(U)$ 
is the endomorphism space of $U$ as a supermodule and $\iEnd(U)$ its inner endomorphism 
space etc. Given an $R$-superbimodule $U$, 
we set $T_R U:=\oplus_{n\geq 0}{U^{\otimes_R n}}$ (with $U^{\otimes 0}:=R$), the tensor 
algebra of $U$, viewed as an $\N\times \Z_2$ graded algebra.

{\bf Note} Throughout this paper, an $A_\infty$ algebra or category will mean 
an $A_\infty$ algebra/category whose sequence of defining products $(r_n)$ terminates 
(i.e. $r_n=0$ for sufficiently large $n$).  This condition is purely technical, 
being needed if one wishes to pass 
to the dual of a coalgebra in naive manner. The condition can be 
removed in standard fashion, by considering profinite modules and formal algebras and 
replacing the relevant maps by continuous maps; then the non-commutative geometric 
objects mentioned above can  be understood in the formal sense. Because most of our 
considerations generalize straightforwardly to formal case, we adopted the convention of 
sometimes writing finite sums without indicating the upper bound; in this paper, it is 
understood that all such sums terminate. When we write a finite sum this manner, it is 
implied that the sum can be extended to a series in the formal theory.

\section{Algebraic description of finite topological D-brane systems} 
\label{algebraic}

In this section, we consider finite D-brane systems in an open
topological string theory, i.e.  finite collections of D-branes,
together with the spaces of boundary observables of all topological
strings stretched between them. We will assume that the total space of
zero-form boundary observables is finite-dimensional, which is the usual case
for topological strings. Using a description derived in  \cite{CIL5, HLL}, we 
will encode the information of all tree-level boundary string amplitudes into 
an algebraic structure on a certain superbimodule built out of the total space 
of zero-form boundary observables.

The results of  \cite{CIL5, HLL} imply that D-brane systems in topological string theory are
described by $A_\infty$ categories.  The precise statement is 
as follows (see Appendix \ref{data} for mathematical background):

\

{\em A topological D-brane system is described by a weak, cyclic and
unital $A_\infty$ category ${\cal A}$.}

\

\noindent 

In general, the $A_\infty$ category is only $\Z_2$-graded. This
grading can be lifted to a $\Z$-grading provided that the relevant
$U(1)$ symmetry of the worldsheet theory is non-anomalous.  The
objects of ${\cal A}$ are the D-branes themselves, while the morphism
space $ \Hom_{\cal A}(u,v)$ for two objects $u,v$ is the complex
supervector space of boundary zero-form observables for the
topological string stretching from $u$ to $v$. The degree of such
observables is given by worldsheet Grassmann parity, and will be
denoted by $|.|$. With respect to this grading, the $A_\infty$
products have degrees $[n]$, where the square bracket indicates mod 2
reduction.  As in Appendix \ref{data}, it is convenient to work with
the parity-changed spaces $\Hom_{\cal A}(u,v)[1]$, the degree of whose
elements we denote by a tilde. Hence ${\tilde x}=[|x|+1]$ for all
homogeneous morphisms $x$.  The unitality condition involves even
elements $1_u$ of the endomorphism spaces $\Hom_{\cal A}(u,u)$
(equivalently, their odd suspensions $\lambda_u:=\Sigma 1_u\in \Hom_{\cal
A}(u,u)[1]$), while the cyclicity constraint involves non-degenerate
$\Z_2$-homogeneous pairings (the boundary topological metrics)
$\rho_{uv}:\Hom_{\cal A}(u,v)\times \Hom_{\cal A}(v,u)\rightarrow \C$
of common degree ${\tilde \omega}\in \Z_2$, obeying the
graded-symmetry condition $\rho_{uv}(x,y)=(-1)^{|x||y|}\rho_{vu}(y,x)$
for homogeneous elements $x\in \Hom_{\cal A}(u,v)$ and $y\in
\Hom_{\cal A}(v,u)$. The latter can also be expressed in
terms of the graded-antisymmetric forms $\omega_{uv}:=\rho_{uv}\circ
\Sigma^{\otimes 2}:\Hom_{\cal A}(u,v)[1]\times \Hom_{\cal
A}(v,u)[1]\rightarrow \C$ obtained from the topological metrics via
suspension. The detailed formulation of this structure (which arises
by introducing boundary sectors in the derivation of \cite{HLL}) is
given in Appendix \ref{data}.

\paragraph{Observation}

It was argued in \cite{CIL5} (see \cite{CIL2, CIL3} for the dG case) that the $A_\infty$ category
${\cal A}_{full}$ obtained by considering `all' D-branes of a
topological string theory must be endowed with a parity-change functor
and be equivalent with its category of twisted complexes. This encodes
the physical requirement that the collection of all topological
D-branes of a given theory is closed under formation of topological
D-brane composites.  This `quasiunitarity constraint' implies that the
cohomology category $H^0({\cal A}_{full})$ is an enhanced triangulated
category (with an $A_\infty$ enhancement).  In the present paper, we
work with a fixed $A_\infty$ sub-category ${\cal A}$ of the full
D-brane category ${\cal A}_{full}$, so ${\cal A}$ need not obey this
supplementary condition (which is impossible to satisfy with a finite collection of objects).

\

For the remainder of this paper, we focus on finite D-brane systems,
which means that $Ob{\cal A}$ is a finite set and all morphism spaces
$\Hom(u,v)$ are finite-dimensional. In this case, we say that the $A_\infty$ category ${\cal A}$ is {\em finite}.

\subsection{Encoding the boundary sector decomposition}

One can encode the categorical data of ${\cal A}$ 
in an equivalent, but more amenable form. Setting ${\cal Q}_0:=Ob{\cal A}$, 
consider the $\Z_2$-graded vector space 
$E:=\oplus_{u,v\in {\cal Q}_0}{\Hom(u,v)}$, which is known as 
the total boundary space of the D-brane system \cite{CIL1}.

\

\noindent {\bf Definition}
A {\em binary homogeneous decomposition} over the set ${\cal Q}_0$ is
a pair $(U, (U_{uv})_{u,v\in {\cal Q}_0})$ such that $U, U_{uv}$ are
finite-dimensional complex supervector spaces and $U=\oplus_{(u,v)\in {\cal Q}_0\times {\cal
Q}_0}{U_{uv}}$ is a homogeneous decomposition of $U$ indexed by the
Cartesian product ${\cal Q}_0\times {\cal Q}_0$. The {\em opposite} of the binary decomposition 
$(U, (U_{uv})_{(u,v)\in {\cal Q}_0\times {\cal Q}_0})$ is the binary homogeneous decomposition 
$(U, (U^{opp}_{uv})_{(u,v)\in {\cal Q}_0\times{\cal Q}_0})$, where $U^{opp}_{uv}=U_{vu}$.

\

A morphism of binary homogeneous decompositions over ${\cal Q}_0$ from
$(U, (U_{uv})_{(u,v)\in {\cal Q}_0\times {\cal Q}_0})$ to 
$(U',(U'_{uv})_{(u,v)\in {\cal Q}_0\times {\cal Q}_0})$
is a degree zero linear map $\phi:U\rightarrow U'$ such that
$\phi(U_{uv})\subset U'_{uv}$ for all $(u,v)\in {\cal Q}_0\times {\cal Q}_0$. 
With this definition, binary
homogeneous decompositions over ${\cal Q}_0$ form a category. A morphism
$\phi$ in this category is an isomorphism iff it is a bijective
map. In this case, we have $\phi(U_{uv})=U'_{uv}$ for all $u,v\in
{\cal Q}_0$.

A topological D-brane system determines two opposite decompositions of its total boundary space, 
namely $E=\oplus_{(u,v)\in {\cal Q}_0\times {\cal Q}_0}{\Hom_{\cal A}(u,v)}$ and 
$E=\oplus_{(u,v)\in {\cal Q}_0\times {\cal Q}_0}{\Hom_{\cal A}(v,u)}$. We will view 
the first of these  as fundamental, so we set $E_{uv}:=\Hom_{\cal A}(u,v)$; then 
$E_{uv}^{opp}:=\Hom_{\cal A}(v,u)$. This is consistent with the convention that morphisms 
compose forward in the definition of an $A_\infty$ category given in Appendix \ref{data}.  
The binary homogeneous decomposition $(E, (E_{uv})_{(u,v)\in {\cal Q}_0\times {\cal Q}_0})$
is  the so-called {\em boundary sector decomposition} of the topological D-brane system \cite{CIL1}. 

Let us consider the finite-dimensional semisimple commutative algebra $R=\oplus_{u\in {\cal Q}_0}{\C}$,
where the right hand side is a direct sum of copies of $\C$, viewed as
an algebra over itself.  We let $\epsilon_u$ $(u\in {\cal Q}_0$) be
the commuting idempotents of $R$ corresponding to the canonical basis
elements of the vector space $\C^{{\cal Q}_0}$, so $R=\oplus_{u\in
{\cal Q}_0}{\C \epsilon_u}$ with
$\epsilon_u\epsilon_v=\delta_{uv}\epsilon_u$.  The unit of $R$ is
$1_R=\sum_{u\in {\cal Q}_0}{\epsilon_u}$. It is not hard to
see that a binary
homogeneous decomposition $U=\oplus_{(u,v)\in {\cal Q}_0\times {\cal Q}_0 }{U_{uv}}$
amounts to giving an $R$-superbimodule structure on $U$. 
Indeed, such a decomposition amounts to giving degree
zero commuting idempotents $\epsilon_u^l, \epsilon_u^r\in \End_\C(U)$
for all $u\in {\cal Q}_0$, namely the projectors on the subspaces
$\oplus_{v}{U_{uv}}$ and $\oplus_{v}{U_{vu}}$ respectively (then
$U_{uv}=\epsilon_u^l\epsilon_v^r(U)=\epsilon_v^r\epsilon_u^l(U)$). Notice
that $U_{uv}$ can be recovered as $U_{uv}=\epsilon_uU\epsilon_v$ from
knowledge of the $R$-superbimodule structure.  
Moreover, a
morphism of binary homogeneous decompositions over ${\cal Q}_0$
amounts to a morphism of $R$-superbimodules.  Hence the category of
binary homogeneous decompositions over ${\cal Q}_0$ is equivalent with
the category of superbimodules over $R$.

Since $R$ is commutative, any $R$-superbimodule $U$ defines another $R$-superbimodule $U^{opp}$ 
whose underlying supervector space coincides with that of $U$ but whose external multiplications 
are given by:
\be
\nn
\alpha*x*\beta=\beta x\alpha~~~~~~\forall \alpha,\beta\in R~~\forall x\in U~~.
\ee
The relations $\epsilon_u*E*\epsilon_v=\epsilon_v E \epsilon_u$ show that the binary decomposition 
determined by $U^{opp}$ is the opposite of the binary decomposition determined by $U$:
\be
\nn
(U^{opp})_{uv}=U_{uv}^{opp}=U_{vu}~~.
\ee

Applying this to a topological D-brane system, we find that the boundary 
sector decomposition $E=\oplus_{(u,v)\in {\cal Q}_0\times {\cal Q}_0}{E_{uv}}=
\oplus_{(u,v)\in {\cal Q}_0\times {\cal Q}_0}{\Hom_{\cal A}{(u,v)}}$ 
is encoded by an $R$-superbimodule structure on $E$. Moreover, the opposite decomposition 
$E=\oplus_{(u,v)\in {\cal Q}_0\times {\cal Q}_0}{\Hom_{\cal A}{(v,u)}}$ is encoded by the opposite 
superbimodule $E^{opp}$. These observations allow one to encode the category-theoretic
structure of $A_\infty$ products into compatibility with the
$R$-superbimodule structure of $E$. Before stating the relevant
result, we need a few more preparations.

Given an $R$-superbimodule $U$, its dual $U^{\rm v}:=\Hom_{\rm R-mod} (U,R)=\iHom_{\rm R-Smod}(U,R)$ 
as a left $R$-module becomes an $R$-superbimodule 
with respect to the external multiplications defined through:
\be
\label{dual_multiplications}
(\alpha f \beta)(x):=f(\alpha x \beta)=\alpha f(x\beta) ~~.
\ee
We warn the reader that the usual definition of an $R$-superbimodule structure on $U^{\rm v}$ corresponds 
to the opposite of that given in (\ref{dual_multiplications}). We adopted the convention 
(\ref{dual_multiplications}) in order to avoid notational morass later on. With this definition, some of 
the usual isomorphisms involve taking the opposite of certain superbimodules, as we explain in Appendix 
\ref{isomorphisms}; in return, the formulas of Section 4,5 and 6 simplify considerably.

It is not hard to see that the binary homogeneous decomposition
$U^{\rm v}=\oplus_{(u,v)\in {\cal Q}_0\times {\cal Q}_0}{(U^{\rm v})_{uv}}$ determined by
the $R$-superbimodule structure (\ref{dual_multiplications})  has components: 
\be
\label{dual_spaces}
(U^{\rm v})_{uv}=(U_{uv})^*~~,
\ee
where $(U_{uv})^*:=\iHom_\C(U_{uv},\C)$ is the linear dual of $U_{uv}$ viewed as a supervector space.
Notice that there is no reversal of the positions of $u$ and $v$ in relation (\ref{dual_spaces}).

With the definition (\ref{dual_multiplications}), a homogeneous
$R$-bilinear form\footnote{Recall that a multilinear map $f$ of $R$-superbimodules is required to satisfy
$f(\alpha x_1, x_2\ldots x_{n-1}, x_n\beta)=\alpha f(x_1\ldots
x_n)\beta$ as well as the balance condition $f(x_1\ldots
x_{j-1}\alpha, x_j\ldots x_n)=f(x_1\ldots x_{j-1}, \alpha x_j\ldots
x_n)$ for all $j$, where $\alpha,\beta\in R$. 
In particular, a bilinear map $\sigma:U\times U\rightarrow R$ 
satisfies $\sigma (x\alpha,y)=\sigma (x,\alpha y)$ and $\sigma (\alpha x, y\beta)=\alpha\sigma(x,y)\beta$.}
 $\sigma:U\times U\rightarrow R$ of degree ${\tilde \sigma}$ induces
an $R$-superbimodule map $U^{opp}\stackrel{j_\sigma}{\rightarrow}U[{\tilde \sigma}]^{\rm v}$ given
by $x\rightarrow f_x(.):=\sigma(\cdot, x)$.  
The form is
called {\em graded-symmetric} if
$\sigma(x,y)=(-1)^{\deg x \deg y}\sigma(y,x)$ for all homogeneous $x,y\in U$
and {\em graded-antisymmetric} if
$\sigma(x,y)=(-1)^{1+\deg x \deg y}\sigma(y,x)$.  In any of these case, it is called {\em
nondegenerate} if the map $x\rightarrow f_x$ is an isomorphism of vector spaces.

A graded-symmetric form $\rho$ on
$U$ induces a graded-antisymmetric form $\omega=\rho\circ
\Sigma^{\otimes 2}$ on $U[1]$, given explicitly by:
\be
\nn
\omega(\Sigma x,\Sigma y)=(-1)^{\tilde x}\rho(x,y)~~~~~~\forall x,y \in U~~,
\ee
where the sign prefactor is due to the Koszul rule (the
suspension operator $\Sigma$ is odd). If $\rho$ is homogeneous, then
$\omega$ is homogeneous of the same $\Z_2$ degree. Hence giving a
graded-symmetric form on $U$ is equivalent to giving a graded-antisymmetric 
form on $U[1]$. Moreover, $\rho$ is non-degenerate iff
$\omega$ is.  A non-degenerate, homogeneous and graded-symmetric
$R$-bilinear form will be called a {\em metric}, while a
non-degenerate, homogeneous and graded-antisymmetric R-bilinear form
will be called a {\em symplectic form}.  A $R$-superbimodule is called
a {\em metric superbimodule} if it is endowed with a metric, and a {\em
symplectic superbimodule} if it is endowed with a symplectic form.  Metric
and symplectic $R$-superbimodules form categories if one defines
morphisms in the obvious fashion.  Notice that parity change induces
an idempotent equivalence between these categories. This reflects the
general principle that metric and symplectic superdata are related
through parity change.  In particular, $(U[1],\omega)$ is a symplectic
R-superbimodule iff $(U,\rho)$ (with $\omega=\rho\circ \Sigma^{\otimes
2}$) is a metric R-superbimodule.

Giving an $R$-bilinear form $\sigma$ on $U$ amounts to giving $\C$-bilinear forms 
$\sigma_{uv}:U_{uv}\times U_{vu}\rightarrow \C$ for all $u,v$. Indeed, $R$-bilinearity 
of $\sigma$ implies:
\be
\nn
\sigma(x\epsilon_v, y)=\sigma(x, \epsilon_vy)~~{\rm and}~~\sigma(\epsilon_u x, y\epsilon_w)=
\sigma(x,y)\epsilon_u\epsilon_w~~.
\ee
Writing $x=\sum_{u,v}{x_{uv}}$ and $y=\sum_{u,v}{y_{uv}}$ with $x_{uv}:=\epsilon_u x \epsilon_v\in U_{uv}$ and 
$y_{uv}:=\epsilon_u y\epsilon_v\in U_{uv}$, this shows that $\sigma(x_{uv}, y_{u'v'})$ vanishes unless $v=u'$ and 
$u=v'$. Hence $\sigma$ is completely determined by its restrictions $\sigma'_{uv}$ to the subspaces 
$U_{uv}\times U_{vu}$. Explicitly, we have $\sigma(x,y)=\sum_{u,v}\sigma'_{uv}(x_{uv},y_{vu})$. 
Each restriction takes $U_{uv}\times U_{vu}$ into the one-dimensional subspace $\C \epsilon_u$ of $R$, so 
we can write $\sigma'_{uv}=\sigma_{uv}\epsilon_u$ for some complex-linear maps 
$\sigma_{uv}:U_{uv}\times U_{vu}\rightarrow \C$. Then: 
\be
\label{sigma_dec}
\sigma(x,y)=\sum_{u,v}\sigma_{uv}(x_{uv},y_{vu})\epsilon_u
\ee
and $\sigma$ is completely determined by $\sigma_{uv}$. Conversely,
any family of complex-bilinear maps $\sigma_{uv}:U_{uv}\times
U_{vu}\rightarrow \C$ determines an $R$-bilinear map $\sigma$ through
relation (\ref{sigma_dec}).  The map $U\stackrel{j_\sigma}{\rightarrow} U^{\rm v}$ induced by $\sigma$ 
has the property $j_\sigma(U^{\rm opp}_{uv})\subset (U^{\rm v})_{uv}$ i.e. 
$j_\sigma(U_{vu})\subset (U_{uv})^*$, and its restrictions to the subspaces $U_{vu}$ can be identified 
with the maps $U_{vu}\rightarrow (U_{uv})^*$ induced by $\sigma_{uv}$. 
Thus $\sigma$ is non-degenerate iff $\sigma_{uv}$ are (the later
means that all maps $U_{vu}\rightarrow (U_{uv})^*$ determined by
$\sigma_{uv}$ are linear isomorphisms). It is also clear that $\sigma$ is
graded-symmetric iff $\sigma_{uv}(x,y)=(-1)^{{\tilde x}{\tilde
y}}\sigma_{uv}(y,x)$ for all homogeneous $x\in U_{uv}$ and $y\in
U_{vu}$; a similar statement holds for graded-antisymmetric forms.

For a topological D-brane system, we have $E_{uv}=\Hom_{\cal
A}(u,v)$ and the topological metrics $\rho_{uv}:\Hom_{\cal
A}(u,v)\times \Hom_{\cal A}(v,u)\rightarrow \C$ are homogeneous of common
degree ${\tilde \omega}$, so they determine an $R$-bilinear form
$\rho$ on $E$ of the same degree.  Moreover, the graded-symmetry and
non-degeneracy conditions for $\rho_{uv}$ amount to the condition that
$\rho$ is a superbimodule metric on $E$.  Similarly, the symplectic
forms $\omega_{uv}=\rho_{uv}\circ \Sigma^{\otimes 2}$ determine a
superbimodule symplectic form on $E[1]$ of degree ${\tilde \omega}$.

A {\em weak $A_\infty$ structure} on a $R$-superbimodule $U$ is a
countable family of odd $R$-linear maps $r_n:U[1]^{\otimes
n}\rightarrow U[1]$ (equivalently, odd $R$-multilinear maps
$r_n:U[1]^n\rightarrow U[1]$, which we denote by the same letters)
with $n\geq 0$, which satisfy the conditions\footnote{There is
some ambiguity in the sign conventions for various objects
related to $A_\infty$ algebras and categories. In this paper, we view the homological
derivation $Q$ discussed in Section \ref{geometrization} as the
fundamental object, and have defined $r_n$ such that most signs
simplify. We refer the reader to \cite{Penkava} for a discussion of
other conventions.}:
\be
\label{ainf}
\sum_{0\leq i+j\leq n} (-1)^{{\tilde x}_1+\ldots +{\tilde x}_i} r_{n-j+1}(x_1\ldots x_i, r_j(x_{i+1}
\ldots x_{i+j}), x_{i+j+1}\ldots x_n)=0~~
\ee
for all $n\geq 0$. In these relations, it is understood that $x_1,
\ldots, x_n$ are arbitrary homogeneous elements of $U[1]$.  The
structure is called {\em strong}, if $r_0=0$ and {\em minimal} if
$r_0=r_1=0$.

We say that a weak $A_\infty$ structure on $U$ is {\em
unital} if there exists an even element $1\in U^R$ such
that its odd suspension $\lambda =\Sigma 1\in U[1]^R$ satisfies the
following conditions:
\begin{eqnarray}
\label{unitality}
r_n(x_1\ldots x_{j-1},\lambda,x_{j+1}\ldots x_n)&=&0~~{\rm for~all}~~~~ n\neq 2 ~{\rm~and~all}~ j~~\nn\\
-r_2(\lambda,x)=(-1)^{{\tilde x}} r_2(x,\lambda)&=&x~~,
\end{eqnarray}
for all homogeneous elements $x,x_j$ of $U$.  In this case, $1$ is
called the {\em even unit} of the $A_\infty$ structure, while
$\lambda$ will be called the {\em odd unit}. It is not hard to
see that the unit of a unital $A_\infty$
structure is unique. Indeed, given another unit $1'$, set $\lambda':=\Sigma
1'$. Then $r_2(\lambda,\lambda')=-\lambda=-\lambda'$, where we used
the second row in (\ref{unitality}) by viewing either $\lambda$ or
$\lambda'$ as the unit. This implies $1=1'$. 

Recall that the {\em center} $U^R$ of an $R$-superbimodule $U$ 
(a.k.a the centralizer of $R$ in $U$) is the 
homogeneous sub-bimodule consisting of all central elements $x\in U$, i.e. those
elements which satisfy $\alpha x=x\alpha$ for all $\alpha\in R$.  In
terms of the homogeneous binary decomposition, we have
$U^R=\oplus_{u\in {\cal Q}_0}{U_{uu}}$.  Thus a central element has
the form $x=\oplus_{u\in {\cal Q}_0}{x_u}$, with
$x_u=\epsilon_ux\epsilon_u=\epsilon_u x=x\epsilon_u\in U_{uu}$. Since
$R$ is semisimple, we have a direct sum decomposition of vector spaces:
\be
\label{CQ_decomp}
U=U^R\oplus [R,U]~~,
\ee
where $[R,U]$ is the complex supervector space generated by the
commutators $[\alpha,x]=\alpha x-x\alpha$ with $\alpha\in R$ and $x\in
U$. This follows as in \cite{CQ} due to the fact that $\oplus_{u\in
Q_0}{\epsilon_u\otimes \epsilon_u}$ is a so-called separability
element. It can also be seen directly by using the identities
$x-\sum_{u}{\epsilon_u
x\epsilon_u}=\frac{1}{2}\sum_{u}{[\epsilon_u,[\epsilon_u, x]]}$ and
$\epsilon_u[\epsilon_v, x]\epsilon_u=0$, which
hold for any element $x\in U$.

Semisimplicity of $R$ implies that the unit of a unital $A_\infty$ structure 
on an $R$-superbimodule $U$ must be central. To see this, let $\lambda$ be the odd 
$A_\infty$ unit and consider the central element $\lambda'=\sum_{u}{\epsilon_u \lambda \epsilon_u}$.
Then the last row in (\ref{unitality}) implies: 
\be
\nn
r_2(x,\lambda')=\sum_{u} r_2(x,\epsilon_u \lambda \epsilon_u)=
\sum_{u} r_2(x\epsilon_u, \lambda )\epsilon_u=(-1)^{\tilde x} \sum_u x\epsilon_u =(-1)^{\tilde x}x~~
\ee
and: 
\be
\nn
r_2(\lambda',x)=\sum_{u} r_2(\epsilon_u \lambda \epsilon_u, x)=
\sum_{u} \epsilon_u r_2(\lambda, \epsilon_u x)=- \sum_u \epsilon_u x=-x~~,
\ee
where we used $R$-bilinearity of $r_2$ and the equation $\sum_u \epsilon_u=1$. These two equations 
imply $\lambda=\lambda'$ by the argument used above to show unicity of the $A_\infty$ unit. 
This shows that $\lambda$ must be central.

A weak $A_\infty$ structure $(r_n)$ on $U$ is
called {\em cyclic} if $U[1]$ is endowed with a homogeneous symplectic form
$\omega$, such that the following relations are satisfied:
\be
\label{rrcyc}
\omega(x_0,r_n(x_1\ldots x_n))=(-1)^{{{\tilde x}_0+{\tilde x}_1+\tilde x}_0
({\tilde x}_1+\ldots +{\tilde x}_n)}\omega(x_1,r_n(x_2\ldots x_n, x_0))~~.
\ee
In terms of the metric $\rho$ determined by $\omega=\rho\circ \Sigma^{\otimes 2}$,
these relations take the following form, which might be more
familiar to some readers: \be
\label{rcyc}
\rho(x_0,r_n(x_1\ldots x_n))=(-1)^{{\tilde x}_0({\tilde x}_1+\ldots
+{\tilde x}_n)}\rho(x_1,r_n(x_2\ldots x_n, x_0))~~.  \ee

\

\noindent We can now state a basic equivalence: 
\paragraph{\bf Proposition}
Giving a finite weak cyclic and unital $A_\infty$ category with object
set ${\cal Q}_0$ amounts to giving a weak, cyclic and unital
$A_\infty$ structure on a $\Z_2$-graded $R$-superbimodule $E$ of finite
complex dimension, over the finite-dimensional semisimple commutative
algebra $R=\oplus_{u\in {\cal Q}_0}\C \epsilon_u$.

\

\noindent 

In view of this proposition, one can {\em define} finite tree-level
topological D-brane systems to be cyclic and unital weak $A_\infty$
structures on some $R$-superbimodule of finite dimension over $\C$, where $R$ is a 
finite-dimensional semisimple commutative $\C$-algebra. The decomposition of the
system into constituent D-branes and the decomposition of the total
boundary space $E$ into boundary sectors
$E_{uv}$ are both encoded by the $R$-superbimodule structure.

\paragraph{Sketch of proof}
As explained above, the superbimodule structure on $E$ amounts to a
homogeneous decomposition $E=\oplus_{(u,v)\in {\cal Q}_0\times {\cal
Q}_0}{E_{uv}}$, where $E_{uv}:=\Hom_{\cal A}(u,v)$.  The rest of the
proof is a lengthy but straightforward check of conditions, showing
that compatibility of various maps with the $R$-superbimodule
structure of $E$ allows one to translate superbimodule $A_\infty$ data
into the categorical $A_\infty$ data listed in Appendix \ref{data}.
We already showed above that giving the categorical bilinear forms $\rho_{uv}$
amounts to giving a metric $\rho$ on this superbimodule.  Similarly,
giving an $R$-multilinear map $r_n:E^n\rightarrow E$ amounts to giving
$\C$-multilinear maps $r_{u_1\ldots u_{n+1}}: E_{u_1 u_2}\times E_{u_2
u_3}\times \ldots \times E_{u_n u_{n+1}}\rightarrow E_{u_1u_{n+1}}$,
and it is clear that the weak $A_\infty$ constraints for $r_n$ amount
to the categorical weak $A_\infty$ constraints for these maps.
Moreover, the cyclicity conditions for $r_n$ with respect to
$\omega=\rho\circ\Sigma^{\otimes 2}$ amount to the categorical
cyclicity constraints with respect to $\omega_{uv}:=\rho_{uv}\circ
\Sigma^{\otimes 2}$.  Finally, the $A_\infty$ units $1_u$ of ${\cal
A}$ give an even central element $1=\oplus_{u}{1_u}$ in the
superbimodule $E$, which is a unit for the superbimodule 
$A_\infty$ structure. Conversely, giving such a unit amounts
to giving elements $1_u$ in each `diagonal' subspace $E_{uu}$, since 
the unit of an $R$- superbimodule $A_\infty$ structure must be central. 
The unitality constraints for $r_n$ amount to the categorical unitality constraints
for $r_{u_1\ldots u_{n+1}}$.

\paragraph{Observation} When the $A_\infty$ structure $(r_n)$ is minimal, the first 
non-trivial $A_\infty$ constraint implies that the product
$\cdot=\Sigma\circ r_2\circ \Sigma^{\otimes 2}$ is associative.
Moreover, cyclicity and unitality imply that the triple $(E,\cdot
,\rho)$ is a (non-commutative) Frobenius algebra, whose multiplication and 
bilinear form are $R$-bilinear. On the other hand, forgetting all higher products of ${\cal A}$ gives a usual
(i.e. associative and unital) category endowed with non-degenerate and
graded symmetric bilinear pairings $\rho_{uv}$ between opposite spaces
of morphisms. As explained in \cite{CIL1}, such a category describes
the boundary part of a two-dimensional topological field theory
defined on bordered Riemann surfaces.  This gives the following:

\paragraph{\bf Corollary} The boundary sector of a topological field theory in
two-dimensions, in the presence of a finite system of topological D-branes,  
is described by a noncommutative Frobenius structure on
an $R$-superbimodule $E$, i.e. a unital noncommutative
Frobenius algebra on the vector superspace $E$, whose associative
product and pairing are $R$-bilinear. 

\

\subsection{The geometrization strategy}

The algebraic formulation of the previous subsection 
allows one to avoid the notational morass of the category-theoretic description. One
is still left with the rather complicated data of a cyclic and unital
weak $A_\infty$ structure on $E$.  To express this synthetically, we will 
 use a $\Z_2$-graded version of the non-commutative symplectic
geometry of quivers developed in \cite{Ginzburg, LB_quivers}. Much of the content of the next two sections is a
relatively straightforward, though tedious, superextension of the construction 
of \cite{Ginzburg, LB_quivers}, so it might be useful to
summarize the main points. Start with a cyclic and unital $A_\infty$
structure on the $R$-superbimodule $E$. To encode this geometrically,
we will proceed as follows:

(1) Giving a weak
$A_\infty$ structure on the superbimodule $E$ amounts to giving an odd
derivation $Q$ of the tensor algebra $A=T_R E[1]^{\rm v}$, satisfying
the condition $Q^2=0$. Hence $(A,Q)$ can be viewed as a noncommutative version of 
the $Q$-manifolds considered in \cite{Konts_Schwarz}.

(2) The symplectic form on $E[1]$ induces a non-commutative symplectic
form on $A$, and one can develop the non-commutative symplectic
supergeometry of this algebra by extending the approach of
\cite{Ginzburg, LB_quivers}. This gives notions of symplectic and
Hamiltonian superderivations having the classical properties. The
Karoubi complex $C_R(A)$ is acyclic in positive degrees so all symplectic
superderivations are Hamiltonian. There is a $\Z_2$-graded version
$\{.,.\}$ of the Kontsevich bracket, a super-Lie bracket on
$C^0_R(A)[{\tilde \omega}]$, where ${\tilde \omega}$ is the
$\Z_2$-degree of the symplectic form.

(3) Cyclicity of $(r_n)$ amounts to the condition that $Q$ be a
symplectic derivation, i.e. $L_Q\omega=0$, where $L_Q$ is the Lie
superderivative along $Q$. Thus $(A,Q,\omega)$ is a noncommutative 
version of the QP-manifolds of \cite{Konts_Schwarz}.

(4) The non-commutative generating function $W$ of the D-brane system
is the Hamiltonian of the homological derivation $Q$; to determine this uniquely, we 
require that it `vanishes at zero' in an appropriate sense.   Thus $W$ is an
element of degree ${\tilde \omega}+1$ of the supervector space
$C^0_R(A)=A/[A,A]$. The weak $A_\infty$ constraint $Q^2=0\Leftrightarrow
[Q,Q]=0$ is equivalent with the condition $\{W,W\}=0$. The superderivation $Q$ can be
reconstructed as the Hamiltonian derivation determined by $W$, so the $A_\infty$
structure defined by $W$ is automatically cyclic.

(6) Unitality of the weak $A_\infty$ structure is equivalent to the
condition $\frac{1}{2}\ld_\lambda W=\mu$, where $\mu$ is a superized version of
the moment map of \cite{BEV} and $\ld_\lambda$ is the cyclic
derivative of $W$ with respect to the odd $A_\infty$ unit
$\lambda$. The noncommutative generating function determines the
symplectic form through this relation.

(7) Since $R=\oplus_{u\in {\cal Q}_0}{\C \epsilon_u}$, the tensor algebra $A$ can be
viewed as the path algebra of a superquiver ${\cal Q}$, a quiver on the
vertex set ${\cal Q}_0$ endowed with a $\Z_2$-valued map on its set of
arrows. The quiver presentation arises by choosing homogeneous bases
of the vector space $E$ which are adapted to its binary homogeneous
decomposition. Using quiver language amounts to working in `special'
coordinates on the non-commutative space determined by $A$, where
`special' means that the coefficients of the non-commutative
symplectic form in such coordinates are complex numbers.

(8) All constructions have natural quiver interpretations. For
example, $C^0_R(A)$ can be described in terms of necklaces, which in our
case are $\Z_2$-graded. Cyclic derivatives with respect to elements of
the adapted basis translate into cyclic derivatives with respect to
the quiver's arrows $a$.  Relative to an appropriate adapted basis,
the unitality constraint amounts to the requirement that $W$ has a certain
dependence on an odd element $\sigma$ of $E[1]^{\rm v}$ determined by
the odd $A_\infty$ unit $\lambda$.

(9) One can view $A$ as the coordinate ring of a non-commutative
supermanifold $\A_{\cal Q}$. Imposing the relations $\ld_a W=0$ gives the
non-commutative extended vacuum space ${\cal Z}$, a `noncommutative subscheme'
of $\A_{\cal Q}$. The non-commutative extended moduli space ${\cal M}$ is obtained by modding 
${\cal Z}$ (in the GIT sense) through those
symplectomorphisms which correspond to unital and cyclic $A_\infty$ automorphisms 
of the underlying D-brane category.  

We now proceed with the detailed discussion of these points.

\section{Noncommutative symplectic geometry of $R$-superalgebras}
\label{sg}

In this section, we extend the construction of
\cite{Ginzburg,LB_quivers} to the case of superalgebras.  The proofs
of most statements are straightforward adaptations of those given in
\cite{Ginzburg,LB_quivers}, so I will only indicate the points where
our conventions are important or something interesting happens.

Let $R$ be a unital and commutative algebra over $\C$.  An {\em
$R$-superalgebra} is a unital superalgebra $A$ containing $R$ as a
subalgebra in even degrees (notice that this is stronger than 
requiring that $A$ be a superalgebra over $R$, since 
we require that $R$ sits inside the degree zero subalgebra of $A$).  
A morphism of $R$-superalgebras is a morphism of
superalgebras whose restriction to $R$ equals the identity map.

\subsection{Noncommutative differential superforms}

Given an $R$-superalgebra $A$, consider the $R$-superbimodule
$A_R:=A/R$, where $A/R$ stands for the vector space quotient. 
We define the {\em space of relative noncommutative
forms of $A$ over $R$} by $\Omega_R A:=\oplus_{n\geq 0}\Omega^n_R A$, where:
\be
\nn
\Omega^n_R A= A\otimes_R T_R^n (A_R)=A\otimes_R A_R^{\otimes_R n}~~.
\ee
We write the elements of this space as $w=a_0da_1\ldots da_n$, where $da$ is the
image of $a\in A$ under the projection $A\stackrel{d}{\rightarrow}
A_R=A/R$ and juxtaposition stands for the tensor product over $R$.
The space $\Omega_R A$ is given a differential algebra structure with product: 
\begin{eqnarray}
&&(a_0 da_1\ldots  da_n) (b_0 db_1 \ldots db_m):=a_0da_1\ldots da_n d(a_nb_0)db_1\ldots db_m\nn\\
&+&\sum_{i=1}^{n-1}{(-1)^{i}a_0  
da_1\ldots  d(a_{n-i}a_{n-i+1}) \ldots 
da_n db_0}\ldots  db_m + (-1)^n a_0 a_1 da_2\ldots da_n db_0\ldots db_m\nn
\end{eqnarray}
and differential $d (a_0da_1\ldots d a_n)=da_0da_1\ldots da_n$. 
In this paper, we view $\Omega_R A$ as an $\N\times \Z_2$ graded algebra,
whose $\N$-grading (with components $\Omega^n_R A$) is given by the `rank' of forms, and whose 
$\Z_2$ grading is induced from $A$. We denote the bidegree of homogeneous elements by $\deg w=({\bar
w},{\tilde w})\in \N\times \Z_2$.  As explained in the introduction, we
always work with the pairing (\ref{pairing}), and will require that $d$
has bidegree $(1,0)\in \Z\times \Z_2$. This means that $d$ has the derivation property:
\be
\nn
d(w_1w_2)=(dw_1)w_2+(-1)^{{\bar w_1}}w_1\cdot d w_2~~{\rm for~all~}~~w_1,w_2\in \Omega_R A~~.
\ee
We stress that in our conventions $d$ has degree {\em zero} with
respect to the $\Z_2$-grading.  The detailed construction of $\Omega_R
A$ is given in Appendix \ref{envelope}.  
The space $\Omega^1_R A$ has an $A$-superbimodule structure with multiplications: 
\be
\nn
\alpha(adb)\beta=(\alpha a)d(b\beta) -(\alpha a b)d\beta~~~~~~~~\forall a,b\in A,~~\alpha,\beta\in R~~.
\ee
As in \cite{CQ}, one has an isomorphism of bigraded 
algebras $\Omega_R A\approx T_A (\Omega^1_R A)$, which induces an $A$-superbimodule 
structure on $\Omega_R A$.  In particular, $\Omega_R A$ is
an $R$-superalgebra (since $R\subset A=\Omega^0_R A \subset \Omega_R A$).

As usual, the pair $(\Omega_R A, d)$ has a universality property.  To
formulate it, we define an $R$-{\em differential superagebra} to be
an $\N\times \Z_2$-graded unital differential algebra $(\Omega,d)$
such that $\deg(d)=(1,0)$, $R\subset \Omega^{0,0}$ and $d(R)=0$, where
$\Omega^{0,0}$ is the subspace of elements of vanishing bidegree. If
$(\Omega_j,d_j)$ are two such algebras, a map
$\phi:\Omega_1\rightarrow \Omega_2$ is called a morphism of
$R$-differential superalgebras if:

\noindent (1) $\phi$ is a morphism of unital $\N\times \Z_2$-graded
algebras (in particular, $\phi$ has vanishing bidegree)

\noindent (2) $\phi|_R=\id_R $

\noindent (3) $d_2\circ \phi=\phi\circ d_1$~~.

\noindent With this definition, $R$-differential superalgebras form a
category. The universality property of $(\Omega_R A, d)$ is as follows.  
Given any $R$-differential superalgebra $(\Omega, d)$ and a morphism of $R$-superalgebras
$\rho:A\rightarrow \Omega$ such that $\rho(A)\subset \Omega^{0}$, there exists a unique morphism
$u:\Omega_R A\rightarrow \Omega$ of $R$-differential superalgebras such
that $\rho=uj$, where $j:A\rightarrow \Omega_R A$ is the
inclusion. Hence $(\Omega_R A,j)$ is an initial object among the pairs
$(\Omega,\rho)$.  In fact, the correspondence $\Omega\rightarrow
\{$ $R$-superalgebra morphisms $\rho:A\rightarrow
\Omega$ with $\rho (A)\subset \Omega^{0}~ \}$ defines a
functor from the category of $R$-differential superalgebras to
the category of sets. The universality property means that $(\Omega_R
A, d)$ represents this functor, so it is the
superdifferential envelope of $A$. In particular, any morphism
$\phi:A_1\rightarrow A_2$ of $R$-superalgebras extends uniquely to a
morphism $\phi^*:\Omega_R A_1\rightarrow \Omega_R A_2$ of $R$-differential
superalgebras.

\subsection{Super Lie derivatives and contractions}

Recall that a left (right) derivation of $A$ is a derivation of $A$
viewed as a left (right) supermodule over $A\otimes A^{op}$. Thus a homogeneous 
left derivation $D$ satisfies $D(ab)=(Da)b+(-1)^{{\tilde a}{\tilde D}}a Db$, while 
a homogeneous right derivation satisfies $(ab)D=a(bD)+(-1)^{{\tilde b}{\tilde D}}(aD) b$, where ${\tilde D}$ is the 
degree of $D$ and we use the convention that left derivations are written to the left, and right derivations 
are written to the right. We will sometimes also indicate this by writing arrows above $D$.

A relative derivation of $A$ is a derivation which is $R$-linear,
i.e. vanishes on the subalgebra $R$.  We let $\Der_l(A) $ and
$\Der_r(A)$ be the complex supervector spaces of {\em relative} left
and right derivations of $A$, viewed as Lie superalgebras with respect to the
supercommutator $[D_1,D_2]=D_1\circ D_2-(-1)^{{\tilde D}_1 {\tilde
D_2}}D_2\circ D_1$, which satisfies $[D_1,D_2]=(-1)^{1+{\tilde D}_1{\tilde D}_2}[D_2,D_1]$.  
We let $\Der_{l,r}^\alpha (A)$ be the subspaces
consisting of left and right relative derivations of degree $\alpha$.

Similarly, let $\Der_{l,r}(\Omega_R A)$ be the $\Z\times \Z_2$-graded
complex vector spaces of {\em relative} left/right derivations of $\Omega_R A$,
i.e. those left/right derivations of $\Omega_R A$ which vanish on $R$. In this
definition, we view $\Omega_R A$ as a $\Z\times \Z_2$ graded algebra
with the degree pairing (\ref{pairing}); thus a bihomogeneous left derivation $D$ of
$\Omega_R A$ satisfies:
\be
\nn
D(w_1w_1)=Dw_1 w_2+(-1)^{\deg D \cdot \deg w_1}w_1 Dw_2~~,
\ee
while a bihomogeneous right derivation obeys:
\be
\nn
(w_1w_2)D=w_1 (w_2D)+(-1)^{\deg D\cdot \deg w_2}(w_1D)w_2~~.
\ee
The spaces $\Der_{l,r}(\Omega_R A)$ are bigraded Lie superalgebras
with respect to the bigraded supercommutator $[D_1,D_2]=D_1\circ
D_2-(-1)^{\deg D_1 \cdot \deg D_2} D_2 \circ D_1$, which satisfies
$[D_1,D_2]=(-1)^{1+\deg D_1\cdot \deg D_2} [D_2,D_1]$.  We let
$\Der_{l,r}^\alpha (\Omega_R A)$ be the subspaces consisting of left
relative derivations of bidegree $\alpha$.

Let $\theta\in \Der_l(A)$ be a homogeneous relative left derivation of
degree ${\tilde \theta}$.  The {\em contraction by $\theta$} is the
unique relative left derivation $i_\theta \in \Der_{l}^{-1, {\tilde
\theta}}(\Omega_R A)$ which satisfies $i_\theta a =0$ and $i_\theta (da)=\theta(a)$
for all $a\in A$. The {\em Lie derivative along $\theta$} is
the unique left derivation $L_\theta \in \Der_l^{0, {\tilde
\theta}}(\Omega_R A)$ which satisfies $L_\theta a=\theta(a)$ and $L_\theta(da)=d\theta(a)$ 
for all $a$. There are obvious versions of these definitions for right derivations. 
It is easy to check that $i_\theta$ and $L_\theta$ are well-defined.
Let $\Aut_R(A)$ be the space of automorphisms of $A$ as an
$R$-superalgebra (in particular, all maps $\phi\in \Aut_R(A)$ are even and restrict to the identity on the subalgebra $R$). 
For any $\theta,\gamma \in \Der_l(A)$ and $\phi\in \Aut_R(A)$, we have the identities:
\begin{eqnarray}
\label{Cartan}
L_\theta&=&\left[i_\theta, d\right]\nn\\
\left[L_\theta, i_\gamma\right]&=&i_{\left[\theta, \gamma\right]}\nn\\
\left[L_\theta, L_\gamma\right]&=&L_{\left[\theta, \gamma\right]}\nn\\
\left[i_\theta, i_\gamma\right]&=&0\\
\left[L_\theta, d\right]&=&0 \nn\\
L_{\phi \theta \phi^{-1}}&=&\phi^*L_\theta \phi^{* -1}\nn\\
i_{\phi \theta \phi^{-1}}&=&\phi^*i_\theta \phi^{* -1}\nn~~.
\end{eqnarray}
As usual, these follow by noticing that all left and right hand sides
are derivations of the same bi-degree on $\Omega_R A$, and checking agreement on the 
generators $a$ and $da$ ($a\in A$). Given $w=a_0da_1\ldots da_n$ with $a_i\in A$, we have: 
\begin{eqnarray}
\label{ids}
i_\theta w &=& \sum_{i=1}^n{
(-1)^{i-1+{\tilde \theta}({\tilde
    a}_0+\ldots +{\tilde a}_{i-1})}
a_0da_1\ldots da_{i-1} \theta(a_i)da_{i+1}\ldots da_n}\nn\\
L_\theta w&=& \theta(a_0) da_1\ldots da_n+
\sum_{i=1}^n{(-1)^{{\tilde \theta}({\tilde
    a}_0+\ldots +{\tilde a}_{i-1})}
a_0da_1\ldots da_{i-1} d\theta(a_i)da_{i+1}\ldots da_n}~~.\nn
\end{eqnarray}

\subsection{The bigraded Karoubi complex}

Consider the $\N\times \Z_2$-graded vector space: 
\be
C_R(A):=\Omega_R A/[\Omega_R A,\Omega_R A]~~,
\ee
where $[\Omega_R A,\Omega_R A]\subset \Omega_R A$ is the image of the bigraded 
commutator map $[.,.]:\Omega_R A\times \Omega_R A\rightarrow \Omega_R A$:
\be
\nn
[w_1,w_2]=w_1w_2-(-1)^{\deg w_1\cdot \deg w_2}w_2 w_1~~.
\ee
Notice that $[\Omega_R A,\Omega_R A]$ is a homogeneous subspace of $\Omega_R A$ but not a
subalgebra. We let $\pi:\Omega_R A\rightarrow C_R(A)$ be the
projection, and use the notation:
\be
\nn
\pi(w):=(w)_c~~{\rm for}~~w\in \Omega_R A~~.
\ee
We also let $C_R^n(A)$ be the $\N$-homogeneous components of $C_R(A)$. 

Any relative derivation of $\Omega_R A$ preserves the subspace $[\Omega_R A, \Omega_R A]$,
so it descends to a well-defined linear operator in $C_R(A)$.  In
particular, $d$ induces a differential ${\bar d}$ on $C_R(A)$. The bigraded 
vector space $(C_R(A),{\bar d})$ is the {\em relative Karoubi (or
non-commutative de Rham) complex} of $A$ over $R$. This differential space is
$\N\times \Z_2$-graded, and we have $\deg {\bar d}=(1,0)$.  The
supervector spaces $H^n_R(A):=H_{\bar d}^n(C_R(A))$ are called the
{\em relative de Rham cohomology groups of $A$}.

Given a morphism of $R$-superalgebras, the induced map 
$\phi^*:(\Omega_R A_1,d_1)\rightarrow (\Omega_R A_2, d_2)$ of $R$-differential superalgebras 
satisfies $\phi^*([\Omega_R A_1,\Omega_R A_1])\subset [\Omega_R A_2,\Omega_R
A_2]$, so it descends to a morphism of bigraded complexes
${\bar \phi}^*:(C_R(A_1),{\bar d}_1)\rightarrow (C_R(A_2),{\bar
d}_2)$.  In particular, any endomorphism $\phi$ of $A$ induces an
endomorphism ${\bar \phi}^*$ of $(C_R(A),{\bar d})$, which is an automorphism 
if $\phi$ is. We let ${\bar
\phi}$ be the restriction of ${\bar \phi}^*$ to $C^0_R(A)=A/[A,A]$ (here $[A,A]\subset A$ is 
the image of the supercommutator $[.,.]:A\times A\rightarrow A$).

The contraction and Lie operators $i_\theta$, $L_\theta$ also descend
to well-defined $\C$-linear maps on $C_R(A)$, which we denote by ${\bar
i}_\theta$, ${\bar L}_\theta$. It is clear that the induced operators
satisfy all properties listed in eqs. (\ref{Cartan}):
\begin{eqnarray}
\label{Cartan_c}
{\bar L}_\theta &=&\left[{\bar i}_\theta, {\bar d}\right]\nn\\
\left[{\bar L}_\theta, {\bar i}_\gamma\right]&=&{\bar i}_{\left[\theta, \gamma\right]}\nn\\
\left[{\bar L}_\theta, {\bar L}_\gamma\right]&=&{\bar L}_{\left[\theta, \gamma\right]}\nn\\
\left[{\bar i}_\theta, {\bar i}_\gamma\right]&=&0\\
\left[{\bar L}_\theta, {\bar d}\right]&=&0 \nn\\
{\bar L}_{\phi \theta \phi^{-1}}&=&{\bar \phi}^*{\bar L}_\theta {\bar \phi}^{* -1}\nn\\
{\bar i}_{\phi \theta \phi^{-1}}&=&{\bar \phi}^*{\bar i}_\theta {\bar \phi}^{* -1}\nn~~,
\end{eqnarray}
where $\phi\in \Aut_R(A)$~~.

\subsection{Noncommutative supersymplectic forms}

An element $\omega\in C^2_R(A)$ is called non-degenerate if the following complex-linear map is bijective: 
\be
\nn
\theta\in \Der_l(A)\rightarrow {\bar i}_\theta \omega\in C^1_R(A)~~.
\ee
A {\em relative non-commutative symplectic form} on $A$ is 
a $\Z_2$-homogeneous element $\omega\in C^2_R(A)$ which is closed (${\bar d}\omega=0$) and non-degenerate.

Given a symplectic form $\omega$ of $\Z_2$-degree ${\tilde \omega}$, a relative derivation $\theta\in \Der_l(A)$
is called {\em symplectic} if ${\bar L}_\theta\omega=0$.  Let
$\Der_l^\omega(A)\subset \Der_l(A)$ be the subspace of all symplectic
derivations. By the third property in (\ref{Cartan_c}), this is a (super) Lie subalgebra of $\Der_l(A)$. 
As in the even case, it is easy to see that the following map is an isomorphism of supervector spaces:
\be
\label{isom_0}
\theta\in \Der_l^\omega(A)\rightarrow {\bar i}_\theta\omega\in C^1_R(A)_{\rm closed}[{\tilde \omega}]~~.
\ee 
Here $C^1_R(A)_{\rm closed}=\{\eta\in C^1_R(A)|{\bar d} \eta=0\}$, a homogeneous subspace of $C^1_R(A)$.

This implies that any $f\in C^0_R(A)$ defines a unique
element $\theta_f\in \Der_l^\omega(A)$, determined by the equation
${\bar i}_{\theta_f}\omega={\bar d}f$. Let
$\psi_\omega:C^0_R(A)\rightarrow \Der_l^\omega(A)$ be the complex-linear map
given by $\psi_\omega(f):=\theta_f$. The relation ${\bar
i}_{{\theta}}\omega={\bar d}f$ implies ${\tilde \theta}_f={\tilde
\omega}+{\tilde f}$ for any homogeneous $f$, so the map $\psi_\omega$
is homogeneous of degree ${\tilde \omega}$.  It is clear that the
following sequence of supervector spaces is exact:
\be
\label{ex_0}
0\rightarrow
H^0_R(A)\hookrightarrow C^0_R(A)\stackrel{\psi_\omega}{\rightarrow}\Der_l^\omega(A)[\tom]~~, 
\ee
where the map in the middle is the inclusion. Elements $\theta\in {\rm
im}\psi_\omega$ are called {\em Hamiltonian derivations}.  Given a
Hamiltonian derivation $\theta$, an element $f\in C^0_R(A)$ such that
$\psi_\omega(f)=\theta\Leftrightarrow \theta=\theta_f$ is called a
{\em Hamiltonian} associated with $\theta$. The sequence (\ref{ex_0})
shows that the Hamiltonian of a Hamiltonian derivation is determined
up to addition of elements of $H^0_R(A)$.

\subsection{$\Z_2$-graded version of the Kontsevich bracket}
\label{sec:bracket}

We have the following generalization of an operation introduced in \cite{Konts_formal}. 

\paragraph{\bf Definition}
The {\em  Kontsevich bracket} induced by $\omega$ is the $\Z_2$-homogeneous 
complex-linear map $\{.,.\}:C^0_R(A)\otimes_\C C^0_R(A)\rightarrow C^0_R(A)$ of degree $\tom$ defined through: 
\be
\label{bdef}
\{f,g\}:={\bar i}_{\theta_f}{\bar i}_{\theta_g}\omega~~~~~\forall f,g \in C^0_R(A)~~,\nn
\ee
where $\theta_f=\psi_\omega(f)$ and $\theta_g=\psi_\omega(g)$. 

\

\noindent Notice the relation:
\be
\label{bL}
\{f,g\}={\bar L}_{\theta_f}g~~.
\ee
The following result gives the basic properties of the bracket in the $\Z_2$-graded case.
\paragraph{\bf Proposition}
The Kontsevich bracket satisfies the identities: 
\be
\label{gasym}
\{g,f\}=(-1)^{1+({\tilde f}+{\tilde \omega})({\tilde g}+{\tilde \omega})}\{f,g\}~~.
\ee
and:
\be
\label{gjacobi}
(-1)^{(\tf_1+\tom)(\tf_3+\tom)}
\{f_1,\{f_2,f_3\}\}+(-1)^{(\tf_2+\tom)(\tf_1+\tom)}\{f_2,\{f_3,f_1\}\}+
(-1)^{(\tf_3+\tom)(\tf_2+\tom)}\{f_3,\{f_1,f_2\}\}=0~~.
\ee
Hence $(C^0_R(A)[\tom],\{.,.\})$ is a Lie superalgebra. 

\begin{proof}

The first property follows immediately from $[{\bar i}_{\theta_f},{\bar i}_{\theta_g}]=0$. 
For the second property, let $f_1,f_2,f_3\in C^0_R(A)$ and
$\theta_i:=\phi_\omega(f_i)$. Identities (\ref{Cartan_c}) give:
\begin{eqnarray}
0={\bar i}_{\theta_2}{\bar i}_{\theta_1}
{\bar i}_{\theta_0}d\omega&=&(-1)^{{\tilde
    \theta}_0({\tilde \theta_1}+{\tilde
    \theta}_2)}{\bar L}_{\theta_0}{\bar i}_{\theta_2}{\bar i}_{\theta_1}\omega+
(-1)^{1+{\tilde \theta}_1 {\tilde
    \theta}_2}{\bar L}_{\theta_1}{\bar i}_{\theta_2}{\bar i}_{\theta_0}\omega+
{\bar L}_{\theta_2}{\bar i}_{\theta_1}{\bar i}_{\theta_0}
\omega\nn\\
&+&
    (-1)^{1+\th_0\th_1}{\bar i}_{\theta_2}{\bar i}_{[\theta_0,\theta_1]}\omega+(-1)^{\th_2(\th_0+\th_1)}
{\bar i}_{\theta_1}
    {\bar i}_{[\theta_0,\theta_2]}\omega+(-1)^{1+\th_0\th_1+\th_2(\th_0+\th_1)}{\bar i}_{\theta_0}
{\bar i}_{[\theta_1,\theta_2]}\omega~~.\nn
\end{eqnarray}
Using the relations
$[\theta_i,\theta_j]=(-1)^{1+\th_i\th_j}[\theta_j,\theta_i]$ and the
properties of ${\bar L}_{\theta_i}$, the
right hand side can be brought to the form:
\be
2\left[(-1)^{1+\th_0\th_1+\th_1\th_2+\th_2\th_0}\{f_0,\{f_1,f_2\}\}+(-1)^{1+\th_1\th_2}
\{f_1,\{f_2,f_0\}\}+(-1)^{1+\th_0\th_1}\{f_2,\{f_0,f_1\}\}\right]\nn
\ee
Multiplying with $(-1)^{1+\th_0\th_1+\th_1\th_2}$ leads to the
identity: 
\be
(-1)^{\th_0\th_2}\{f_0,\{f_1,f_2\}\}+(-1)^{\th_1\th_0}\{f_1,\{f_2,f_0\}\}+
(-1)^{\th_2\th_1}\{f_2,\{f_0,f_1\}\}=0~~.\nn
\ee
This implies equation (\ref{gjacobi}) upon changing the indices
$0,1,2$ into $1,2,3$ and using $\th_i={\tilde
  f}_i+{\tilde \omega}$. 

\end{proof}

\noindent The map $\psi_\omega$ has another property which parallels classical behavior. 

\paragraph{\bf Proposition}
{\em We have: 
\be
\label{Lie_morphism}
\theta_{\{f,g\}}=[\theta_f, \theta_g]~~~~~~\forall f,g\in C^0_R(A)~~.
\ee
Thus $\psi_\omega:(C^0_R(A)[\tom],\{.,.\}) \rightarrow (\Der_l^\omega(A),[.,.])$ is a morphism of Lie
superalgebras over $\C$. }

\begin{proof}
Compute: 
\be
{\bar d}\{f,g\}={\bar d}{\bar i}_{\theta_f}{\bar i}_{\theta
g}\omega={\bar L}_{\theta_f}{\bar i}_{\theta_g}\omega+{\bar i}_{\theta_f}
{\bar d}({\bar i}_{\theta_g}\omega)={\bar L}_{\theta_f}{\bar i}_{\theta_g}\omega=[{\bar L}_{\theta_f},
{\bar i}_{\theta_g}]\omega={\bar i}_{[\theta_f,\theta_g]}\omega~~.\nn
\ee
In the third equality, we used ${\bar i}_{\theta_g}\omega={\bar d}g$ and ${\bar d}^2=0$,
while in the fourth we used ${\bar L}_{\theta_f}\omega=0$.
\end{proof}

An $R$-superalgebra automorphism $\phi\in \Aut_R(A)$ is called a {\em relative
symplectomorphism} if ${\bar \phi}^*(\omega)=\omega$.  We let
$\Aut^\omega_R(A)\subset \Aut_R (A)$ be the subgroup of relative
symplectomorphisms of $A$.  By the sixth property in (\ref{Cartan_c}),
the obvious action of $\Aut^\omega_{alg}(A)$ on $\Der_l(A)$ preserves
the Lie subalgebra $\Der_l^\omega(A)$. Given a symplectomorphism
$\phi$, the last property in (\ref{Cartan_c}) implies $\phi\circ
\theta_f\circ \phi^{-1}=\theta_{{\bar \phi}(f)}$ for $f\in C^0_R(A)$,
i.e.  $\psi_\omega\circ {\bar \phi} ={\rm Ad}_{\phi}\circ \psi_\omega$
for all $\phi\in \Aut^\omega_R(A)$. In turn, this gives $\{{\bar
\phi}(f),{\bar \phi}(g)\}={\bar \phi}(\{f,g\})$.  Hence
$\Aut^\omega_R(A)$ acts on $(A[{\tilde \omega}],\{.,.\})$ by Lie
algebra automorphisms.

\section{Noncommutative calculus for finite D-brane systems}

Consider a finite D-brane system with object set ${\cal Q}_0$. As in
Section \ref{algebraic}, we set $R=\oplus_{u\in {\cal Q}_0}{\C \epsilon_u}$ and let $E$ be the
$R$-superbimodule of boundary sectors.  Recall that $E[1]$ carries a
symplectic form $\omega=\rho\circ \Sigma^{\otimes 2}$, the parity change of the topological metric $\rho$.
Setting $V=E[1]^{\rm v}$, we consider the $\N\times \Z_2$ graded tensor
algebra $A=T_R (V)$ whose $\N$-homogeneous subspaces we denote by $A_n=T_R^n (V)$.  
With respect to its $\Z_2$-grading, $A$ is an $R$-superalgebra with  $R=A_0$. 

\subsection{Generalities}
\label{tensor}

The symplectic form $\omega$ on $E[1]$ induces a
relative noncommutative symplectic form $\omega_{form}$ on $A$ as
follows. Recall from Section 2 that the symplectic form $\omega$ defines 
a map $j_\omega: E[1]^{opp}\rightarrow E[1]^{\rm v}$ which is a morphism of $R$-bimodules, i.e. 
an element of $\iHom(E[1]^{opp},E[1]^{\rm v})$. 
As explained in Appendix \ref{isomorphisms}, there exists 
an isomorphism of $R$-superbimodules between 
$\iHom(E[1]^{opp},E[1]^{\rm v})$ and the center of the superbimodule 
$E[1]^{\rm v}\otimes_R E[1]^{\rm v}$,
which allows us to view $\omega$ as an element ${\hat \omega}$ of degree ${\tilde \omega}$
of the space $(V\otimes_R V)^R$. If ${\hat \omega}=\sum_{i}{f_i\otimes_R g_i}$ with $f_i,g_i\in V$, 
then it is shown in Appendix \ref{isomorphisms} that $\omega$ can be recovered as: 
\be
\label{omega_rec}
\omega(x,y)=\sum_i f_i(xg_i(y))~~.
\ee
Using the element ${\hat \omega}$, we define a noncommutative two-form on $A=T_R V$ through the relation:
\be
\label{omega_form_gen}
\omega_{form}=-\frac{1}{2}\sum_{i}{(df_i dg_i)_c}\in C^2_R(A)~~,
\ee
where the minus sign is introduced for later convenience. 
It is easy to see that $\omega_{form}$ is well-defined
and of the same $\Z_2$-degree as $\omega$.  Moreover, it is not hard to check that this two-form is 
symplectic.

Since $A=T_R V$, the algebra of noncommutative forms $\Omega_RA$ has a
second $\N$-grading, which is induced from the $\N$-grading of $A$
(with respect to this grading, we have $\deg a =\deg da=1$ for all
$a\in V$, while $R$ sits in degree zero).  This induces a similar
$\N$-grading on $C_R(A)$.  We let $(\Omega_RA)_n$ and $C_R(A)_n$ be
the homogeneous subspaces determined by this grading. A {\em
constant} noncommutative two-form \footnote{The notion of constant
noncommutative symplectic form depends on the specific realization of
$A$ as a tensor algebra $T_R V$. In particular, this concept is not
invariant under the $R$-superalgebra automorphism group $\Aut_R(A)$
(because a superalgebra automorphism need not preserve the
$\N$-grading of $A$).} on $A$ is an element of $C^2_R(A)_2$. Thus
(\ref{omega_form_gen}) is a constant symplectic form, and any constant
symplectic form on $A$ has such an expansion.  We let $CNS^{\tilde \omega}(V)$ be the
vector space of all constant noncommutative symplectic forms on $T_R V$ of degree ${\tilde \omega}$. In
what follows, we often denote $\omega_{form}$ by $\omega$; which of
the two is meant should be clear from the context.

Notice that $[A,A]\cap A_0=[A_0,A_0]=[R,R]=0$ since $A_0=R$ and $R$ is commutative.  In particular, 
we have $C^0_R(A)_0=R$. The following result follows as in \cite{Ginzburg}, by
considering the Euler derivation associated with the $\N$-grading induced from $A$.

\paragraph{\bf Proposition} {\em We have $H^0_R(A)=C^0_R(A)_0=R$ and $H^n_R(A)=0$ for all $n\geq 1$.}

\

\noindent Using this in (\ref{ex_0}) gives a short exact sequence: 
\be
\label{ex}
0\rightarrow
R\rightarrow C^0_R(A)\stackrel{\psi_\omega}{\rightarrow}\Der_l^\omega(A)[\tom]\rightarrow 0~~,
\ee
where surjectivity of $\psi_\omega$ follows by using $H^1_R(A)=0$
in (\ref{isom_0}). In particular, {\em any relative symplectic derivation of $A$
is Hamiltonian}. The sequence (\ref{ex}) also shows that the
Hamiltonian of a symplectic derivation is determined up to addition of
elements of $R$, which can be viewed as the subspace $C^0_R(A)_0$ of $C^0_R(A)$.

Any element $f$ of $C^0_R(A)$ has a decomposition $f=\sum_{n\geq
0}{f_n}$, with $f_n\in C^0_R(A)_n$ (in particular, $f_0\in C^0_R(A)_0=R$).  
We say that $f$ {\em has order $k$
at zero} if $f_0= f_1= \ldots =f_{k-1}=0$ and $f_k\neq 0$. We say that
$f$ {\em vanishes at zero} if $f_0=0$, i.e. $f$ has order at least one
at zero. We define the {\em canonical Hamiltonian} of a symplectic
derivation $\theta$ to be that Hamiltonian of $\theta$ which vanishes
at zero. The sequence (\ref{ex}) shows that the canonical
Hamiltonian exists and is unique.

\subsection{Adapted bases and superquivers} 
\label{coord_description}

Using the binary decomposition $E=\oplus_{(u,v)\in {\cal Q}_0\times {\cal Q}_0}{E_{uv}}$, 
consider a homogeneous basis $(\psi_a)$ of the supervector space $E$ having the following properties:

\

(1)   $a=(u,v, j)$ is a multi-index with $u,v \in {\cal Q}_0$ and $j=1\ldots \dim_\C E_{uv}$

\

(2) $\psi_{uvj}$ for $j=1\ldots \dim_\C E_{uv}$ is a 
homogeneous basis of $E_{uv}$ for all $u,v\in {\cal Q}_0$.

\

\noindent We say that such a homogeneous basis is {\em adapted} to the binary decomposition of $E$. 
Setting $e_a:=\Sigma \psi_a$, we let $(s^a)$ be the basis of the super-vector space 
$V=E[1]^{\rm v}$ dual to $(e_a)$: 
\be
\nn
s^a(e_b)=\delta^a_b~~.
\ee
Relation (\ref{dual_spaces}) shows that 
$s^{uvj}$ are bases of $V_{uv}=(E[1]^{\rm v})_{uv}=(E_{uv}[1])^*$, odd dual to the 
bases  $\psi_{uvj}$  of $E_{uv}$. We set ${\tilde
a}:=\deg s^a=\deg e_a=[|a|+1]$, where $|a|:=|\psi_a|$.

It is convenient to keep track of indices by considering a quiver
${\cal Q}$ determined by the multi-indices $a$. Specifically, the {\em index quiver} ${\cal Q}$ is the quiver 
on the vertex set ${\cal Q}_0$ obtained by drawing an arrow from $u$ to $v$ for
each $j=1\ldots \dim_\C V_{uv}$. With this construction, we can identify each
multi-index $a$ with the corresponding arrow of ${\cal Q}$. We let ${\cal Q}_1$ be the set of all
arrows and ${\cal Q}_1(u,v)$ the subset of arrows going from $u$ to $v$.  We
also let $h,t:{\cal Q}_1\rightarrow {\cal Q}_0$ be the head and tail maps of ${\cal Q}$.

The index quiver is in fact a {\em superquiver},
being endowed with a map $\deg: {\cal Q}_1\rightarrow \Z_2$ 
given by $\deg (a)={\tilde a}$.  An arrow $a$
is called even if ${\tilde a}=0$ and odd if ${\tilde a}=1$.  The path algebra
$\C {\cal Q}$ becomes a superalgebra by declaring a path $p$ to be even
or odd if it contains an even or odd number of odd arrows. That is, we
define the degree of $p$ by the formula:
\be
\nn
{\tilde p}:=\sum_{j=1}^k{\tilde a_j}~~,
\ee 
where $p=a_1\ldots a_k$ is the arrow decomposition. The trivial paths
are taken to be even.  The path algebra is in fact $\N\times
\Z_2$-graded, where the $\N$-grading is induced by the length of paths. 
We let $(\C {\cal Q})_n$ be the components of degree $n$ with respect to the length grading.

As usual, the subspace spanned by the trivial paths forms a finite-dimensional semisimple commutative algebra. 
We identify this with the boundary sector algebra $R$ by sending the trivial path at $u$ 
into the idempotent $\epsilon_u$.  On the other hand, the subspace spanned by
the arrows is isomorphic with the supervector space $V$:
\be
\nn
(\C {\cal Q})_1\approx \oplus_{u,v\in {\cal Q}_0}\oplus_{a\in {\cal Q}_1(u,v)}\C[{\tilde a}]\approx V~~
\ee
via the identification $a\equiv s^a$. It is also clear that the
$R$-superbimodule structure of $V$ coincides with the
$(\C{\cal Q})_0$-superbimodule structure induced on $(\C{\cal Q})_1$ by multiplication in
the path algebra. In fact,  the entire path algebra is isomorphic
with the tensor algebra $T_R (\C {\cal Q})_1$ as an $\N\times
\Z_2$-graded algebra. Combining these observations, we find an isomorphism of bigraded algebras: 
\be
\nn
A\approx \C{\cal Q}
\ee
which extends the isomorphism $V \approx (\C {\cal Q})_1$. Hence:

\

{\em Choosing an adapted basis of $E$ identifies the tensor algebra
$A=T_R V$ with the path algebra of the index superquiver ${\cal Q}$.}

\

\noindent In the next subsections, we explore the consequences of this identification. 

\subsection{Quiver description of the symplectic structure}
\label{symplectic_structure}

Recall that the symplectic form $\omega$ on $E[1]$ corresponds to bilinear forms $\omega_{uv}:E[1]_{uv}\times
E[1]_{vu}\rightarrow \C$ of common degree ${\tilde \omega}$, such that $\omega_{uv}(x,y)=(-1)^{{\tilde
x}{\tilde y}+1}\omega_{vu}(y,x)$.  We define coefficients $(\omega_{ab})_{a,b\in {\cal Q}_1}$ 
through:
\be
\nn
\omega_{ab}=-\omega_{t(a)h(a)}(e_a,e_b)\in \C~~{\rm~if~}~~t(a)=h(b)~~{\rm~and~}~~h(a)=t(b)
\ee
and zero otherwise. The minus sign in this expression is introduced in 
order to simplify certain formulas which will appear in the next sections. 
In terms of the superbimodule symplectic form, we have 
$\omega(e_a,e_b)=-\omega_{ab}\epsilon_u$ where $u:=t(a)=h(b)$. 
The coefficients have the graded-antisymmetry properties:
\be
\label{omega_symm}
\omega_{ab}=(-1)^{{\tilde a}{\tilde b}+1}\omega_{ba}~~
\ee
and satisfy the selection rule:
\be
\label{omega_sel}
\omega_{ab}=0~~{\rm~unless~}~~{\tilde a}+{\tilde b}={\tilde \omega}~~.
\ee
It is not hard to see that the inverse matrix $(\omega^{ab})_{a,b\in {\cal Q}_1}$ also satisfies the selection
rule:
\be
\nn
\omega^{ab}=0~~{\rm unless}~~{\tilde a}+{\tilde b}={\tilde \omega}
\ee 
and graded antisymmetry property:
\be
\nn
\omega^{ab}=(-1)^{{\tilde a}{\tilde b}+1}\omega^{ba}~~.
\ee
Moreover, $\omega^{ab}$ vanishes unless $h(a)=t(b)$ and $t(a)=h(b)$. 

Let ${\hat \omega}$ be the central element of $V\otimes_R V$ determined by $\omega$. Since $s^a$ is a vector 
space basis of $V$, we can expand ${\hat \omega}=\sum_{a,b}{{\hat \omega}_{ab}s^a\otimes_R s^b}=
\sum_{a}{s^a\otimes_R (s^a)'}$, where we set $(s^a)':=\sum_{b}{\hat \omega}_{ab}s^b$. 
Then equation (\ref{omega_rec}) gives $\omega(e_a,e_b)=\sum_{c}s^c(e_a (s^c)'(e_b))={\hat \omega}_{ab}$. 
Thus ${\hat \omega}_{ab}=-\omega_{ab}\epsilon_{t(a)}$ and we find that our definition of coefficients 
corresponds to the expansion: 
\be
\nn
{\hat \omega}=-\sum_{a,b\in {\cal Q}_1}\omega_{ab}s^a\otimes_R s^b\in V\otimes_R V~~,
\ee
where we used the relation $\epsilon_{t(a)} s^a=s^a$, which holds because $s^a\in V_{t(a)h(a)}$. 
Using equation (\ref{omega_form_gen}), 
it follows that the non-commutative symplectic form induced on $A$  is given by:
\be
\label{omega_form}
\omega_{form}=\frac{1}{2}\sum_{\tiny 
\begin{array}{c}a,b\in {\cal Q}_1 \\h(a)=t(b),  h(b)=t(a)\end{array}}\omega_{ab}(ds^a ds^b)_c\equiv
\frac{1}{2}\sum_{a,b\in Q_1}\omega_{ab}(da db)_c~~.
\ee
In the last equality, notice that $(dadb)_c$ vanishes unless
$h(a)=t(b)$ and $h(b)=t(a)$, which can be seen immediately by
inserting idempotents $\epsilon_u$ in the appropriate places.  As in
\cite{Ginzburg, LB_quivers}, it is easy to check that 
non-degeneracy of the constant two-form (\ref{omega_form}) 
amounts to non-degeneracy of the matrix $(\omega_{ab})_{a,b\in {\cal Q}_1}$.  A {\em symplectic
superquiver} is a superquiver whose path algebra is endowed with a
constant symplectic form of type (\ref{omega_form}). Hence picking an
adapted homogeneous basis allows us to encode the information of the
symplectic superbimodule $(E[1],\omega)$ into a symplectic
superquiver. Of course, the inverse correspondence also holds.

Recall that the topological metric $\rho$ on $E$ is given by
$\omega=\rho\circ \Sigma^{\otimes 2}$. We define its coefficients
trough:
\be
\nn
\rho_{ab}=\rho(\psi_a,\psi_b)~~,
\ee
{\em without} a minus sign insertion. Equation $\omega(e_a,e_b)=(-1)^{\tilde a}\rho(\psi_a,\psi_b)$ gives: 
\be
\label{omega_rho}
\omega_{ab}=(-1)^{{\tilde a}+1}\rho_{ab}~~.
\ee
The coefficients of $\rho$ have the properties:
\be
\nn
\rho_{ab}=(-1)^{{\tilde a}{\tilde b}+{\tilde \omega}+1}\rho_{ba}=(-1)^{|a||b|}\rho_{ba}
\ee 
and: 
\be
\nn
\rho_{ab}=0{\rm~unless~}{\tilde a}+{\tilde b}={\tilde \omega}\Longleftrightarrow |a|+|b|={\tilde \omega}~~.\nn
\ee
Relation (\ref{omega_rho}) shows that the inverse of the matrix $(\rho_{ab})$ takes the form:
\be
\rho^{ab}=(-1)^{{\tilde b}+1}\omega ^{ab}~~.\nn
\ee
It is clear that the inverse matrix satisfies the relations:
\be
\label{oinv_symm}
\rho^{ab}=(-1)^{{\tilde a}{\tilde b}+{\tilde \omega}+1}\rho^{ba}=(-1)^{|a||b|}\rho^{ba}~~.
\ee
and
\be
\label{oinv_sel}
\rho^{ab}=0{\rm~unless~}{\tilde a}+{\tilde b}={\tilde \omega}\Longleftrightarrow |a|+|b|={\tilde \omega}~~.
\ee

The structure theorem for graded antisymmetric matrices implies the following: 

\noindent (1) If ${\tilde \omega}=0$, then we can find an adapted basis and an ordering  
$a_1,\ldots, a_{2m},a_{2m+1},\ldots, a_N$ of the arrows ($N={\rm Card} {\cal Q}_1$) 
such that $a_1,\ldots, a_{2m}$ are even, $a_{2m+1},\ldots, a_N$ are odd and:
\be
\nn
\omega=\sum_{i=1}^m{(da_id a_{i+m})_c}+\frac{1}{2}\sum_{j=2m+1}^N{(da_j da_j)_c}~~.
\ee
Setting $p_i:=a_i$ and $q_i=a_{i+m}$ for $i=1\ldots m$ and $\xi_\alpha:=a_{2m+\alpha}$ for $\alpha=1\ldots N-2m$, 
we can write this in the form: 
\be
\label{e_can}
\omega=(dp_idq_i)_c+\frac{1}{2}(d\xi_\alpha d\xi_\alpha)_c
\ee
with even $p_i,q_i$ and odd $\xi_\alpha$. Hence $\omega_{p_iq_j}=-\omega_{q_jp_i}=\delta_{ij}$ 
and $\omega_{\xi_\alpha\xi_\beta}=\delta_{\alpha\beta}$. 

\noindent (2) If ${\tilde \omega}=1$, then ${\rm Card} {\cal Q}_1=2m$ for some integer $m$ and 
we can find an adapted basis and an ordering 
$a_1,\ldots,a_{2m}$ of the arrows such that $a_1,\ldots, a_{m}$ are odd, $a_{m+1},\ldots, a_{2m}$ are even and:
\be
\nn
\omega=\sum_{i=1}^m{(da_id a_{i+m})_c}~~.
\ee
Setting $p_i:=a_i$ and $q_i=a_{i+m}$ for $i=1\ldots m$, this becomes: 
\be
\label{o_can}
\omega=(dp_idq_i)_c
\ee
with even $p_i$ and odd $q_i$. Hence $\omega_{p_iq_j}=-\omega_{q_jp_i}=\delta_{ij}$.

In general, one can set $a^*=\sum_{b\in {\cal Q}_1}\omega_{ab}b$, 
which brings $\omega$ to the form $\omega=\frac{1}{2}(dada^*)_c$. 
For an even $\omega$ in the canonical basis (\ref{e_can}), we have $p_i^*=q_i$, $q_i^*=-p_i$ and 
$\xi_\alpha^*=\xi_\alpha$. In this case, $*$ squares to minus the identity on the subspace spanned by $q_i, p_i$ 
but to plus the identity on the subspace spanned by $\xi_\alpha$. 
For odd $\omega$ in the basis (\ref{o_can}), we have $p_i^*=q_i$ and $q_i^*=-p_i$, so 
$*$ squares to minus the identity on the entire subspace $A_1=V$. 

It is clear from the above that a D-brane system has different behavior 
depending on the parity of $\omega$. We say that the system 
is {\em even} or {\em odd} if ${\tilde \omega}=0$, respectively $1$.

\subsection{Quiver description of $C^0_R(A)$}
\label{quivc0}

Any element $f\in A$ has an expansion: 
\be
\label{f_path}
f=\sum_{p=path}{f_p p}
\ee
where the sum is over all paths $p$ of ${\cal Q}$ (including the
trivial paths) and where $f_p\in \C$. In this and subsequent
relations, we agree that only a finite number of coefficients 
are nonzero, so that all sums are finite. We can also write (\ref{f_path}) as:
\be
\label{f}
f=\sum_{n\geq 0}{f_{a_1\ldots a_n}a_1\ldots a_n}~~,
\ee
where we use implicit summation over the arrows $a_j$ and we agree
that $f_{a_1\ldots a_0}a_1\ldots a_0$ stands for the sum
$c(f):=\sum_{u\in {\cal Q}_0}{f_u\epsilon_u}$ (with $f_u\in \C$).  
The product $a_1\ldots a_n$ vanishes unless it is a path. This is seen by
inserting idempotents:
\be
\nn
a_1\ldots a_ia_{i+1}\ldots a_n=a_1\ldots a_i\epsilon_{h(a_i)}\epsilon_{t(a_{i+1})}a_{i+1}\ldots a_n~~,
\ee
and noticing that the right hand side vanishes unless
$h(a_i)=t(a_{i+1})$. Thus only the coefficients $f_{a_1\ldots a_n}$
which correspond to paths $a_1\ldots a_n$ are defined; for convenience, we 
define $f_{a_1\ldots a_n}$ to vanish if the word $a_1\ldots a_n$ is not a path. 

The obvious relations
\be
\label{lindep}
({a_1}\ldots {a_n})_c=(-1)^{({\tilde a}_1+\ldots +
  {\tilde a}_i)({\tilde a}_{i+1}+\ldots +{\tilde a}_n)}({a_{i+1}}\ldots
  {a_n}{a_1}\ldots {a_i})_c~~.
\ee 
show that $({a_1}\ldots {a_n})_c$ vanishes unless $a_1\ldots
a_n$ is a cycle of ${\cal Q}$ (as we will see below, it can still
vanish even for a cycle). For a general element (\ref{f}), relations (\ref{lindep}) give:
\be
\nn
f_c:=\pi(f)=\sum_{n\geq 0}{f_{a_1\ldots a_n}({a_1}\ldots
  {a_n}})_c=\sum_{n\geq 0}{f_{(a_1\ldots a_n)}({a_1}\ldots
  {a_n})_c}~~,
\ee
where we introduced the `cyclicized coefficients': 
\be
\nn
f_{(a_1\ldots a_n)}:=\frac{1}{n}\sum_{i=0}^{n-1}{(-1)^{({\tilde a}_1+\ldots +{\tilde
      a}_i)({\tilde a}_{i+1}+\ldots +{\tilde a}_n)}f_{a_{i+1}\ldots
    a_n, a_1\ldots a_i}}~~,
\ee
which satisfy $f_{(a_1\ldots a_n)}=(-1)^{({\tilde a}_1+\ldots +
  {\tilde a}_i)({\tilde a}_{i+1}+\ldots +{\tilde a}_n)} f_{(a_{i+1}\ldots
      a_n, a_1\ldots a_i)}$.
For $n=0$, we set $f_{(u)}=f_{u}$ for all $u\in {\cal Q}_0$.

The observations made above show that any element $f\in C^0_R(A)$ can be expanded as: 
\be
\label{fc_expansion}
f=\sum_{n\geq 0}{f_{a_1\ldots a_n}(a_1\ldots a_n)_c}
\ee
where the coefficients are taken to be graded cyclic: 
\be
\nn
f_{a_1\ldots a_n}=(-1)^{({\tilde a}_1+\ldots +
  {\tilde a}_i)({\tilde a}_{i+1}+\ldots +{\tilde a}_n)} f_{a_{i+1}\ldots
      a_n, a_1\ldots a_i}~~
\ee
and the term $n=0$ in the sum stands for $\sum_{u\in {\cal Q}_0}{f_u\epsilon_u}\in R$. 
We also define the {\em strict coefficients} of $f\in C^0_R(A)$ by: 
\be
\nn
{\bar f}_{a_1\ldots a_n}:=nf_{a_1\ldots a_n}~~{\rm if}~~ n\neq 0~~
\ee
and ${\bar f}_{u}:=f_u$ for $n=0$. Then the expansion of $f_c$ becomes:
\be
\nn
f_c=c(f)+\sum_{n\geq 1}{\frac{{\bar f}_{a_1\ldots a_n}}{n}({a_1}\ldots {a_n})_c}~~.
\ee 

Consider the set ${\cal C}({\cal Q})$ of cycles of ${\cal Q}$. We say
that two cycles $\gamma_1$ and $\gamma_2$ are {\em equivalent}, and
write $\gamma_1\sim \gamma_2$, if they have the same length and differ
by a cyclic permutation of their arrows (i.e. they differ only in the
choice of the initial=terminal point of the cycle).  This is an
equivalence relation on ${\cal C}({\cal Q})$, whose equivalence classes are
known as {\em necklaces}. We let $N({\cal Q}):={\cal C}({\cal Q})/\sim$ denote the
set of necklaces, and write $[\gamma]$ for the equivalence class of a
cycle $\gamma$. We define the length $l([\gamma])$ to be the length of
any representative cycle. The $\Z_2$ degree of a necklace is the
degree of any of its representatives.  This gives a well-defined map
from $N({\cal Q})$ to $\Z_2$.

A cycle $\gamma$ is called primitive  if it cannot be
written in the form $\gamma=u^k$ with $u$ a non-trivial cycle and
$k\geq 2$ (with this definition, the trivial paths $\epsilon_u$ are
primitive).  Any non-trivial cycle $\gamma$ can be written uniquely in
the form $\gamma=r(\gamma)^{p(\gamma)}$ where $p(\gamma)\in \N^*$ and
$r(\gamma)$ is a primitive cycle. This representation is called the
primitive decomposition of $\gamma$.  The integer $p(\gamma)$ is
called the period, while the path $r(\gamma)$ is called the primitive
root of $\gamma$.  Given a necklace $\nu$ and representatives
$\gamma_1,\gamma_2\in \nu$, we have $p(\gamma_1)=p(\gamma_2)$ and
$r(\gamma_1)\sim r(\gamma_2)$. This allows us to define the period and
primitive root of necklaces through $p([\gamma])=p(\gamma)$ and
$r([\gamma])=[r(\gamma)]$.

\paragraph{\bf Definition}
A {\em null necklace} is a necklace $\nu$ such that $p(\nu)$ is
even and $\deg~r(\nu)=1\in \Z_2$. 

\

\paragraph{\bf Proposition}
Let $\gamma$ be a cycle of the quiver ${\cal Q}$. The vector
$\gamma_c=\pi(\gamma)\in C^0_R(A)$ vanishes if and only if the necklace
$\nu:=[\gamma]$ is null.

\

\noindent We let $N_\bullet({\cal Q}):=\{ \nu\in N({\cal Q}) |\nu~{\rm~is~{\rm
not}~null} \}$ be the set of non-null necklaces. Notice that the
trivial paths $\epsilon_u$ are not null, and thus belong to $N_\bullet({\cal Q})$.
\begin{proof}

Let $\gamma=r^p$ be the primitive decomposition of $\gamma$. Commuting
one copy of $r$ to the right gives:
\be 
\gamma_c=\pi(r^p)=(-1)^{(p-1){\tilde r}}\gamma_c,\nn 
\ee 
where we noticed that $\deg (r)\deg(r^{p-1})={\tilde r}(p-1)$ in
$\Z_2$.  Thus $\gamma_c=0$ if $p$ is even and ${\tilde r}$ is odd. The
converse follows from the description of $\pi$ given in Section
\ref{tensor} and the fact that $r$ is the period of $\gamma$.
\end{proof}

\noindent

Consider a necklace $\nu=[a_1\ldots
  a_n]$. Relations (\ref{lindep}) show that the subspace $V_\nu:=\C({a_1}\ldots
  {a_n})_c\subset C^0_R(A)$ depends only on $\nu$, while the proposition implies that $V_\nu=0$ if $\nu$ is null
  and $V_\nu\approx \C$ otherwise.  This gives a direct sum
  decomposition: \be C^0_R(A)=\oplus_{\nu \in N_\bullet({\cal Q})}{V_\nu}~~,\nn
  \ee where $V_{u}:=\C\epsilon_u$ for the null paths $\epsilon_u$.
  Accordingly, any element $f\in C^0_R(A)$ decomposes as: \be
  f=\sum_{\nu \in N_\bullet({\cal Q})}{f(\nu)}~~,\nn \ee where all but a finite number
  of the vectors $f(\nu)\in V_\nu$ vanish.  If $l(\nu)=n>0$, then: \be
  f(\nu)=\sum_{a_1a_2\dots a_n\in \nu}f_{a_1a_2\ldots
  a_n}({a_1}\ldots {a_n})_c~~,\nn \ee where the coefficients are
  defined as in (\ref{fc_expansion}). Since $f_{a_1\ldots a_n}$ are
  cyclic, all vectors $f_{a_1\ldots a_n}({a_1}\ldots {a_n})_c\in
  C^0_R(A)$ for which the cycle $a_1\ldots a_n$ belongs to a given
  necklace $\nu$ are equal. Thus: \be\nn f(\nu)={\bar f}_{a_1\ldots
  a_n}({a_1}\ldots {a_n})_c~~~{\rm for~any~fixed~cycle~}~~a_1\dots a_n
  \in \nu~~.  \ee (in this relation, no summation over $a_j$ is
  implied).

In view of these observations, the coefficients in the expansion
(\ref{fc_expansion}) associated to null necklaces are not defined.
For simplicity, we will define all such coefficients to be zero. Thus
all coefficients associated with words $a_1 \ldots a_n$ which fail to
correspond to a cycle or correspond to a cycle in a null necklace are
set to zero by definition. We will use this convention repeatedly in
what follows.

\paragraph{Observation}
$C^0_R(A)$ can be identified with the vector space $\C^{N_{\cal Q}}=\oplus_{\nu
\in N_{\cal Q}}\C\nu $ generated by the set of non-null necklaces $N_{\cal Q}$.  For
this, pick an enumeration of ${\cal Q}_1$, let $c_\nu:=\min \nu$ be the
minimal representative of $\nu$ with respect to the induced
lexicographic order on the set of paths of ${\cal Q}$, and identify $V_\nu$ with $\C$
by sending $\pi(c_\nu)$ into the complex unit.  With this
identification, we have $f(\nu)\equiv {\bar
f}_{c_\nu}$ and $f\equiv \sum_{\nu \in N_\bullet({\cal Q})} {\bar f}_{c_\nu}
\nu$.  Notice that such an identification requires that we pick an enumeration of
${\cal Q}_1$.

\subsection{Cyclic derivatives and loop partial derivatives}
\label{cycloop}

The isomorphism $\Omega^1_RA\approx A\otimes V$ of Section \ref{tensor} 
shows that any one-form $w\in \Omega_R^1 A$ has well-defined coefficients 
$w_a\in A$ determined by:
\be
\nn
w=\sum_{a\in {\cal Q}_1}{(w_a d a)_c}=\sum_{a\in {\cal Q}_1}{ (-1)^{{\tilde a}({\tilde w}+1)}((d a) w_a)_c}~~,
\ee
where we noticed that ${\tilde w}_a={\tilde w}+{\tilde a}$. Inserting
idempotents shows that each $w_a\in A$ must be a linear combination of
paths starting at $h(a)$ and ending at $t(a)$.

We define the {\em left and right quiver cyclic derivatives} $\ld_a f, f\rd_a\in A$ of 
an element $f\in C^0_R(A)$ through:
\be
\label{df}
{\bar d} f=\sum_{a} ((f\rd_a)d a)_c =\sum_{a} (d a (\ld_a f))_c~~.
\ee
This gives linear maps $\ld_a, \rd_a: C^0_R(A)\rightarrow A$, satisfying $f\rd_a=
(-1)^{{\tilde a}({\tilde f}+1)}\ld_a f$. 
Equation (\ref{df}) gives: 
\begin{eqnarray}
\ld_a f&=&\sum_{n} {\bar f}_{a a_1\ldots a_n}a_1\ldots a_n ~~\nn\\
f\rd_a &=&\sum_{n} {\bar f}_{a_1\ldots a_n a }a_1\ldots a_n~~.\nn
\end{eqnarray}
Notice that our convention (\ref{pairing}) is crucial for these simple formulas. 
In particular, for any cycle $\gamma=a_1\ldots a_n$ of ${\cal Q}$: 
\begin{eqnarray}
\label{cd_props_general}
\ld_a({a_1}\ldots {a_n})_c &=&\sum_{i=1}^n{(-1)^{({\tilde a}_1+\ldots
    +{\tilde a}_{i-1})({\tilde a}+{\tilde
      a}_{i+1}+\ldots +{\tilde a}_{n})}\delta_{a,a_i}{a_{i+1}}\ldots
  {a_n} {a_1}\ldots {a_{i-1}}}\\
({a_1}\ldots {a_n})_c\rd_a &=&\sum_{i=1}^n{(-1)^{({\tilde a}+{\tilde a}_1+\ldots +{\tilde a}_{i-1})({\tilde
      a}_{i+1}+\ldots +{\tilde a}_{n})}\delta_{a,a_i}{a_{i+1}}\ldots
  {a_n} {a_1}\ldots {a_{i-1}}}~~.\nn
\end{eqnarray}
These equations imply:
\begin{eqnarray}
\label{cd_proj}
\pi(\ld_a({a_1}\ldots {a_n})_c) &=&\sum_{i=1}^n(-1)^{{\tilde a}({\tilde a}_1+\ldots
    +{\tilde a}_{i-1})}\delta_{a,a_i}(a_1\ldots a_{i-1}a_{i+1}\ldots a_n)_c\\
\pi(({a_1}\ldots {a_n})_c \rd_a) &=&\sum_{i=1}^n (-1)^{{\tilde a}({\tilde a}_{i+1}+\ldots
    +{\tilde a}_n)}\delta_{a,a_i}(a_1\ldots a_{i-1}a_{i+1}\ldots a_n)_c~~.\nn
\end{eqnarray}
It is clear that all terms in right hand side vanish 
unless $a_i$ is a loop.  Hence the
linear operators $\pi\circ \ld_a$ and $\pi\circ \rd_a$ induced
on $C^0_R(A)$ are non-trivial only when  $a$ is a loop.

We also define {\em loop partial derivatives} $\lp_a,\rp_a$ if $a$ is
a (non-trivial) {\em loop} of the quiver.  Given such a loop of
${\cal Q}$, the loop derivatives $\lp_a\in \Der_l(A)$ and $\rp_a\in
\Der_r(A)$ are the unique $R$-linear left and right derivations of
$A$ of degree ${\tilde a}$ such that:
\be
\label{lp_def}
\lp_a b:=b\rp_a=\delta^b_a \epsilon_u ~~{\rm~for~all}~b\in {\cal Q}_1~~,
\ee
where $u=h(a)=t(a)$. 
It is not hard to see that $\lp_a$ and $\rp_a$ are well-defined and $\lp_a f=(-1)^{{\tilde a}({\tilde f}+1)}f\rp_a$.
It is clear from  (\ref{lp_def}) that loop partial derivatives supercommute:
\be
\nn
[\lp_a,\lp_b]=[\rp_a,\rp_b]=0~~{\rm for~all~loops}~~a,b\in {\cal Q}_1.
\ee

Let ${\cal Q}_1(u)$ be the subset of loops at the vertex $u$ and 
let $a^*=\sum_{b\in {\cal Q}_1}{\omega_{ab}b}=\sum_{b\in {\cal Q}_1(u)}{\omega_{ab}b} \in A$  be the 
conjugate element of the loop $a$ introduced in Subsection \ref{symplectic_structure}. 
A simple computation gives the relation: 
\be
\nn
{\bar i}_{\lp_a}\omega={\bar d}\pi(a^*)~~,
\ee
which shows that $\lp_a$ is the symplectic relative left derivation
with Hamiltonian $\pi(a^*)\in C^0_R(A)$. Similarly, one can
view $\rp_a$ as the symplectic relative right derivation with Hamiltonian
$\pi(a^*)$. Notice that the necklace of a loop $a$ has only one
representative, so one can identify loops with their projections
through $\pi$; we will sometimes do so in what follows.

Let $\partial^r_a$ and $\partial_a^l$ be the complex-linear maps
induced by $\lp_a$ and $\rp_a$ on $C_R^0(A)=A/[A,A]$.  For any
cycle $\gamma=a_1\ldots a_n$, we have:
\begin{eqnarray}
\partial^l_a\gamma_c&=&\pi(\lp_a \gamma)=\sum_{i=1}^n(-1)^{{\tilde a}({\tilde a}_1+\ldots
    +{\tilde a}_{i-1})}\delta_{a,a_i}(a_1\ldots a_{i-1}a_{i+1}\ldots a_n)_c~~\nn\\
\gamma_c\partial^r_a &=&\pi(\gamma\rp_a)=\sum_{i=1}^n(-1)^{{\tilde a}({\tilde
      a}_{i+1}+\ldots +{\tilde a}_{n})}\delta_{a,a_i}(a_1\ldots a_{i-1}a_{i+1}\ldots a_n)_c~~.\nn
\end{eqnarray}
Comparing with (\ref{cd_proj}) gives: 
\be
\nn
\partial^l_a=\pi\circ \ld_a~~,~~\partial^r_a=\pi\circ \rd_a~~.
\ee
Hence the cyclic and loop partial derivatives induce the same complex-linear operators on $C^0_R(A)$. 

Given an element $x\in E[1]$, we expand $x=\sum_{a\in {\cal Q}_1}{x^a e_a}$ with
$x^a\in \C$ and define the (relative) left cyclic derivative along $x$
via:
\be
\nn
\ld_x f:=\sum_{a\in {\cal Q}_1}{x^a\ld_a f}~~~~\forall f\in C^0_R(A)~~.
\ee
If $x$ is central in $E$, then $x^a$ vanish unless $a$ is a loop. In
this case, we define the (relative) loop left partial derivative along
$x$ via:
\be
\nn
\lp_x f=\sum_{a=loop}{x^a\lp_a f}~~.
\ee
For a central element, these two notions induce the same map on $C^0_R(A)$:
\be
\nn
\partial^l_x f=\sum_{a=loop}{x^a\partial^l_a f}~~.
\ee
These definitions are well-behaved with respect to changes of adapted bases. 
It is clear that $\partial_{e_a}^l=\partial_a^l$ etc. 

\paragraph{Observations}

(1) Loop partial derivatives and quiver cyclic derivatives are related to certain double 
derivations introduced in \cite{Bergh}. Consider 
the $A$-superbimodule $A\otimes_\C A$, where we use the so-called outer superbimodule structure: 
\be
\nn
\alpha (a\otimes b)\beta:=(\alpha a)\otimes (b\beta)~~~~\forall \alpha,\beta, a, b \in A~~. 
\ee
Since $R$ is a subalgebra of $A$, this is also an $R$-superbimodule. A {\rm relative double derivation} of $A$ 
is an $R$-linear derivation of the $A$-superbimodule $A\otimes A$. 
As in \cite{Bergh}, consider the double left derivations determined by: 
\be
\nn
{\bf D}_a(b)=\epsilon_{t(a)}\otimes \epsilon_{h(a)}\delta^b_a~~~~~\forall a,b\in {\cal Q}_1~~.
\ee
Then the loop partial derivatives can be recovered as:
\be
\nn
\lp_a={\bf m}\circ {\bf D}_a~~{\rm for~~}a{\rm ~~a~loop}
\ee
where ${\bf m}:A\otimes A\rightarrow A$ is the $A$-superbimodule morphism: 
\be
\nn
{\bf m}(a\otimes b)=ab~~.
\ee
Similarly, relations (\ref{cd_props_general}) show that the cyclic
derivatives are induced by the maps $m\circ {\bf D}_a$, where
$m:A\otimes A\rightarrow A$ is the linear map ({\em not} a bimodule
morphism !): \be \nn m(a\otimes b)=(-1)^{{\tilde a}{\tilde b}}ba~~.
\ee Indeed, it is not hard to see that $m\circ {\bf D}_a$ vanishes on
$[A,A]$, so it induces a map from $C^0_R(A)$ to $A$ which coincides
with the cyclic derivative $\ld_a$.

(2) Let ${\cal Q}$ be a quiver with a single vertex $u$. Then all
arrows are loops and $A$ is the free superalgebra $\C\langle
\{a\}\rangle$ generated by these loops.  In this case, $\lp_a$ and
$\rp_a$ reduce to the standard left and right partial derivatives of
the free algebra $A$. Moreover, the observations above show that
$\ld_a$ are a superized version of the objects considered in
\cite{RSS, Voiculescu}. This justifies our terminology.

\subsection{Description of one-forms and closed two-forms}
\label{12forms}

Consider the reduced tensor algebra $A_{\geq 1}:=V\otimes_R
A=T_R^{\geq 1} V=\oplus_{n\geq 1}{V^{\otimes_R n}}$, which coincides
with the subspace of $A$ spanned by all paths of length at least one. As in \cite{Ginzburg,LB_quivers} 
(see also \cite{Lazarev}), it is easy to see that the super-vector space $C^1_R(A)$ is isomorphic
with the center $A_{\geq 1}^R=(V\otimes_R A)^R$, the space spanned by the non-trivial cycles of the quiver. 
The isomorphism $\Xi:A_{\geq 1}^R\rightarrow C^1_R(A)$ has the form: 
\be
\label{Xi}
\sum_{i}{x_i\otimes_R f_i}\stackrel{\Xi}{\rightarrow} \sum_{i}{(dx_i f_i)_c}\in C^1_R(A)~~.
\ee
This follows from the observation that any element of $C^1_R(A)$ can be
written uniquely as: 
\be
\nn
w=\sum_{n\geq 0}f_{a a_1\ldots a_n}(da a_1\ldots a_n )_c=\sum_{a}(da f_a)_c=\Xi(\sum_a af_a)~~,
\ee
where $f_a:=\sum_{n\geq 0}f_{a a_1\ldots a_n }a_1\ldots a_n$ and 
the complex coefficients $f_{a_1\ldots a_n}$ are taken to vanish unless $a_1\ldots a_n$ is 
a cycle. Then $\Xi^{-1}(w)=\sum_{n\geq 0} f_{a_1\ldots a_n}a_1\ldots a_n:=f=\sum_a af_a$. 
For $g=\sum_{n\geq 0}g_{a_1\ldots a_n} a_1\ldots a_n \in A^R_{\geq 1}$, we have 
${\bar d}\pi(g)={\bar d}g_c=(da\ld_a g_c)_c=
\sum_{n\geq 0} {\bar g}_{(a a_1\ldots a_n)}(da a_1\ldots a_n)_c$, where 
${\bar g}_{(a_1\ldots a_n)}:=n g_{(a_1\ldots a_n)}$ are the strict coefficients of $g_c$. 
Thus $w={\bar d}(g)_c$ iff 
$f_{a_1\ldots a_n}={\bar g}_{(a_1\ldots a_n)}$. It is clear from these observations that 
$w$ is exact iff its coefficients $f_{a_1\ldots a_n}$ are graded-cyclic; in this case, we 
can take $g=\frac{1}{n} f_{a_1\ldots a_n}a_1\ldots a_n$ and we have $f_a=\ld_a g$. 

Thus $\Xi(f)\in C^1_R(A)_{\rm closed}= C^1_R(A)_{\rm exact}$. We let $A_{\geq 1}^c$ be the {\em
cyclic subspace} of $A_{\geq 1}^R$, i.e. the subspace 
consisting of elements with graded-cyclic coefficients. The space $A_{\geq 1}^c$
consists of linear combinations of non-trivial cycles, such
that cycles belonging to the same necklace of length $n$ appear with
coefficients related by the action of $\Z_n$. This subspace provides
an embedding of $C^0_R(A)/R$ into $A_{\geq 1}$. In fact, the projection
$\pi:A\rightarrow C^0_R(A)$ induces an isomorphism $A_{\geq
1}^c\approx C^0_R(A)/R=A_{\geq 1}/[A,A]$ (recall that
$[A,A]_0=[R,R]=0$ so $[A,A]\subset A_{\geq 1}$). Thus we have a vector
space decomposition $A_{\geq 1}=[A,A]\oplus A_{\geq 1}^c$, as well as 
$A_{\geq 1}^R=[A,A]^R\oplus A_{\geq 1}^c$. In particular, the subspace $A_{\geq 1}^c$ gives a 
natural complement of $[A,A]^R$ inside $A_{\geq 1}^R$.

Since the Karoubi complex is acyclic in positive degrees, we have $C^2_R (A)_{\rm closed}=
{\bar d}(C^1_R (A))$ and the isomorphism $\Xi$ shows that any closed 
two-form can be written as: 
\be
\label{dXi}
u=-\sum_{n\geq 0}{(da df_a)_c}={\bar d}\Xi(f)~~.
\ee
where $f=af_a\in A_{\geq 1}^R$ is a combination of cycles of 
length at least one. Since $\Xi(A_{\geq 1}^c)=C^1_R(A)_{\rm closed}$, the two-form (\ref{dXi}) 
vanishes precisely when $f$ belongs to the cyclic subspace $A_{\geq 1}^c$. Thus $\ker(d\Xi)=A_{\geq 1}^c$, 
and the map $d\Xi$ induces an isomorphism:
\be
\label{kappa_gen}
\kappa:[A,A]^R\stackrel{\approx}{\rightarrow} C^2_R (A)_{\rm closed}
\ee
between the complement $[A,A]^R$ of this subspace in $A_{\geq 1}^R$ and the space 
of closed two forms\footnote{This result is also discussed in \cite{VG}. I thank V. Ginzburg for pointing this 
out.}.

\subsection{Quiver description of the Kontsevich bracket}

Consider the non-commutative symplectic form (\ref{omega_form}) on $A$.
For $\theta\in \Der_l(A)$, we set $\theta(a):=\theta^a\in A$. $R$-linearity of $\theta$ implies 
that $\theta^a$ is a linear combination of paths which start at $t(a)$ and end at $h(a)$.
Equation (\ref{df}) gives:  
\be
\nn
{\bar L}_\theta f={\bar i}_\theta {\bar d} f=
\pi(i_\theta\sum_{a\in {\cal Q}_1} (da) \ld_a f)=\sum_{a\in {\cal Q}_1}{(\theta(a )\ld_a f)_c}~~.
\ee
Thus: 
\be
\label{dftheta}
{\bar L}_\theta f=\sum_{a\in {\cal Q}_1}(\theta^a \ld_a f)_c~~~~\forall \theta\in \Der_l(A)~~.
\ee
If $\theta$ is homogeneous of degree ${\tilde \theta}$, we set: 
\be
\theta_a:=\theta^b\omega_{ba}~~.\nn
\ee
Notice that $\tilde{\theta_a}={\tilde a}+{\tilde \omega}+{\tilde \theta}$. Expanding 
$\theta^a=\sum_{n\geq 0}\theta_{a_1\ldots a_n}\,\,^a {a_1}\ldots {a_n}$ with $\theta_{a_1\ldots a_n}\,\,^a\in \C$, 
we find  $\theta_a=\sum_{n\geq 0}\theta_{a_1\ldots a_n a}{a_1}\ldots {a_n}$, where: 
\be
\nn
\theta_{a_1\ldots a_na}=\theta_{a_1\ldots a_n}\,\,^b\omega_{ba}~~.
\ee
As usual,  $\theta_{a_1\ldots a_n}\,\,^a$ are taken to vanish  unless $a_1\ldots a_n$ is a path. Also notice 
that  $\theta_{a_1\ldots a_n}\,\,^a$ vanishes automatically unless this path starts at $t(a)$ and ends at $h(a)$.
Similarly, $\theta_{a_1\ldots a_n}$ vanishes unless $a_1\ldots a_n$ is a cycle of ${\cal Q}$. 

An easy computation gives:
\be
\label{hash}
{\bar i}_\theta \omega=(\theta_a da)_c=\sum_{n\geq 0}\theta_{a_1\ldots a_n a}({a_1}\ldots {a_n}da)_c~~.
\ee
Given $f\in C^0_R(A)$, we have ${\bar d}f=(f\rd_a
da)_c$. Comparing with (\ref{hash}) gives $(\theta_f)_a=f\rd_a$, where
$\theta_f$ is the Hamiltonian vector field of $f$.  Hence the map
$\psi_\omega:C^0_R(A)\rightarrow \Der_l(A)$ of (\ref{ex}) is
given by:
\be
\label{phiomega}
\theta_f(a):=\theta_f^a=f\rd_b\omega^{ba}=\sum_{n\geq 0}{\bar f}_{a_1\ldots a_n b}\omega^{ba} s^{a_1}\ldots s^{a_n}~~,
\ee
where ${\bar f}_{a_1\ldots a_n}$ are the strict 
coefficients of $f$. This allows us to write the Kontsevich bracket in more familiar form. 

\paragraph{\bf Proposition} 
\label{coord_bracket}
We have $\{f,g\}= (f\rd_a\omega^{ab}\ld_b g)_c$ for all $f,g\in
C^0_R(A)$. 

\

\begin{proof}

Using (\ref{bL}), (\ref{dftheta}) and (\ref{phiomega}), we compute
$\{f,g\}={\bar L}_{\theta_f}(g)=(\theta_f^a\ld_a
g)_c=(f\rd_b\omega^{ba}\ld_a g)_c$.

\end{proof}

\subsection{Some canonical forms and coefficient expressions}
\label{canforms}

In this subsection, we  give some expressions which are useful in applications. 
Let $W\in C^0_R(A)$ be an element of degree ${\tilde \omega}+1$. As we will see in 
the next section, the boundary generating function of a topological D-brane system 
is such an element.

For ${\tilde \omega}=0$, let us choose an adapted basis as in (\ref{e_can}). 
Then $W$ is odd, and one finds:
\be
\nn
\{W, W\}=(W\rd_{p_i}\ld_{q_i}W-W\rd_{q_i}\ld_{p_i}W)_c+(W\rd_{\xi_\alpha}\ld_{\xi_\alpha}W)_c=
2(W\rd_{p_i}\ld_{q_i}W)_c+(W\rd_{\xi_\alpha}\ld_{\xi_\alpha}W)_c
\ee
since $(W\rd_{p_i}\ld_{q_i}W)_c=-(W\rd_{q_i}\ld_{p_i}W)_c$. Also notice that $W\rd_{\xi_\alpha}=\ld_{\xi_\alpha} W$. 

Now let ${\tilde \omega}=1$ and choose an adapted basis as in (\ref{o_can}). 
Then $W$ is even and we have:
\be
\nn
\{W, W\}=(W\rd_{p_i}\ld_{q_i}W-W\rd_{q_i}\ld_{p_i}W)_c=2(W\rd_{p_i}\ld_{q_i}W)_c~~,
\ee
since again $(W\rd_{p_i}\ld_{q_i}W)_c=-(W\rd_{q_i}\ld_{p_i}W)_c$. 

One can also extract the coefficient expression of the cyclic bracket by
direct computation. The case relevant for us is as follows.  For $W$ as above, notice that $\rho^{ab}$ can be used to
raise and lower indices `from the left':
\begin{eqnarray}
\label{lifts}
{\bar W}_{a_1\ldots a_{i-1}} {\,\, }^a_{\,\,a_{i+1}\ldots
a_{n}}&:=&\rho^{ab}{\bar W}_{a_1\ldots a_{i-1}b a_{i+1}\ldots
a_n}~~.\nn\\ {\bar W}_{a_1\ldots a_{i-1}b a_{i+1}\ldots
a_n}&:=&\rho_{ab}{\bar W}_{a_1\ldots a_{i-1}} {\,\, }^b_{\,\,
a_{i+1}\ldots a_{n}}\nn~~.
\end{eqnarray}
Then it is shown in Appendix \ref{coeffs} that that bracket of $W$ with itself takes the form: 
\be
\frac{1}{2}\{W,W\}=\frac{1}{2}{\bar W}_a{\bar W}^a+
\sum_{n\geq 1}\frac{1}{n}\left(\sum_{0\leq i+j\leq n}(-1)^{{\tilde a_1}+\ldots +{\tilde a}_i}
{\bar W}_{a_1\ldots a_i a a_{i+j+1}\ldots a_n} {\bar W}^a_{\,\,a_{i+1}\ldots
      a_{i+j}}\right)({a_1}\ldots {a_n})_c~~,\nn
\ee
which is valid irrespective of the degree of $\omega$.

\section{Geometry of finite D-brane systems}
\label{geometrization}

Consider a finite topological D-brane system with total boundary space
$E$ and boundary algebra $R$.  As before, we let $V=E[1]^{\rm v}$ and
consider the tensor algebra $A=T_R V$. As explained in Section
\ref{algebraic}, the data of all integrated boundary correlators on
the disk is encoded by an $R$-superbimodule structure on $E$, together with
a cyclic and unital weak $A_\infty$ structure on this
superbimodule. We will use the machinery developed in the previous two
sections to encode this into a `noncommutative
generating function' $W\in C^0_R(A)$ subject to simple constraints. To
this end, we pick an adapted basis of $E$ and let ${\cal Q}$ be its
index superquiver.

\subsection{Geometric description of cyclic weak $A_\infty$ structures}

It turns out that a weak $A_\infty$
structure on the $R$-superbimodule $E$ is the same as an {\em odd}
relative derivation $Q$ of the tensor algebra $A$.
With our conventions, the relation is as follows.  Picking adapted
coordinates, we define the coefficients of $Q$ through:
\be
\label{Q_exp}
Q(a)=\sum_{n\geq 0} Q_{a_1\ldots a_n}{\,}^a a_1\ldots a_n
\ee
and construct odd linear maps $r_n:E[1]^{\otimes n}\rightarrow E[1]$ via: 
\be
\label{rQ}
r(e_{a_1}\ldots e_{a_n})=Q_{a_1\ldots a_n}{\,}^a e_a~~. 
\ee
Thus $Q(s^a)=\sum_{n\geq 0} s^a(r(e_{a_1}\ldots e_{a_n})) s^{a_1}\otimes_R \ldots \otimes_R s^{a_n}$.
Thinking in terms of arrows, it is clear that $r_n$ are $R$-multilinear.
Since $Q$ is odd, we have $[Q,Q]= 2 Q^2$, which implies that $Q^2$ is
a derivation of $A$. Since $a$ generate the algebra, this means that the condition $Q^2=0$ is equivalent with
$Q^2(a)=0$ for all $a$. Using expansion (\ref{Q_exp}), one finds
that this amounts to the relations:
\be
\nn
\sum_{0\leq i+j\leq n}(-1)^{{\tilde a_1}+\ldots +{\tilde a}_i}
Q_{a_1\ldots a_i b a_{i+j+1}\ldots a_n}{\,}^a Q_{\,\,a_{i+1}\ldots
      a_{i+j}}{\,}^b=0~~{\rm for~all~}~~ n\geq 0~~,
\ee
which are the $A_\infty$ constraints (\ref{ainf}).

It is also not hard to check that the nilpotent derivation $Q$ is
symplectic iff the associated $A_\infty$ structure is cyclic.  An easy
way to see this is as follows. By the exact sequence (\ref{ex}), we
have that $Q$ is symplectic iff it is Hamiltonian, which via equation
(\ref{phiomega}) amounts to the existence of a $W\in C^0_R(A)$ such
that:
\be
Q_{a_1\ldots a_n}{\,}^a={\bar W}_{a_1\ldots a_nb}\omega^{ba}~~
\ee
or, equivalently:
\be
\label{QW}
Q_{a_1\ldots a_n}={\bar W}_{a_1\ldots a_n}~~.
\ee
As usual, we have set $Q_{a_1\ldots a_n a}:=Q_{a_1\ldots a_n}{\,}^b\omega_{ba}$. 
It is clear that a $W$ exists if and only if 
the coefficients $Q_{a_1\ldots a_n}$ are cyclic. We have:
\begin{eqnarray}
\label{cyc_arg}
\rho(e_{a_0},r_n(e_{a_1}\ldots e_{a_n}))&=&(-1)^{{\tilde a}_0}\omega(e_{a_0},
r_n(e_{a_1}\ldots e_{a_n}))=(-1)^{1+{\tilde a}_0{\tilde \omega}}
\omega(r_n(e_{a_1}\ldots e_{a_n}), e_{a_0})\nn\\
&=&(-1)^{{\tilde a}_0 {\tilde \omega}}{\bar W}_{a_1\ldots a_n a_0}={\bar W}_{a_0\ldots a_n}~~,
\end{eqnarray}
where we used the superselection rules for $\omega$ and $W$.
Thus:
\be
\label{QW1}
\rho(e_{a_0},r_n(e_{a_1}\ldots e_{a_n}))={\bar W}_{a_0a_1\ldots a_n}~~,
\ee
and we see that $L_Q \omega=0$ implies that the left hand side is
cyclic, which is the cyclicity constraint (\ref{rrcyc}).  Conversely,
if the LHS is cyclic then we define $W$ through equation
(\ref{QW1}). Then relations (\ref{cyc_arg}) show that $Q_{a_1\ldots
a_n a_0}={\bar W}_{a_1\ldots a_n a_0}$, i.e. $Q$ is symplectic with
Hamiltonian $W$.  Combining everything and  noticing that ${\tilde W}={\tilde \omega}+1$ (because
$Q$ is odd), we have:

\

{\em Giving a cyclic weak $A_\infty$ structure on $E$ amounts to giving an element $W\in C^0_R(A)$, 
of degree ${\tilde \omega}+1$,  such that $\{W,W\}=0$.}

\

\paragraph{Observation} The triplet $(A,Q,\omega_{form})$ can be viewed as a noncommutative generalization 
of the so-called $QP$-manifolds of \cite{Konts_Schwarz}, while the doublet $(A,Q)$ generalizes the concept of 
$Q$-manifold discussed in the same paper (notice, though, that we consider both even and odd symplectic forms, 
so we generalize the work of \cite{Konts_Schwarz} in two directions). It was shown in  \cite{Konts_Schwarz} that 
$QP$-manifolds give the general geometric setting of the classical BV-formalism. 
Accordingly, for odd symplectic forms, the 
triplet $(A,Q,\omega_{form})$ defines a noncommutative version of that formalism.

\subsection{The unitality constraint}
\label{sec:unitality}

We saw that a cyclic weak $A_\infty$ structure on $E$ is the same as
an $R$-linear symplectic derivation $Q\in \Der_l^\omega(A)$ such that
$Q^2=0$.  We let $W$ be the {\em canonical} Hamiltonian of $Q$, i.e. that Hamiltonian which vanishes at zero 
(see Section \ref{tensor}). Explicitly, equations (\ref{QW}) and (\ref{QW1}) give: 
\be
\nn
W=\sum_{n\geq 0}{\frac{1}{n+1}\rho(e_{a_0}, r_n(e_{a_1}\ldots e_{a_n}))(s^{a_0}\ldots s^{a_n}})_c~~,
\ee
which allows us to reconstruct $r_n$ from $W$ provided that we know $\omega=\rho\circ \Sigma^{\otimes 2}$.
The homological derivation $Q$ can be recovered as the Hamiltonian derivation defined by  $W$, 
which amounts to relations (\ref{QW}). 

For a topological D-brane system, the underlying weak $A_\infty$ structure should be unital.
To formulate this condition in terms of $W$, we write the unitality constraints (\ref{unitality}) as:
\begin{eqnarray}
\label{ueqs}
r_n(e_{a_1}\ldots e_{a_{j-1}},\lambda,e_{a_{j+1}}\ldots e_{a_n})&=&0~~{\rm for~all}~~~~ n\neq 2 ~{\rm~and~all}~
j=1\ldots n~~\nn\\
-r_2(\lambda,e_{a})=(-1)^{{\tilde a}} r_2(e_{a},\lambda)&=&e_a~~,
\end{eqnarray}
where $\lambda$ is an odd central element of $E$. 
Given $\lambda=\oplus_{u\in {\cal Q}_0}{\lambda_u}\in E[1]^R$ with $\lambda_u\in E_{uu}[1]$, we
can choose adapted coordinates such that each $\lambda_u$ is one of the odd basis elements $\{e_a\}$. 
We then let $\sigma_u\in E[1]^{\rm v}$ be the corresponding elements of the dual basis $\{s^a\}$ of $V$ (those 
dual basis elements which satisfy $s^{\sigma_u}(e_a)=\delta_{e_a,\lambda_u}$). 
It is clear that each $\sigma_u$ is an odd loop of the quiver starting and ending at the vertex $u$. With such 
a choice of adapted basis, we have $\lambda_u=e_{\sigma_u}$ and the first row in (\ref{ueqs}) is equivalent with:
\be
\label{un1}
W_{a_1\ldots a_n}=0 {\rm ~if~}n\neq 3{\rm ~and~any~of~the~arrows~}
a_j{\rm ~coincides~with~any~of~the~loops}~\sigma_u~~
\ee
while the second row amounts to: 
\be
\label{un2}
{\bar W}_{\sigma_u ab}=-\omega_{ab}\Longleftrightarrow   W_{\sigma_u ab}=-\frac{1}{3}\omega_{ab}
~~~{\rm~for~}t(a)=h(b)=u~~{\rm~and~~}h(a)=t(b)~~. 
\ee
Hence unitality of $(r_n)$ boils down to the requirement that
the adapted basis $\{\psi_a\}$ can be chosen such that the vertices 
the index quiver carry distinguished odd loops $\sigma_u$ satisfying
(\ref{un1}) and (\ref{un2}).

The two conditions above say that $W$ takes the form:
\be
\label{W_decomp}
W=W_g +W_d
\ee
where the `generic' contribution is given by:
\be
\label{Wg}
W_g:=-\omega_{ab}(\sigma a b)_c=
-\sum_{\tiny \begin{array}{c}a,b\in {\cal Q}_1, u\in {\cal Q}_0\\t(a)=h(b)=u\\h(a)=t(b)
\end{array}}{\omega_{ab}(\sigma_u ab)_c}~~
\ee
while the `deformation part'  
$W_d$ vanishes at zero and is independent of all $\sigma_u$. In the first form of the last expression, 
we used Einstein summation over $a$ and $b$ and have set $\sigma=\sum_{u\in {\cal Q}_0}{\sigma_u}\in A_1$. 
Notice the lack of a $1/3$ prefactor in (\ref{Wg}); 
this is because we brought all terms to a form in which a $\sigma_u$ insertion appears in the first position.

To describe this more elegantly, notice\footnote{For an 
element $f=\sum_{n\geq 0} f_{a_1\ldots a_n}(a_1\ldots a_n)_c\in C^0_R(A)$, the condition $\ld_a f=0$ amounts 
to vanishing of all cyclic coefficients $f_{a_1\ldots a_n}$ for which one of the $a_j$ coincides with $a$. 
Further, the cyclic derivative $\ld_{\lambda }W$ determines $\ld_{\sigma_u}W=\epsilon_u \ld_{\lambda} W$. 
Thus (\ref{W_decomp}) and (\ref{Wg}) amount to $\ld_\lambda\left(W+\omega_{ab}(\sigma a b)_c\right)=0$, 
which gives the desired statement.} 
that (\ref{W_decomp}) together with (\ref{Wg}) amounts to the condition: 
\be
\nn
\ld_\lambda W=-\sum_{a,b\in {\cal Q}_1}\omega_{ab} ab~~
\ee
where $\lambda:=\sum_{u\in {\cal Q}_0}{\lambda_u}$ and 
$\ld_\lambda:=\sum_{u\in {\cal Q}_0} \ld_{\sigma_u} $ as in Subsection \ref{cycloop}. 
Using the graded antisymmetry of $\omega_{ab}$, the last relation takes the form:
\be
\label{ldcond1}
\ld_\lambda W=-\frac{1}{2}\sum_{a\in {\cal Q}_1}[a,a^*]\in [A,A]_2^R~~,
\ee
where we introduced the conjugate variables
\be
\label{adual_expansion}
a^*:=\sum_{b\in {\cal Q}_1}\omega_{ab}b=
\sum_{\tiny \begin{array}{c}b\in {\cal Q}_1 \\h(a)=t(b),  h(b)=t(a)\end{array}}{\omega_{ab}b}~~
\ee 
as in Subsection \ref{symplectic_structure}.  Here $[A,A]_2$ is the subspace of $[A,A]$ 
consisting of elements of degree two with respect to the
$\N$-grading, while $[A,A]_2^R\subset [A,A]^R$ is the centralizer of $[A,A]_2$ in $R$.  

Relation (\ref{ldcond1}) allows one reconstruct $\omega$ from $W$. To
formulate this invariantly, remember from Subsection \ref{12forms}
that the space of closed noncommutative two-forms $C^2_R
(A)_{\rm closed}$ is isomorphic with $[A,A]^R$.  Restricting the map
(\ref{kappa_gen}) to the subspace $C^2_R (A)_2 \subset C^2_R(A)_{\rm
closed}$ of constant two-forms gives an isomorphism
$[A,A]_2^R\stackrel{\approx}{\rightarrow} C^2_R (A)_2$ whose
explicit form is given by relation (\ref{dXi}):
\be
\label{kappa_0}
\sum_{a,b\in {\cal Q}_1}{f_{ab}ab}\stackrel{\kappa}{\rightarrow} -\sum_{a, b\in {\cal Q}_1}f_{ab}(dadb)_c~~.
\ee
Here $f=\sum_{a,b}f_{ab}ab$ is the general element of $[A,A]_2^R$, with graded-antisymmetric complex coefficients 
$f_{ab}=(-1)^{1+{\tilde a}{\tilde b}}f_{ba}$, so we can also write $f=\frac{1}{2}\sum_{a,b}f_{ab}[a,b]$. 
Applying this to the noncommutative symplectic form, we find:
\be
\kappa^{-1}(\omega)=-\frac{1}{2}\omega_{ab}ab=-\frac{1}{4}\sum_{a\in {\cal Q}_1}[a,a^*]~~.
\ee
Thus relation (\ref{ldcond1}) can be written as either of the following equivalent conditions: 
\be
\label{diff_unitality}
\ld_\lambda W=2\kappa^{-1}(\omega)\Leftrightarrow \omega=\frac{1}{2}\kappa(\ld_\lambda W)~~.
\ee
These observations allow us to write the unitality constraint
(\ref{ldcond1}) as condition (\ref{diff_unitality}). In
particular, the element $\ld_\lambda W\in [A,A]_2^R $ must belong to the 
subspace spanned by the $\kappa$-preimages of quiver symplectic forms.  To
describe this space, notice that any element $\mu\in [A,A]_2^R$ can
be expanded uniquely as:
\be
\mu=-\frac{1}{4}\sum_{a\in {\cal Q}_1}{[a,a^*]}~~,
\ee
where each $a^*$ is a linear combination of arrows going from $h(a)$
to $t(a)$.  We say that $\mu$ is {\em non-degenerate} if the elements
$(a^*)_{a\in {\cal Q}_1}$ form a basis of $V$; in this
case, we can expand $a^*$ as in (\ref{adual_expansion}), with
coefficients $\omega_{ab}$ forming the entries of a
graded-antisymmetric non-degenerate matrix. Moreover, it is clear that $\mu$ has 
$\Z_2$-degree ${\tilde \omega}$ iff this matrix satisfies the selection 
rules (\ref{omega_sel}). We let  $\Mom_V\subset [A,A]_2^R$ 
be the $\Z_2$-homogeneous subspace of non-degenerate elements in $[A,A]_2^R$ and 
let $\Mom^0_V$ and $\Mom^1_V$ be its homogeneous components.
The observations made above show that $\kappa$ induces an isomorphism between 
the space $CNS^{\tilde \omega}(V)$ of constant 
non commutative symplectic forms on $A$ having degree ${\tilde \omega}$ and the space 
$\Mom^{\tilde \omega}_V$. It follows from Proposition 8.1.1 of \cite{BEV} that
${\rm Mom}_V$ is the space of noncommutative moment maps
associated to quiver symplectic forms on the path algebra $A$.

\

\noindent We can now formulate the unitality criterion as follows: 

\paragraph{\bf Proposition}
Let $E[1]$ be a symplectic $R$-superbimodule of finite complex
dimension whose symplectic form has degree ${\tilde \omega}$, let 
$A:=T_R E[1]^{\rm v}$ and let $W$ be an element of $C^0_R(A)$
which has degree ${\tilde \omega}+1$ and vanishes at zero. Let $\omega$ be the noncommutative symplectic
form induced on $A$ and assume that $\{W,W\}=0$, where $\{.,.\}$ is the Kontsevich bracket defined by
$\omega$.  Then the following statements are equivalent:

(1) The cyclic weak $A_\infty$ structure determined by $W$ on $E[1]$
    is unital

(2) There exists an odd central element $\lambda\in E^R$ such that
    $\frac{1}{2}\ld_\lambda W=\kappa^{-1}(\omega)$.

\noindent In this case:

(a) $\Sigma \lambda$ is the unit of the $A_\infty$ structure.

(b) The element $\mu:=\frac{1}{2}\ld_\lambda W\in A$ belongs to the
subspace $\Mom_V$ of $A$

(c) The non-commutative symplectic form can be recovered via the
relation $\omega=\kappa (\mu)$.

(d) Let $\lambda=\sum_{u\in {\cal Q}_0}{\lambda_u}$ ($\lambda_u\in
E_{uu}$) be the decomposition of $\lambda$, and choose an adapted
basis $e_a$ of $E[1]$ containing $\lambda_u$ among the basis
elements. Let $s^a\equiv a$ be the dual basis, and let $\sigma_u$ its
elements associated with $\lambda_u$.  Then $\sigma_u$ correspond to odd
loops of the associated superquiver and $W$ takes the form given in
eqs. (\ref{W_decomp}) and (\ref{Wg}), where $W_d$ vanishes at zero and
is independent of all $\sigma_u$.

\

\subsection{Noncommutative geometry of D-brane systems}

Combining the discussion of the previous subsections, we have the following non-commutative geometric description 
of finite topological D-brane systems:

\

{\em Let $R$ be a finite-dimensional semisimple commutative algebra over $\C$ 
and $E$ an $R$-superbimodule which is finite-dimensional over $\C$. 
Giving a finite topological D-brane system with boundary
decomposition described by ($R$, $E$) and topological metrics of  $\Z_2$-degree ${\tilde \omega}$ amounts to giving
a `noncommutative function' $W\in C^0_R(A)$ on the tensor algebra $A=T_R E[1]^{\rm v}$ 
and an odd central element $\lambda \in E^R$ such that
the following conditions are satisfied:

\

\noindent (1) $W$ vanishes at zero and has $\Z_2$-degree ${\tilde \omega}+1$.

\noindent (2) The element $\mu:=\frac{1}{2}\ld_\lambda W$ belongs to $\Mom_V^{\tilde \omega}$ 

\noindent (3) We have $\{W,W\}=0$, where $\{.,.\}$ is the Kontsevich bracket determined on $C^0_R(A)$ by 
the constant noncommutative symplectic form  $\omega:=\kappa (\mu)$. 
}

\paragraph{Observations} 

(1) The coefficient expressions given in Section
    \ref{canforms} show that equation $\{W,W\}=0$ is
    equivalent with:
\be
\label{W_inf}
\sum_{0\leq i+j\leq n}(-1)^{{\tilde a_1}+\ldots +{\tilde a}_i}
{\bar W}_{a_1\ldots a_i a a_{i+j+1}\ldots a_n} {\bar W}^a_{\,\,a_{i+1}\ldots
      a_{i+j}}=0~~{\rm for~all~}~~ n\geq 0~~,
\ee
where lifting of indices is done from the left  with $\rho^{ab}=(-1)^{\tilde
b+1}\omega^{ab}$, and $\omega^{ab}$ is the inverse of the matrix $\omega_{ab}=-{\bar W}_{\sigma ab}$.  
The first equation (for $n=0$) is ${\bar W}_a{\bar
W}^a=0$.  Notice that $\omega$ is determined by $W$, so the equations are not quadratic. 
This countable system of nonlinear algebraic conditions is
a non-commutative analogue of the WDVV equations \cite{WDVV}.

(2) It was shown in \cite{HLL} that the background satisfies the string equations of motion iff 
the underlying $A_\infty$ algebra is minimal. It is clear that 
this amounts to the requirement that $W$ has order at least $3$ at the origin.

(3) The structure given above can be viewed as an 'off-shell extension' of the 'boundary part' of the data described in 
\cite{CIL1, Moore_Segal, Moore}. The latter arises in the particular case when $W$  has degree three at the origin
(i.e. the underlying $A_\infty$ structure is minimal), and can be recovered 
by forgetting all terms of $W$ of order higher than $3$. 
In physics language, the structure of \cite{CIL1, Moore_Segal, Moore} corresponds to keeping only the 
boundary three-point functions on the disk, thereby forgetting all {\em integrated} amplitudes. As explained in the 
introduction to \cite{CIL1}, this reflects the difference between two-dimensional topological field theory and 
topological string theory, namely in the topological field theory one does not consider integration of amplitudes 
over the moduli space of the underlying Riemann surface (since by definition 
the worldsheet metric is not a dynamical variable).  

\subsection{Deformations of the underlying string theory}
\label{theory_deformations}

Each solution of the constraints described in the previous subsection represents the tree-level boundary data of 
an open topological string theory. We would like to make some basic observations about the space of such 
theories.

It is instructive to consider the trivial approximation $W_d=0$ i.e. $W=W_g$, which --- as we shall see in 
a moment --- is appropriate under certain assumptions. Starting from $W_g=-\omega_{ab}(\sigma ab)_c$, we 
compute: 
\be
\label{rdWg}
-W_g\rd_a=\omega_{\alpha a}\sigma \alpha+(-1)^{{\tilde \beta}{\tilde \omega}} \omega_{a\beta}\beta\sigma+
(-1)^{\tilde \omega}\delta^\sigma_ a\omega_{\alpha\beta}\alpha\beta~~
\ee
and: 
\be
\label{ldWg}
-\omega^{ab}\ld_b W_g=(-1)^{\tilde \omega}a\sigma +(-1)^{{\tilde a}+{\tilde \omega}}\sigma a+\omega^{a\sigma}
\omega_{\gamma\delta}\gamma\delta~~.
\ee
Combining these equations and using appropriate cyclic permutations gives: 
\be
\nn
\{W_g,W_g\}_c=(W_g\rd_a\omega^{ab}\ld_bW_g)_c=(-1)^{\tilde \omega}\omega^{\sigma\sigma}\omega_{\alpha\beta}
\omega_{\gamma\delta}(\alpha\beta\gamma\delta)_c~~.
\ee
Since $\sigma$ is odd, $\omega^{\sigma\sigma}$ vanishes for degree reasons unless ${\tilde \omega}=0$. When 
$\omega$ is even, the term in the right hand side need not vanish.  

Let us assume that ${\tilde \omega}=1$ or that ${\tilde \omega}=0$ but 
$\omega^{\sigma\sigma}\omega_{\alpha\beta}\omega_{\gamma\delta}(\alpha\beta\gamma\delta)_c$ vanishes. 
In this case, we have $\{W_g,W_g\}=0$ and $W_g$ gives a marked point in the space of open 
string theories with underlying supermodule $E$, unit $\lambda$ and symplectic form $\omega$. The cyclic unital 
$A_\infty$ algebra corresponding to this solution has a single product $r_2^g$, which is given by:
\be
\nn
r_2^g(e_a,e_b)=(-1)^{{\tilde \omega}+1}\omega_{ab}\omega^{\lambda c}e_c~~{\rm for}~~a,b\neq \sigma
\ee
and by the unitality constraint $-r_2^g(\lambda,e_a)=(-1)^{{\tilde a}} r_2^g(e_a,\lambda)=e_a$ for the 
remaining combinations of basis elements. The $A_\infty$ constraints (\ref{ainf}) reduce to: 
\be
\nn
r_2^g(r_2^g(x,y),z)+(-1)^{{\tilde x}}r_2^g(x,r_2^g(y,z))=0~~,
\ee
which means that $m_g:=\Sigma\circ r_2\circ \Sigma^{\otimes 2}:E^{\otimes 2}\rightarrow E$ satisfies the associativity condition: 
\be
\nn
m_g(m_g(x,y),z)=m_g(x,m_g(y,z))~~.
\ee
Moreover, the unitality constraint for $r_2^g$ amounts to $m_g(\lambda, x)=m_g(x,\lambda)=x$.  Hence the distinguished
solution $W=W_g$ corresponds to an associative superalgebra
structure on $E$, and the underlying $A_\infty$ category reduces to
an ordinary (i.e. associative) $\Z_2$-graded category.  In the
topological string theory, all integrated boundary correlators on the
disk vanish and the entire information is contained in the boundary
three-point functions. Fixing $\lambda$ and $\omega$, other string
theories with the same units and topological metrics are given by
solutions of the equations $\{W_g+W_d, W_g+W_d\}=0$ and $\ld_\lambda W_d=0$, the first of which reduces to:
\be
\nn
\{W_g, W_d\}+\frac{1}{2}\{W_d,W_d\}=0~~,
\ee
where we used the graded antisymmetry property of the Kontsevich bracket. Letting $Q_g=\theta_{W_g}$ 
be the (odd) Hamiltonian vector field defined by $W_g$ (i.e. $dW_g=i_{Q_g} \omega$), this equation takes the form:
\be
\label{MC}
L_{Q_g}W_d+\frac{1}{2}\{W_d,W_d\}=0~~.
\ee
Notice that $L_{Q_g}$ squares to zero (since $L_{Q_g}^2=\frac{1}{2}[L_{Q_g}, L_{Q_g}]=\frac{1}{2}L_{[Q_g,Q_g]}=
\frac{1}{2}L_{\theta_{\{W_g,W_g\}}}=0$) and that it acts as an odd derivation of the Lie superalgebra 
$(C^0_R(A)[{\tilde \omega}],\{.,\})$ (due to the Jacobi identity for the Kontsevich bracket):
\be
\nn
L_{Q_g}\{f,g\}=\{L_{Q_g}f, g\}+(-1)^{{\tilde f}+{\tilde \omega}}\{f, L_{Q_g} g\}~~.
\ee 
Hence $(C^0_R(A)[{\tilde \omega}], L_Q, \{.,.\})$ is a differential Lie superalgebra, and (\ref{MC}) is its 
Maurer-Cartan equation. This means that one can study the moduli space of boundary string {\em theories}
with fixed units and topological metrics by using the deformation theory of Lie superalgebras.

\section{The noncommutative moduli space}

In this section, we use the formalism developed above to construct a
noncommutative version of the extended moduli space of finite D-brane
systems (the boundary part of the extended moduli space of topological
strings).

\subsection{Symmetries}
\label{automf_group}

Let us fix a D-brane system described by the noncommutative
generating function $W$, with Hamiltonian derivation $Q$ and
symplectic form $\omega$. We assume given adapted coordinates
including odd loops $\sigma_u$ associated with the units of the
underlying $A_\infty$ structure. It is clear from the categorical
formulation of Appendix \ref{data} that a symmetry of the D-brane
system amounts to an automorphism of the underlying cyclic and unital
weak $A_\infty$ category, called a cyclic and unital
$A_\infty$ automorphism (an automorphism is a strict autoequivalence, as 
appropriate for a finite category). In this subsection, we describe such symmetries 
as symplectomorphisms of $A$ which obey certain supplementary properties. 

Given a relative automorphism $\phi$ of $A$, we set
$a':=\phi(a)$ for all $a\in {\cal Q}_1$, and let $V'\subset A$ be the $R$-sub-bimodule
spanned by the elements $a'$.  Then $A$ is isomorphic {\em as a superalgebra} with the tensor
algebra $T_R V'$, and $\phi$ can be viewed as a change of 
coordinates from $a$ to $a'$.  More precisely, the restriction of $\phi$ to $A_1=V$ 
gives an isomorphism of $R$-superbimodules $\phi_1:V\rightarrow V'$ and $\phi$ can be identified 
with the isomorphism of bigraded $R$-superalgebras  
$T_R(\phi)=\oplus_{n\geq 0}{\phi^{\otimes_R n}}:T_RV\rightarrow T_R V'$ induced by $\phi_1$. 
Of course, we have $a'=\phi_1(a)=\phi(a)$ and $a'$ is an adapted basis for the $R$-superbimodule $V'$.

$R$-linearity of $\phi$ implies that
each $a'$ is a linear combination of paths starting at $t(a)$ and
ending at $h(a)$. However, $\phi$ need not be homogeneous with respect
to the $\N$-grading of $A$, so generally each $\phi(a)$ can be a
linear combination of paths of different length. Defining $A'_n$ to be
the subspace spanned by $n$-factor monomials in $a'$, we have
$A'_n\approx T_R^n V'$ and a new decomposition:
\be
\nn
A=\oplus_{n\geq 0}{A'_n}~~
\ee
with $A'_0=R$. In particular, a generic $R$-superalgebra automorphism induces a change of $\N$-grading.

Equation (\ref{df}) implies:
\be
\nn
{\bar d} {\bar \phi}(f)={\bar\phi}^*({\bar d}f)=\sum_{a} (\phi(f\rd_a)d a')_c~~,
\ee
which shows that the cyclic derivatives with respect to the new coordinates are given by:
\be
\label{ld_tf}
{\bar \phi}(f)\rd_{a'}=\phi(f\rd_a)\Leftrightarrow \ld_{a'} {\bar \phi}(f)=\phi(\ld_a f)~~.
\ee

Relative superalgebra endomorphisms $\phi$ of $A$ having the property $\phi\circ Q \circ \phi=Q$ correspond to 
endomorphisms of the underlying weak $A_\infty$ category. The correspondence is obtained by expanding:
\be
\label{phi_expansion}
\phi(a)=\sum_{n\geq 0}\phi_{a_1\ldots a_n}^a a_1\ldots a_n~~,
\ee
where the complex-valued coefficients $\phi_{a_1\ldots a_n}^a$ vanish unless $a_1\ldots a_n$ is a path starting 
at $t(a)$ and ending at $h(a)$. The evenness condition on $\phi$ gives the selection rules:
\be
\nn
\phi^a_{a_1\ldots a_n}=0~~{\rm unless}~{\tilde a}={\tilde a}_1+\ldots +{\tilde a}_n~~.
\ee
The $n=0$ part of (\ref{phi_expansion}) stands for the sum over loops 
$\sum_{a\in Q_1(u,u)}\phi^{a}\epsilon_u$, i.e. we use the convention $\phi_{a_1\ldots a_0}^a:=
\phi^a \delta^u_{h(a)}\delta^u_{t(a)}\epsilon_u$. Then the $A_\infty$ morphism associated with $\phi$ is 
given by the even $R$-multilinear maps $\phi_n:E[1]^n\rightarrow E[1]$ defined through:
\be
\label{phi_maps}
\phi_n(e_{a_1}\ldots e_{a_n})=\phi^a_{a_1\ldots a_n}e_a~~.
\ee
The conditions $\phi(ab)=\phi(a)\phi(b)$ amount to the complicated relations giving the traditional 
definition.  The maps (\ref{phi_maps}) define an endomorphism of the
weak $A_\infty$ structure on the superbimodule $E$; as usual,
$R$-multilinearity allows one to decompose them into
complex-multilinear maps describing an endomorphism of the underlying
weak $A_\infty$ category.  In particular, the map $\phi_0:R\rightarrow
E[1]$ gives even linear maps $\phi_u:\C\rightarrow E_{uu}[1]$ via the
decomposition $\phi_0(\sum_{u}{\alpha_u
\epsilon_u})=\sum_{u}{\epsilon_u \phi_u(\alpha_u)\epsilon_u}$ for
complex $\alpha_u$; these can also be viewed as the odd elements
$\phi_u(1_\C)=\sum_{h(a)=t(a)=u}\phi^a e_a \in E_{uu}$.

An $A_\infty$ endomorphism of $(E,(r_n))$ is called {\em unital} if $\phi_1(\lambda)=\lambda$
and $\phi_n(e_{a_1}\ldots e_{a_n})$ for $n\neq 1$ vanishes when any of 
the elements $e_{a_1}\ldots e_{a_n}$ coincides with the odd $A_\infty$ unit $\lambda$. In terms of the 
coefficients of $\phi$, this means $\phi^a_\sigma=\delta^a_\sigma$ and 
$\phi^{a}_{a_1\ldots a_n}=0$ for all $n\neq 1$, if $\sigma\in \{a_1\ldots a_n\}$. Plugging this into  
expansion (\ref{phi_expansion}), we see that the $A_\infty$ endomorphism is unital iff:
\be
\label{phi_unitality}
\phi(\sigma)=\sigma~~~{\rm and}~~~\phi(a)={\rm ~independent~of~}\sigma {\rm~for~all}~~a\neq \sigma~~.
\ee

The $A_\infty$ endomorphism determined by $\phi$ is {\em cyclic} if $\phi^*(\omega)=\omega$; writing
this condition explicitly gives a series of complicated relations used in the traditional definition.
In particular, a cyclic $A_\infty$ automorphism of $(E,\rho, (r_n)))$ amounts to a 
relative symplectomorphism of $A$ preserving  the homological derivation $Q$. 

Let $\phi\in \Aut_R^\omega(A)$ be a relative symplectomorphism. 
Remember from the end of Subsection \ref{sec:bracket} that 
the map $\psi_\omega:C^0_R(A)[{\tilde \omega}]\rightarrow \Der_l^\omega(A)$  
is equivariant with respect to the action of $\Aut_R^\omega(A)$. Moreover, the exact sequence (\ref{ex}) shows that 
$\psi_\omega$ induces an isomorphism of vector spaces $C^0_R(A)[{\tilde \omega}]/R\approx \Der_l^\omega(A)$.
Thus $Q$ is $\phi$-invariant iff its (canonical) Hamiltonian $W$ is invariant under the action of $\phi$ up to
addition of elements of $R$: 
\be
\label{W_variation}
{\bar \phi}(W)=W+\alpha~~{\rm~for~some~}\alpha\in R~~.
\ee
Moreover, the associated $A_\infty$ endomorphism is unital iff $\phi$ satisfies relations 
(\ref{phi_unitality}).

Combining these observations, we see that a cyclic and unital
endomorphism of the underlying $A_\infty$ structure amounts to a
symplectomorphism of $A$ which satisfies (\ref{phi_unitality}) and
(\ref{W_variation}).  The {\em symmetry group}  ${\cal G}\subset
\Aut^\omega_R(A)$ of the system is the group of all such
symplectomorphisms of $A$.  If $\phi$ belongs to ${\cal G}$, equations
(\ref{ld_tf}) and (\ref{phi_unitality}), (\ref{W_variation})  imply: \be \nn \mu=\frac{1}{2}\ld_\sigma
W=\frac{1}{2}\ld_{\sigma'} {\bar \phi}(W)=
\frac{1}{2}\phi(\ld_\sigma W)=\phi(\mu)~~, \ee so $\phi$ preserves
the moment element $\mu$.

\subsection{Algebraic construction of the noncommutative moduli space}

\label{nc_moduli}

Consider the two-sided ideal $J$ of $A$ generated by the elements: 
\be
\nn
\ld_a W\in A ~~(a\in {\cal Q}_1)~~,
\ee
which we shall call the {\em critical ideal} of the noncommutative generating function. Notice that $J$
is also generated by $W\rd_a$, due to the relations
$\ld_a W=(-1)^{{\tilde a}{\tilde \omega}}W\rd_a$. We let
$\C[{\cal Z}]:=A/J$. Since $J$ is $\Z_2$-homogeneous (being generated by
$\Z_2$-homogeneous relations), the associative algebra $\C[{\cal Z}]$ is $\Z_2$
graded.  Passage to $\C[{\cal Z}]$ implements the conditions:
\be
\nn
\sum_{n\geq 0} {\bar W}_{aa_1\ldots a_n}a_1\ldots a_n=0~~\forall a\in {\cal Q}_1 \Leftrightarrow 
\sum_{n\geq 0} {\bar W}_{a_1\ldots a_n a}a_1\ldots a_n=0~~\forall a \in {\cal Q}_1~~,
\ee
which (in view of the isomorphism $V\otimes_R A\approx C^1_R(A)$) can also be written as:
\be
\nn
{\bar d}W=0~~.
\ee
In particular, the distinguished central element $\lambda=\sum_{u\in
{\cal Q}_0}{\lambda_u}$ gives the relation:
\be
\nn
\ld_\lambda W=0\Leftrightarrow \mu=0\Leftrightarrow \omega_{ab}ab=\frac{1}{2}\omega_{ab}[a,b]=0\Leftrightarrow 
\sum_{a\in {\cal Q}_1}[a,a^*]=0~~,
\ee
where, as usual,  $[a,b]=ab-(-1)^{{\tilde a}{\tilde b}}ba$ is the supercommutator
in $A$. Hence `extremizing' $W$ automatically imposes the zero-level 
constraint for the noncommutative moment  map of \cite{BEV, Bergh}.

For $\phi\in \Aut_R(A)$, relations (\ref{df}) imply $\phi^*({\bar d}W)=(\phi(W\rd_a) da')_c$
and $\phi^*({\bar d}W)={\bar d}({\bar \phi}(W))=({\bar \phi}(W)\rd_a da)_c=({\bar \phi}(W)\rd_a d\phi^{-1}(a'))_c$, 
where $a':=\phi(a)$. Expanding $d\phi^{-1}(a')$ in the second expression and comparing with the first, 
we find that
$\phi(W\rd_a)$ belongs to the ideal generated by ${\bar
\phi}(W)\rd_a$. This shows that relative automorphisms which preserve
$W$ (i.e. ${\bar \phi}(W)=W$) also preserve the ideal $J$, so they
descend to $R$-linear automorphisms of $\C[{\cal Z}]$ (the $R$-superbimodule
structure on $\C[{\cal Z}]$ is induced from its obvious $A$-superbimodule
structure). In particular, the group ${\cal G}$ preserves $J$, and we obtain 
a group morphism $\gamma:{\cal G}\rightarrow \Aut_R(\C[{\cal Z}])$, i.e.  
an action of ${\cal G}$ by $R$-linear automorphisms of the superalgebra $\C[{\cal Z}]$.
The canonical epimorphism $\zeta:A\rightarrow \C[{\cal Z}]=A/J$ is ${\cal G}$-equivariant:
\be
\nn
\zeta\circ \phi=\gamma(\phi)\circ \zeta~~~~~~\forall \phi\in {\cal G}~~.
\ee
We set $\C{\cal M}=\C[{\cal Z}]^{\cal G}$, the homogeneous 
subalgebra of elements invariant under the action of ${\cal G}$. We will view these algebras 
as noncommutative coordinate rings of `noncommutative schemes' ${\cal Z}$, ${\cal M}$, which we 
call the {\em noncommutative extended vacuum space} and {\em noncommutative extended moduli space} 
respectively.

\section{The case of a single D-brane}
\label{single_brane}

Let us illustrate the general discussion with the simple case of a single D-brane. 
Then ${\cal Q}$ consists of $m$ loops at a single vertex, and we 
let $m_\pm$ be the numbers of even and odd loops. 
The space of boundary observables is 
$E=\C^{m_-|m_+}$, with parity-changed dual 
$V=E[1]^*=\C^{m_+|m_-}$. The boundary algebra $R$ coincides with $\C$ while 
the path algebra is the free superalgebra $A=\C\langle\{a\}\rangle $ generated by all loops. The underlying 
$A_\infty$ category has a single object, so it reduces to a weak,
cyclic and unital $A_\infty$ algebra on the supervector space $E$. 
This is the structure found in \cite{HLL}.

It is easy to see that $C^0(A):=C^0_\C(A)$ can be identified with the cyclic subspace $A_{\rm cyclic}$ of $A$, 
defined as the image of the idempotent operator:
\be
\nn
P=\id_{\C}\oplus \oplus_{n\geq 1}\left[\frac{1}{n}\sum_{i=0}^{n-1}{(\gamma_n)^i}\right]\in \End_\C(A)~~.
\ee 
Here $\gamma_n\in \End_\C (A_n)$ are the generators of the obvious $\Z_n$ action on $A_n$:
\be
\nn
\gamma_n(x_1\otimes \ldots \otimes x_n)=(-1)^{{\tilde x}_1({\tilde x}_2+\ldots +{\tilde x}_n)}x_2\otimes \ldots 
\otimes x_n \otimes x_1~~.
\ee
Thus $A_{\rm cyclic}$ consists of all polynomials $f=\sum_{n \geq 0}{f_{a_1\ldots a_n}a_1\ldots a_n}\in A$
whose complex coefficients satisfy the conditions $f_{a_1\ldots a_n}=(-1)^{{\tilde a}_1({\tilde a}_2+\ldots +
{\tilde a}_n)}f_{a_2\ldots a_n a_1}$. Writing $f=\sum_{n\geq 1}{f_n}$ with 
$f_n\in A_n =V^{\otimes n}$, such a polynomial belongs to $A_{\rm cyclic}$ iff $\gamma_n(f_n)=f_n$ for 
all $n$. 

Let $\sigma$ be the dual basis element corresponding to the parity changed $A_\infty$ unit $\lambda$. 
We assume given a basis of $E$ such that $\sigma$ is one of the odd loops. 
The generating function is a constant-free polynomial:
\be
\nn
W= \sum_{n\geq 1}W_{a_1\ldots a_n}a_1\ldots a_n\in A_{\rm cyclic}
\ee
in the non-commuting variables $a$ such that
$W_{a_1\ldots a_n}$ are graded-cyclic and satisfy the conditions
$W_{a_1\ldots a_n}=0$ unless ${\tilde a}_1+\ldots +{\tilde
a}_n={\tilde \omega}+1$, as well as the $A_\infty$ constraints
(\ref{W_inf}). Moreover, we must have $W=W_g +W_d$ with
$W_g=-\frac{1}{3}\omega_{ab}[\sigma ab+(-1)^{{\tilde a}+{\tilde b}}ab\sigma +
(-1)^{{\tilde b}({\tilde a}+1)}b\sigma a]$ 
and where $W_d$ vanishes at zero and is independent of $\sigma$. The
matrix $(\omega_{ab})$ satisfies properties (\ref{omega_symm}) and
(\ref{omega_sel}) of Section \ref{symplectic_structure} and no further
constraints.

Any endomorphism $\phi$ of $A$ is determined by its values on the generators:
\be
\nn
\phi(a):=\phi_{a}=\sum_{n\geq 0} \phi^{a}_{a_1\ldots a_n}a_1\ldots a_n
\ee
and can be viewed as an $m$-tuple $(\phi_{a_1}\ldots \phi_{a_m})$ of polynomials 
in the non-commuting variables $a$. The degree zero condition on $\phi$ gives the constraints
$\deg\phi(a)={\tilde a}$, so  the complex coefficient $\phi^{a}_{a_1\ldots a_n}$ must vanish 
unless ${\tilde a}_1+\ldots +{\tilde a}_n={\tilde a}$. The relative automorphism group is the usual 
group $\Aut(A)$ of superalgebra automorphisms. An automorphism $\phi$ belongs to the symmetry group ${\cal G}$ 
if $\phi(\sigma)=\sigma$, $\phi(a)$ are independent of $\sigma$ for $a\neq \sigma$  
and $W(\{\phi(a)\})$ equals $W(\{a\})$ as a polynomial in the non-commuting variables $\{a\}$. This imposes 
nonlinear algebraic conditions on the coefficients $\phi^a_{a_1\ldots a_n}$. 

\paragraph{Observation} The automorphism group $\Aut(A)$ is a rather exotic object. 
In the even case $m_-=0$, it is known\cite{Umirbaev} (see also \cite{DY}) that
$\Aut(A)$ contains wild automorphisms\footnote{ An automorphism is
called wild if it is not a composition of the so-called elementary
automorphisms $(a_1\ldots a_m)\rightarrow (a_1\ldots a_i, \alpha
a_i+f(a_1\ldots a_{i-1}, a_{i+1} \ldots a_m), a_{i+1}\ldots a_m)$ with
$\alpha\in \C^*$ and $f$ a polynomial independent of $x_j$.} as soon
as $m\geq 3$. Even in the commutative case, there are well-known
open problems about automorphisms of polynomial algebras such as the
Jacobian conjecture. The $\Z_2$-graded, noncommutative case does not
seem to have been studied systematically.

\

\section{Examples} 
\label{examples}

\subsection{Even system with a single boundary degree of freedom}

This is the simplest example relevant for topological sigma models
with target spaces of even complex dimension. In this case, we
have a single D-brane ($R=\C$) with $E=\C$ (concentrated in even degree), $V=\C[1]$ ( a purely odd 
supervector space) and ${\tilde
\omega}=0$.  The canonical forms of Subsection \ref{symplectic_structure} 
show that this is the only possibility when the boundary superspace has dimension one.

Up to rescaling, the boundary sector contains a single boundary observable, 
namely the identity operator $1$, which is the even $A_\infty$ unit.  
We set $\lambda=\Sigma 1$ and let $\sigma$ be dual odd element in $V=\C[1]$ (of course, $\sigma$ can be 
identified with $\lambda$ since we identify $\C^*$ with $\C$ using the canonical basis of $\C$ given by the unit).
The superquiver consists of the single odd loop
$\sigma$, with path superalgebra $A=\C\langle \sigma\rangle$. 
By a change of normalization of the $A_\infty$ products, we can take $\omega_{\sigma \sigma}=1$; 
then $\omega_{form}=\frac{1}{2}(d\sigma^2)_c$, and $\sigma$ can also be viewed as the canonical odd coordinate 
$\xi$ in equation (\ref{e_can}). 
  
The non-commutative generating function is an odd
polynomial $W=\sum_{n=odd}{W_n(\sigma^n)_c}\in C^0(A)=A_{\rm cyclic}=A^{\rm odd}$. The
unitality constraint requires the splitting $W=W_d+W_g$ with $W_d$
a constant.  Since $W$ must vanish at zero, this gives
$W_d=0$.  Thus we must have: \be\nn
W=W_g=-\frac{1}{3} \sigma^3 \ee and the only
non-trivial strict coefficient (see Subsection \ref{quivc0}) is ${\bar W}_{\sigma\sigma\sigma}=-1$. The $A_\infty$
constraint (\ref{W_inf}) is trivially satisfied. Since $W$ is cubic, the $A_\infty$ algebra 
contains only the product $r_2$, which is completely determined by the unitality constraint
$r_2(\lambda, \lambda)=-\lambda$. 
The associative product $\cdot=\Sigma \circ r_2 \circ \Sigma^{\otimes 2}$ on $E$ is given 
by $1\cdot 1=1$, which of course is the unique associative product on $\C$ with unit $1$.

Since elements
$\phi\in {\cal G}$ must preserve $\sigma$, we have ${\cal
G}=\{\id_A\}$. The critical ideal is generated by $\ld_\sigma W=-\sigma^2$, so
$J=A\sigma^2A$ consists of all polynomials of order at least two at the
origin. Thus $\C[{\cal Z}]=A/(\sigma^2)$ is the Grassmann algebra $\C[\sigma]$
on the odd generator $\sigma$. The noncommutative moduli space coincides
with the noncommutative vacuum space, having coordinate ring: \be
\C[{\cal M}]=\C[{\cal Z}]=\C[\sigma]~~.\nn \ee Hence ${\cal M}={\cal
Z}=\C^{0|1}$, the odd point of usual supergeometry.

In this extremely simple example, passage to the noncommutative moduli
space gives nothing new, since supercommutativity is imposed as a
consequence of the equations of motion.

\subsection{Odd system with two boundary degrees of freedom}
This is the simplest example with an odd topological boundary metric, obtained for 
$E=\C^{1|1}$. Choosing canonical coordinates as in (\ref{o_can}), we have:
\be
\nn
\omega_{form}=(dp_0dq_0)_c
\ee  
with odd $q_0$ and even $p_0$, and take $\sigma=q_0$. Since
$\sigma$ must be a loop, the only possibility for the boundary algebra
is $R=\C$, i.e. the index superquiver consists of an even and an
odd loop at a single vertex (indeed, $\omega_{form}$ vanishes for any
other boundary structure). Thus $A=\C\langle p_0,q_0\rangle$ is a free
associative superalgebra. In this case, we have $W_g=2(q_0^2p_0)_c$
and $W_d=W_d(p_0)$ must be a constant-free univariate polynomial in
$p_0$.  It is not hard to see that the Maurer-Cartan equation
(\ref{MC}) for $W_d$ is trivially satisfied. Indeed, it is clear that
$\{W_d,W_d\}=0$, while direct computation gives:
\be
\nn
\{W_g,W_d\}=-(W_g\rd_{q_0}\ld_{p_0}W_d)_c=
2([q_0,p_0]\ld_{p_0}W_d)_c=2(q_0[p_0, \ld_{p_0}W_d])_c=0
\ee
where we noticed that the commutator $[p_0, \ld_{p_0}W_d]$ vanishes
because $W_d$ is a polynomial in $p_0$.  Hence the general system of
this type is described by $W=W_g+W_d$, where $W_d$ is an arbitrary
constant-free polynomial in $p_0$. The defining equations $\ld_a W=0$
for the noncommutative vacuum space take the form:
\be
\label{def_eqs}
[q_0,p_0]=0~~,~~q_0^2=-\frac{1}{2}\ld_{p_0}W_d(p_0)~~.
\ee
Thus $\C[{\cal Z}]=\C\langle q_0,p_0\rangle/([q_0,p_0],
q_0^2+\frac{1}{2}\ld_{p_0}W_d(p_0))$ and ${\cal Z}$ is a bona-fide
noncommutative superspace. It can be viewed as a `fibration' over the
noncommutative affine line $\A^1$ with coordinate $p_0$, where the
(pure fuzz) fiber is a point-dependent deformation of the usual odd point
$\C^{0|1}$.

It is known \cite{Cz, ML} that all automorphisms of a free associative
algebra on two generators are tame, i.e. given by iterated composition
of elementary automorphisms of the form $(q_0,p_0)\rightarrow (q_0,
\alpha p_0+f(q_0))$ and $(q_0,p_0)\rightarrow (\beta q_0+g(p_0),p_0)$
with $\alpha,\beta\in \C^*$ and $f,g$ arbitrary univariate
polynomials. Moreover (see \cite{DY} for a more general result), 
all algebra automorphisms fixing one variable are triangular. In particular, 
automorphisms fixing $q_0$ have the form $(q_0,p_0)\stackrel{\phi}{\rightarrow}(q_0,\alpha p_0+f(q_0))$, 
with {\em super}algebra automorphisms obtained by restricting to even polynomials. 
Obviously $\phi(p_0)$ is independent of $\sigma=q_0$ iff $f$ is the constant polynomial. In this 
case, $\phi$ is a symplectomorphism iff $\alpha=1$.  Hence ${\cal G}$ is 
a subgroup of the one-dimensional translation group:
\be
\nn
{\cal T}:~~q_0\rightarrow q_0~~,~~p_0\rightarrow p_0+t~~~~(t\in \C)~~.
\ee
Such an automorphism preserves $W_g$ up to addition of constants, and takes $W_d(p_0)$ into $W_d(p_0+t)$. 
It follows that $W$ is preserved up to constants iff $W_d$ is linear in $p_0$. 
Thus ${\cal G}=\{\id_A\}$ unless $W_d=w(p_0)_c$ for some $w\in \C$, in which case ${\cal G}={\cal T}$.

Hence the generic case of a nonlinear $W_d$ gives ${\cal M}={\cal Z}$. 
When $W_d=w(p_0)_c$, equations (\ref{def_eqs})
reduce to $[q_0,p_0]=0$ and $q_0^2=-\frac{w}{2}$
 and we find $\C[{\cal M}]=\C[{\cal Z}]^{\cal T}=\C\langle
q_0\rangle/(q_0^2+\frac{w}{2})$, i.e ${\cal M}$ is the quantum
deformation of the odd point $\C^{0|1}$ given by $q_0^2=-\frac{w}{2}$.

\subsection{A family of odd examples}

Consider a theory with ${\tilde \omega}=1$, where we choose adapted
coordinates $p_0\ldots p_m, q_0\ldots q_m$ (with $2(m+1)={\rm Card}
{\cal Q}_0$, odd $q_i$ and even $p_i$) such that that $\omega$ has
canonical form (\ref{o_can}).  Remember from Subsection
\ref{symplectic_structure} that the coordinates $a^*=\omega^{ab}b$ are
given by:
\be
\nn
p_i^*=q_i~~,~~q_i^*=-p_i~~,
\ee
where $*$ can be viewed as an involution on $A_1=V$. Also remember
that $p_i$ are even and $q_i$ are odd.  We assume that $q_0=\sigma$
corresponds to the unit.

Since ${\tilde \omega}=1$, we have $\{W_g,W_g\}=0$ and the discussion
of Subsection \ref{theory_deformations} applies. In particular, a
general theory with the given unit and symplectic form is specified by
a solution $W_d$ of the Maurer-Cartan equation (\ref{MC}).  For
${\tilde \omega}=1$, relation (\ref{rdWg}) gives:
\be
\nn
W_g\rd_a=[\sigma,a^*]+\delta_a^\sigma\sum_{\alpha\in Q_1}{\alpha\alpha^*}
\ee
and the conditions $W\rd_a=0$ take the form: 
\be
\nn
[\sigma,a^*]+\delta_a^\sigma\sum_{\alpha\in Q_1}{\alpha\alpha^*}=-W_d\rd_a~~,
\ee
where, as usual, $[.,.]$ stands for the graded commutator.  In
canonical coordinates, we have $[\sigma,
\sigma^*]=[q_0,q_0^*]=-[q_0,p_0]=[p_0,q_0]$ and $\sum_{\alpha\in
Q_1}{\alpha\alpha^*}=\sum_{i=0}^m{[p_i,q_i]}$. The noncommutative
criticality constraints become:
\be
\label{E1}
2[p_0,q_0]+\sum_{i=1}^m{[p_i,q_i]}=-W_d\rd_{q_0}~~,
\ee
and:
\be
\label{E2}
[q_0,p_i]=W_d\rd_{q_i}~~\forall i=1\ldots m,~~[q_0, q_i]=-W_d\rd_{p_i}~~~~~\forall i=0\ldots m~~.
\ee

Let us assume that we are given a particular solution $W=W_g+W_d$ for
which $W_d=W_d(p_1\ldots p_m)$ depends only on $p_1\ldots
p_m$. Special solutions of this type were found in \cite{Katz} for
systems which obey $\Z$-valued selection rules (for the examples of
\cite{Katz}, equation (\ref{MC}) is trivially satisfied by the ansatz
$W_d=W_d(p_1\ldots p_m)$ due to the integer-valued degree condition
obeyed by $W$ in the case of Calabi-Yau compactifications).  In this
case, eqs. (\ref{E1}) and (\ref{E2}) reduce to the following defining
relations for the noncommutative vacuum space ${\cal Z}$:
\begin{eqnarray}
\label{Z_eqs}
[q_0,q_i]&=&-\ld_{p_i}W_d~~~~~\forall i=1\ldots m\nn\\
\left[q_0,p_i\right]&=&0~~~~~~~~~~~~~~~~\forall i=1\ldots m\\
\left[q_0,p_0\right] &=&\frac{1}{2}\sum_{j=1}^m{[p_j,q_j]}~~\nn\\
q_0^2&=&0~~\nn~~.
\end{eqnarray}
To arrive at this form, we noticed that $\ld_{p_i} W=W\rd_{p_i}$ since
$p_i$ are even. Note that $[q_0,q_i]=q_0q_i+q_iq_0$ since $[.,.]$ is
the graded commutator.

For simplicity, let us consider the case when the underlying quiver
has a single vertex.  Then we can view ${\cal Z}$ as a `fibration'
over the noncommutative affine plane $\A^{m+1}$ with coordinates
$p_0\ldots p_m$, whose `fiber' is a subspace of the noncommutative
affine space $\A^{m+1}$ (with coordinates $q_0\ldots q_m$) determined
by (\ref{Z_eqs}).  The Abelian locus ${\cal Z}_{Ab}$ in ${\cal Z}$ is
obtained by requiring that all variables supercommute. In this case,
eqs. (\ref{Z_eqs}) reduce to the conditions
$\partial_{p_i}W_d^{Ab}(p_1\ldots p_m)=0$ and we find that ${\cal
Z}_{Ab}$ coincides with the critical locus ${\rm
Crit}(W_d^{Ab})\subset \C^{m}$ of $W_d^{Ab}$, which is the usual
vacuum space expected in the supercommutative formulation. The
Abelianization epimorphism $\C\langle q_i, p_i\rangle \rightarrow
\C[q_i,p_i]$ induces an embedding of ${\cal Z}_{Ab}$ into the much
larger noncommutative space ${\cal Z}$. In the noncommutative vacuum
space, one can move away from the critical locus of $W_{Ab}$ at the
price of allowing for a non-vanishing commutator of $q_0$ with $q_i$;
notice that this is possible even along the locus in ${\cal Z}$ where
$p_i$ are required to commute.
Determining the symmetry group ${\cal G}$ and noncommutative moduli
space ${\cal M}$ in this class of examples is rather formidable in
general and will not be attempted here. We only note that the trivial case $m=0$ corresponds 
to the limit $W_d=0$ of the example discussed in the previous subsection.

\section{Conclusions}

We showed that the totality of boundary tree-level data determined by
a topological string theory in a finite D-brane background can be
encoded {\em faithfully} by using the non-commutative algebraic
geometry of a superquiver determined by the boundary decomposition of
the D-brane system. In particular, cyclicity of integrated boundary
amplitudes on the disk and the weak $A_\infty$ constraints on such
amplitudes amount to the condition $\{W,W\}=0$, where the boundary
potential $W$ is a function defined on a {\em noncommutative}
superspace $\A_{\cal Q}$ determined by the quiver. We also found a
differential constraint on $W$ which expresses the presence of unit
boundary observables in the boundary-preserving sector.

Fixing the bulk worldsheet data, but varying the D-brane background,
gives rise to the (extended) boundary moduli space of such a
system. We argued that this moduli space can be viewed as a
noncommutative superspace ${\cal M}$ constructed as an (invariant
theory) quotient of the `noncommutative critical locus' ${\cal Z}$ of
$W$ by a certain group of symplectomorphisms acting on ${\cal Z}$.

One upshot of this analysis is that the complicated structure
determined by {\em all} integrated boundary correlators on the disk is
encoded faithfully by a form of noncommutative geometry. According to
this point of view, the theory of topological D-brane deformations is
intrinsically noncommutative.  This gives a stringy realization of
non-commutativity at the level of such moduli spaces. It should be
compared with the realization of \cite{SW}, which arises by
translating the effective action of open strings into an action
governing objects (such as connections) defined over a noncommutative
space determined by the {\em closed} strings.  In both cases,
non-commutativity originates \cite{Schomerus} from the fact that
insertions of boundary observables on the disk do not commute. Hence
these ostensibly different realizations are related, and it would be
interesting to understand precisely how.

\

\acknowledgments{  I am grateful to V. Ginzburg for pointing out
the relevance of quiver algebras, and to M. Kontsevich for sharing his
ideas. This work originated from discussions during the Third Workshop
on Noncommutative Algebraic Geometry held at the Mittag Leffler
Institute.}

\appendix

\section{Topological D-brane systems as cyclic and unital weak $A_\infty$ categories}
\label{data}

\subsection{Mathematical background}

A weak ($\Z_2$-graded) $A_\infty$ category ${\cal A}$ consists of a
collection of objects $Ob{\cal A}$ and complex supervector spaces
$\Hom_{\cal A}(u,v):=\Hom(u,v)$ for $u,v\in Ob {\cal A}$, together with
odd multilinear maps\footnote{Notice that we take morphisms to compose {\em forward} under the \
$A_\infty$ products.}  $r_{u_1\ldots
u_{n+1}}:\Hom(u_1, u_2)[1]\times \Hom(u_2, u_3)[1]\times\ldots \times \Hom(u_n, u_{n+1})[1]
\rightarrow \Hom(u_1,u_{n+1})[1]$ for all $n\geq 0$.  The
case $n=0$ corresponds to odd maps $r_u:\C\rightarrow \Hom(u,u)[1]$, which
amounts to giving even elements $\theta_u=r_u(1)\in \Hom(u,u)$.  The maps $r$ are
required to satisfy certain conditions called the weak $A_\infty$
constraints.  To formulate them, let $t(x), h(x)\in Ob {\cal A}$
denote the tail and head of a morphism $x$, the unique objects of
${\cal A}$ such that $x\in \Hom(t(x), h(x))$. We say that an ordered
collection of morphisms $(x_1,\ldots, x_n)$ is {\em composable} if
$h(x_j)=t(x_{j+1})$ for all $j=1\ldots n-1$. In this
case, we let:
\be
[x_1\ldots x_n]:=t(x_1)t(x_2)\ldots t(x_n)h(x_n)~~,
\ee
viewed as a word on the set $Ob {\cal A}$. 
In particular, we set $[x]=t(x) h(x)$ for all morphisms $x$. 

We use $|.|$ to denote the degree of homogeneous elements of
$\Hom(u,v)$ and ${\tilde .}$ for the degree of homogeneous elements of
$\Hom(u,v)[1]$.  Then the maps $r_n$ are required to satisfy:
\be
\label{cat_ainf}
\sum_{0\leq i+j\leq n} (-1)^{{\tilde x}_1+\ldots +{\tilde x}_i} 
r_{[x_1\ldots x_i][x_{i+j+1}\ldots  x_{n}]}
(x_1\ldots x_i, r_{[x_{i+1}\ldots x_{i+j}]}(x_{i+1}\ldots x_{i+j}), x_{i+j+1}\ldots x_n)=0~~
\ee
for any system of $\Z_2$-homogeneous and composable morphisms $(x_1, \ldots, x_n)$.

A weak $A_\infty$ category is called {\em strong} if $r_u=0\Leftrightarrow \theta_u=0$ for all
$u\in Ob {\cal A}$ and {\em minimal} if it is strong and $r_{uv}=0$
for all $u,v\in Ob{\cal A}$. It is called {\em unital} if one is given
even elements $1_u\in \Hom(u,u)$ for each object $u$, such that the following conditions are
satisfied:
\begin{eqnarray}
\label{cat_unitality}
r_{[x_1\ldots x_{j-1}] [x_{j+1}\ldots x_n]}(x_1\ldots x_{j-1},e_{u_j},x_{j+1}\ldots x_n)&=&0~~
{\rm for~all}~~~~ n\neq 2 ~{\rm~and~all}~ j~~\nn\\
r_{[\lambda_u,x]}(\lambda_u,x)=-x~~,~~r_{[y,\lambda_u]}(y,\lambda_u)&=&(-1)^{{\tilde y}} y~~,
\end{eqnarray}
where $\lambda_u:=\Sigma 1_u\in \Hom(u,u)[1]$ and $u_j:=h(x_{j-1})=t(x_{j+1})$. In these relations, it is understood 
that $(x_1,\ldots, x_{j-1},x_{j+1},\ldots, x_n)$ is any composable system consisting of $\Z_2$-homogeneous elements. 
The last conditions in (\ref{cat_unitality}) are 
imposed for any homogeneous $x\in \Hom(u,v)$ and $y\in \Hom(v,u)$, with arbitrary $u,v$.  
The elements $1_u, \lambda_u$ are called {\em units} and {\em odd units} respectively. 

The $A_\infty$ category is called {\em cyclic} if one is given
non-degenerate\footnote{This means that all linear maps $\Hom(v,u)\rightarrow \Hom(u,v)^*$ determined by 
$\rho_{uv}$ are bijective.} bilinear forms $\rho_{uv}:\Hom(u,v)\times
\Hom(v,u)\rightarrow \C$, homogeneous of the same $\Z_2$-degree ${\tilde
\omega}$, such that $\rho_{uv}(x,y)=(-1)^{|x||y|}\rho_{vu}(y,x)$ and such that the
following identities are satisfied:
\be
\label{cat_rcyc}
\rho_{t(x_0) h(x_0)}(x_0,r_{[x_1\ldots x_n]}(x_1\ldots x_n))=
(-1)^{{\tilde x}_0({\tilde x}_1+\ldots +{\tilde x}_n)}
\rho_{t(x_1) h(x_1)}(x_1,r_{[x_2\ldots x_0]}(x_2\ldots x_n, x_0))~~,
\ee
whenever $(x_0,x_1 ,\ldots, x_n)$ is a homogeneous composable system 
and $[x_0\ldots x_n]$ is a cyclic word, i.e. $h(x_n)=t(x_0)$. Equivalently,
\be
\label{cat_rrcyc}
\omega_{t(x_0) h(x_0)}(x_0,r_{[x_1\ldots x_n]}(x_1\ldots x_n))=
(-1)^{{{\tilde x}_0+{\tilde x}_1+\tilde x}_0({\tilde x}_1+\ldots +{\tilde x}_n)}
\omega_{t(x_1) h(x_1)}(x_1,r_{[x_2\ldots x_0]}(x_2\ldots x_n, x_0))~~.
\ee
where $\omega_{uv}=\rho_{uv}\circ \Sigma^{\otimes 2}$
(i.e. $\omega_{uv}(x,y)=(-1)^{\tilde x}\rho_{uv}(x,y)$) are the
suspended bilinear forms $\omega_{uv}:\Hom(u,v)[1]\times
\Hom(v,u)[1]\rightarrow \C$, which satisfy
$\omega_{uv}(x,y)=(-1)^{{\tilde x}{\tilde y}+1}\omega_{vu}(y,x)$.

The concept of $A_\infty$ category was introduced in \cite{Fukaya_inf}
as a generalization of the 
notion of $A_\infty$ algebra \cite{Stasheff}. These objects are studied mathematically 
in \cite{Keller, Fukaya_mirror, KS, Hasegawa, Lyubashenko1,Lyubashenko2,Lyubashenko3, Lyubashenko4}. 
It is by now well-understood that they play an important role in homological algebra, in particular 
in giving a natural formulation of derived categories \cite{KS_book}. They also play a crucial 
role in the homological mirror symmetry program \cite{HMS}. 

\subsection{Category-theoretic description of topological D-brane systems}

It was pointed out in \cite{CIL5} (and derived in \cite{HLL} from the
worldsheet perspective, see also \cite{Costello1, Costello2}) that topological D-branes in string theory
form the structure of a weak, cyclic and unital $A_\infty$
category. In this realization, the D-branes of the theory are the
objects of $A$, while each morphism space $\Hom_{\cal A}(u,v)$ is the
space of zero-form topological observables for a string stretching from $u$ to
$v$. The bilinear forms $\rho_{uv}$ are the topological metrics, while
the $A_\infty$ units $1_u$ are the identity observables in the
boundary sectors $\Hom_{\cal A}(u,u)$.  The $A_\infty$ products arise
by dualizing the integrated correlators on the disk. These can be
recovered from the former with the aid of the topological metrics:
\be
\nn
\langle \langle x_0\ldots x_n\rangle \rangle =\rho_{h(x_0) t(x_0)}(x_0,r_{[x_1\ldots x_n]}(x_1\ldots x_n))~~,
\ee
for composable $(x_0\ldots x_n)$ such that $[x_0\ldots x_n]$ is a cyclic word. 
We refer the reader to \cite{HLL, CIL5} for further details.

As explained in \cite{HLL, CIL5}, non-vanishing maps $r_u$ are present when
the background of the topological string theory does not satisfy the string equations of motion. 
In this case, the elements $\theta_u=r_u(1)\in \Hom_{\cal A}(u,u)$ correspond to
tadpoles. This can usually be corrected by shifting the string vacuum
until all tadpoles are eliminated \cite{CIL5, HLL}. One sometimes has an obstruction to reaching a 
solution, a phenomenon which was originally noticed in \cite{Fukaya_book}. 

The structure described above is a homotopy-theoretic generalization
of the boundary data of a topological field theory defined on open
Riemann surfaces, which was studied in \cite{CIL1, Moore_Segal,
Moore}. The latter data arises by forgetting all integrated boundary
correlators, thereby keeping only the information contained in the
boundary units, topological metrics and three-point functions.

The fact that weak cyclic and unital $A_\infty$ categories describe
topological D-brane systems is implicit in the homological mirror
symmetry conjecture \cite{HMS} and in the work of Fukaya and
collaborators \cite{Fukaya_inf, Fukaya_book, Fukaya_mirror}.  The
physical interpretation of this was discussed in \cite{CIL5}, the dG
case having been considered previously in \cite{CIL2, CIL3, CIL6}. A
connection with the D-brane superpotential and Chern-Simons theory was
made in \cite{CIL4, CIL5}, realizing explicitly an observation
originally due to \cite{Witten_CS}.  See \cite{CIL7,CIL8,CIL9,CIL10}
for further physics discussion. $A_\infty$ algebras as descriptions of
open string vertices appeared originally in \cite{Gaberdiel} in the
context of bosonic string field theory, which was further studied in
\cite{Kajiura}. As discussed in \cite{CIL4, CIL5, Kajiura}, the
$A_\infty$ structure describing open topological strings can also be
obtained from the results of \cite{Gaberdiel}, by using the well-known
formal analogy of bosonic and topological string theories.

\section{Some basic isomorphisms}
\label{isomorphisms}

Convention (\ref{dual_multiplications}) for the superbimodule structure of the dual 
affects some standard isomorphisms. Given two $R$-superbimodules $U,V$, the 
supervector space $\Hom_{\rm Mod-R}(U^{opp}, V)$ becomes an $R$- superbimodule
with respect to the external multiplications given by $(\alpha\phi\beta)(x)=\alpha\phi(x\beta)$. 
There exists an isomorphism of $R$-superbimodules 
$\Hom_{\rm Mod-R}(U^{opp}, V) \approx V\otimes_R U^{\rm v}$, whose inverse takes 
$y\otimes_R f\in V\otimes_R U^{\rm v}$
into the right $R$-supermodule morphism $\phi:U^{opp}\rightarrow V$ given by 
$\phi(x):=yf(x)$. This isomorphism maps the $R$-sub-bimodule $\Hom_{\rm R-Mod-R}(U^{opp},V)$
of $\Hom_{\rm Mod-R}(U^{opp}, V)$ into the center of $V\otimes_R U^{\rm v}$:
\be
\label{basic_isomorphism}
\iHom(U^{opp},V)=\Hom_{\rm R-Mod-R}(U^{opp},V)\approx [V\otimes_R U^{\rm v}]^R~~.
\ee
In particular, we have $\iHom(U^{opp},U)\approx [U^{\rm v}\otimes_R U^{\rm v}]^R$, so 
given a bilinear form $\sigma$ on $U$, the map $j_\sigma:U^{opp}\rightarrow U^{\rm v}$ defined in 
Section 2 can be identified with a central element ${\hat \sigma}\in [U^{\rm v}\otimes_R U^{\rm v}]^R$. 
Tracing through the identifications, one finds that $\sigma$ can be recovered from ${\hat \sigma}$ 
as follows. If ${\hat \sigma}=\sum_{i} f_i\otimes_R g_i $ with $f_i,g_i\in U^{\rm v}$, then
$\sigma(x,y)=\sum_{i}f_i(xg_i(y))$.

\section{Explicit construction of the differential envelope}
\label{envelope} 
The construction below is a slight adaptation of that given \cite{CK}. 
Let $A$ be an $R$-superalgebra.
\bd
Consider the $R$-superbimodule $A_R:=A/R$ and let $\nu:A\rightarrow
A_R$ be the natural projection.  We let $\nu(a):={\bar a}$ for all
$a\in A$ and let $K_R =A\oplus A_R $, viewed as an $R$-superbimodule.
Define:
\be
\Omega_R A=T_R K_R /J
\ee  
where $T_R K$ (endowed with its obvious $\Z\times \Z_2$ grading) is
the tensor algebra on $K$ and $J\subset T_R K_R $ is the $\Z\times
\Z_2$-homogeneous two-sided ideal generated by all elements of the
form ${\overline a}\otimes b+a\otimes {\overline b}-\overline{ab}$ and
$a\otimes b-ab$ with $a,b\in A$.  We let $d$ be the unique $R$-linear
derivation of $\Omega_R A$ of bidegree $(1,0)\in \Z\times \Z_2$ (with
respect to the pairing (\ref{pairing})) which satisfies the relations:
\be
da={\bar a}~~,~~d{\bar a}=0~~{\rm for~all~}a\in A~~
\ee
(this derivation is well-defined). The $\Z\times \Z_2$-grading of $\Omega_R A$ is induced by the obvious 
$\Z\times \Z_2$-grading of $T_R K_R$. 
\ed

\noindent If $\cdot$ denotes multiplication in $\Omega_R A$, we find
that $\Omega_R A$ is spanned by all finite products of elements of the
form $a$ and $db={\bar b}$ with $a,b\in A$, with the relations
$d(ab)=(da)\cdot b+a\cdot (db)$ and $a\cdot b=ab$
Notice that the unit element of $\Omega_R A$ coincides with $1_A$ due to
the relation $1_A\cdot a=1_A a=a=a1_A=a\cdot 1_A$ for $a\in A$, and
that $d$ squares to zero due to the relations $d{\bar
a}=0\Leftrightarrow d^2a=0$ for $a\in A$. Moreover, any element of
$\Omega_ A$ can be written as a finite sum of elements of the form
$a_0\cdot da_1\cdot \ldots \cdot da_n$ with $a_i\in A$, by applying
the identity:
\begin{eqnarray}
a_0 \cdot{\bar a}_1\cdot\ldots \cdot{\bar a}_n\cdot b_0 \cdot{\bar
b}_1\cdot \ldots \cdot{\bar b}_m&:=&a_0 \cdot{\bar a}_1 \cdot \ldots
{\bar a}_{n-1}\cdot\overline{a_n b_0}\cdot~{\bar b}_1\cdot\ldots
\cdot{\bar b}_m\nn\\ &&+\sum_{i=1}^{n-1}{(-1)^{n-i}a_0 \cdot{\bar
a}_1\cdot\ldots \cdot\overline{a_{n-i}a_{n-i+1}} \cdot\ldots \cdot
{\bar a}_n\cdot {\bar b}_0}\cdot\ldots \cdot{\bar b}_m \nn\\ &&+(-1)^n
(a_0a_1) \cdot{\bar a}_2 \cdot\ldots \cdot{\bar a}_n \cdot{\bar
b}_0\cdot \ldots \cdot{\bar b}_m~~\nn\\ ~~{\rm for~all}~~ a_0\ldots
a_n, b_0\ldots b_m\in A~~, \nn
\end{eqnarray}
which follows from the relations valid in $\Omega_R A$. Furthermore, we
can denote the product in $\Omega_R A$ by juxtaposition.

\section{Coefficient expression for the cyclic bracket}
\label{coeffs}

\bp
\label{cycbracket}
Let $f, g\in C^0_R(A)$ have the forms: 
\begin{eqnarray}
f&=&\sum_{n\geq 0}{f_{a_1\ldots a_n}({a_1}\ldots {a_n})_c}=c(f)+
\sum_{n\geq 1}{\frac{{\bar f}_{a_1\ldots a_n}}{n}({a_1}\ldots {a_n})_c}
~~\nn\\
g&=&\sum_{n\geq 0}{g_{a_1\ldots a_n}({a_1}\ldots {a_n})_c}= 
c(g)+
\sum_{n\geq 1}{\frac{{\bar g}_{a_1\ldots a_n}}{n}({a_1}\ldots {a_n})_c}~~\nn
\end{eqnarray}
with strict coefficients ${\bar f}_{a_1\ldots a_n}$, ${\bar g}_{a_1\ldots a_n}$
and assume that both $f$ and $g$ are homogeneous of degree
${\tilde \omega}+1\in \Z_2$.  Then the cyclic bracket of $f$ and $g$: 
\be
\{f,g\}=c(\{f,g\})+\sum_{n\geq 1}{\frac{{\bar \phi}_{a_1\ldots a_n}}{n} ({a_1}\ldots {a_n})_c}~~,\nn
\ee
has strict coefficients:
\begin{eqnarray}
\label{strict_bracket}
{\bar \phi}_{a_1\ldots a_n}=\sum_{j=0}^n \sum_{i=0}^{n-j} (-1)^{{\tilde a}_1+\ldots + {\tilde a}_i}
&[&{\bar f}_{a_1\ldots a_i a a_{i+j+1}\ldots a_n} {\bar g}^a_{\,\,a_{i+1}\ldots
      a_{i+j}}\nn\\
&+&{\bar g}_{a_1\ldots a_i a a_{i+j+1}\ldots a_n}{\bar f}^a_{\,\,
      a_{i+1}\ldots a_{i+j}}]~~.
\end{eqnarray}
and: 
\be
\label{c}
c(\{f,g\}):={\bar f}_a{\bar g}^a={\bar f}^a{\bar g}_a~~,
\ee
where we lift coefficients with $\rho^{ab}$ as in (\ref{lifts}).
\ep

\noindent With our conventions, expression (\ref{c}) can be obtained by formally setting 
$n=0$ in equation (\ref{strict_bracket})) and dividing by two. The second equality in (\ref{c}) follows upon using the 
symmetry property of $\rho^{ab}$ and the selection rule ${\tilde a}={\tilde \omega}+1$.

\begin{proof}
By Proposition \ref{coord_bracket}, we have:
\be
\{f,g\}=\sum_{n\geq 0}\sum_{j=0}^n F^{(j)}_{(a_1\ldots a_n)}({a_1}\ldots {a_n})_c\nn
\ee
where: 
\be
F^{(j)}_{a_1\ldots a_n}={\bar f}_{a_1\ldots a_jb}\omega^{ba}{\bar g}_{a a_{j+1}\ldots a_n}~~.\nn
\ee
The case $n=0$ can be checked directly, so we discuss only the case $n\geq 1$. 

\noindent If $i+j\leq n$, we compute: 
\be
F^{(j)}_{a_{i+1}\ldots a_n, a_1\ldots a_i}={\bar f}_{a_{i+1}\ldots a_{i+j}b}\omega^{ba}
{\bar g}_{aa_{i+j+1}\ldots a_na_1\ldots a_i}=
(-1)^{\sigma_1} {\bar g}_{a_1\ldots a_i a a_{i+j+1}\ldots a_n}{\bar f}^a_{\,\, a_{i+1}\ldots a_{i+j}}~~,\nn
\ee
where we used cyclicity of $f$ and $g$ to permute indices. The sign factor is easily determined 
keeping in mind the symmetry property (\ref{oinv_symm}) and the selection rule (\ref{oinv_sel}):
\be
\sigma_1={\tilde \omega}({\tilde a}+1)+({\tilde a}_1+\ldots +{\tilde a}_i)({\tilde a}+
{\tilde a}_{i+j+1}+\ldots +{\tilde a}_n)+({\tilde a}+{\tilde \omega})({\tilde a}_{i+1}+\ldots 
+{\tilde a}_{i+j})~~({\rm mod}~2)~~.\nn
\ee
For $i+j>n$, we find:
\be
F^{(j)}_{a_{i+1}\ldots a_n, a_1\ldots a_i}=
(-1)^{{\tilde a}+{\tilde \omega}+1}{\bar f}_{a_{i+1}\ldots a_na_1\ldots a_{j-n+i}a}{\bar g}^a_{\,\,a_{j-n+i+1}\ldots a_i}=
(-1)^{\sigma_2}{\bar f}_{a_1\ldots a_{j-n+i}aa_{i+1}\ldots a_n}{\bar g}^a_{\,\,a_{j-n+i+1}\ldots a_i}~~,\nn
\ee
with:
\be
\sigma_2={\tilde a}+{\tilde \omega}+1+
({\tilde a}_{i+1}+\ldots +{\tilde a}_n)({\tilde a}+{\tilde a}_1+\ldots +{\tilde a}_{j-n+i})
~~({\rm mod}~2)~~.\nn
\ee

\noindent This allows us to write: 
\be
\label{two_sums}
\sum_{j=0}^n F^{(j)}_{(a_1\ldots a_n)}=\frac{1}{n}\left[\sum_{j=0}^n\sum_{i=1}^{n-j}
(-1)^{\epsilon_1} F^{(j)}_{a_{i+1}\ldots a_n, a_1\ldots a_i}+ \sum_{j=0}^n\sum_{i=n-j+1}^{n}
(-1)^{\epsilon_2} F^{(j)}_{a_{i+1}\ldots a_n, a_1\ldots a_i}\right]\nn
\ee
where:
\begin{eqnarray}
\epsilon_1&=&{\tilde a}+{\tilde \omega}+
({\tilde a}+{\tilde \omega}+{\tilde a}_1+\ldots +{\tilde a}_i)({\tilde a}+{\tilde a}_{i+1}+\ldots 
+{\tilde a}_{i+j})\nn\\
\epsilon_2&=& 1+{\tilde a}+{\tilde \omega}+
({\tilde a}_{i+1}+\ldots +{\tilde a}_n)({\tilde a}+{\tilde a}_{j-n+i+1}+\ldots +{\tilde a}_i)~~.\nn
\end{eqnarray} 
We now perform the replacement $i\rightarrow i'=i+j-n, j\rightarrow j'=n-j$ in the second 
double sum. Denoting the new summation indices $i',j'$ by $i,j$, this gives: 
\be
\sum_{j=0}^n F^{(j)}_{(a_1\ldots a_n)}=\frac{1}{n}\sum_{j=0}^n\sum_{i=1}^{n-j}\left[(-1)^{\epsilon_1}
{\bar g}_{a_1\ldots a_i a a_{i+j+1}\ldots a_n}{\bar f}^a_{\,\, a_{i+1}\ldots a_{i+j}}+(-1)^{\tilde \epsilon_2}
{\bar f}_{a_1\ldots a_i a a_{i+j+1}\ldots a_n}{\bar g}^a_{\,\, a_{i+1}\ldots a_{i+j}}\right]~~,\nn
\ee
where:
\be
{\tilde \epsilon}_2=1+{\tilde a}+{\tilde \omega}+({\tilde a}+{\tilde a}_{i+1}+\ldots +{\tilde a}_{i+j})({\tilde a}_{i+j+1}+
\ldots +{\tilde a}_n)~~.\nn
\ee
The next step is to notice that the first term in square brackets vanishes unless: 
\be
{\tilde a}+{\tilde a}_1+\ldots +{\tilde a}_i={\tilde g}+{\tilde
  a}_{i+j+1}+\ldots +{\tilde a}_n ~~({\rm mod}~2)\nn
\ee
and: 
\be
{\tilde a}+{\tilde a}_{i+1}+\ldots +{\tilde a}_{i+j}={\tilde f}+{\tilde \omega}~~ 
~~({\rm mod}~2)~~,\nn
\ee
which allows us to replace $\epsilon_1$ by:
\be
\epsilon_1'={\tilde a}+{\tilde \omega}+
({\tilde f}+{\tilde \omega})({\tilde g}+{\tilde \omega}+{\tilde a}_{i+j+1}+\ldots +
{\tilde a}_n)~~.\nn
\ee
Similarly, the second term in the square brackets vanishes unless: 
\be
{\tilde a}+{\tilde a}_{i+1}+\ldots +{\tilde a}_{i+j}={\tilde g}+{\tilde \omega}~~,\nn
\ee
which allows us to replace ${\tilde \epsilon}_2$ by: 
\be
\epsilon_2'={\tilde a}+ {\tilde \omega}+1+({\tilde g}+{\tilde \omega})({\tilde a}_{i+j+1}+\ldots +
{\tilde a}_n)~~.\nn
\ee
Finally, using the assumption ${\tilde f}={\tilde g}={\tilde \omega}+1$, we find: 
\be
\epsilon'_1=\epsilon_2'=1+{\tilde \omega}+{\tilde a}+{\tilde a}_{i+j+1}+\ldots +{\tilde a}_n={\tilde a}_1+\ldots +
{\tilde a}_i~~({\rm mod}~2)~~,\nn
\ee
where we used the selection rule ${\tilde a}_1+\ldots +{\tilde a}_i+{\tilde a}+
{\tilde a}_{i+j+1}+\ldots 
+{\tilde a}_n={\tilde f}={\tilde g}={\tilde \omega}+1$. 

\end{proof}

\paragraph{Corollary}
\label{maincor}
Let $f=c(f)+\sum_{n\geq 0}{\frac{{\bar f}_{a_1\ldots a_n}}{n}({a_1}\ldots {a_n}})_c$ be a 
cyclic element of $A$ of degree 
${\tilde \omega}+1$. Then: 
\be
\frac{1}{2}\{f,f\}=\frac{1}{2}{\bar f}_a{\bar f}^a+
\sum_{n\geq 1}\frac{1}{n}\left(\sum_{0\leq i+j\leq n}(-1)^{{\tilde a_1}+\ldots +{\tilde a}_i}
{\bar f}_{a_1\ldots a_i a a_{i+j+1}\ldots a_n} {\bar f}^a_{\,\,a_{i+1}\ldots
      a_{i+j}}\right)({a_1}\ldots {a_n})_c~~.\nn
\ee

%%%%%%%%%%%%%%%%%%%%%%%%%%%%%%%%%%%%%%%%%%%%%

\end{document}